\documentclass[msom,nonblindrev]{informs3}

\OneAndAHalfSpacedXI

\usepackage{natbib}
 \bibpunct[, ]{(}{)}{,}{a}{}{,}%
 \def\bibfont{\small}%

\usepackage[utf8]{inputenc} 
\usepackage[T1]{fontenc}    
\usepackage{hyperref}       
\usepackage{url}            
\usepackage{booktabs}       
\usepackage{amsfonts}       
\usepackage{nicefrac}       
\usepackage{microtype} 
\usepackage{enumerate}
\usepackage{comment} 
\newcommand{\N}{\mathbb{N}}

\newcommand{\I}{\mathbb{I}}
\newcommand{\R}{\mathbb{R}}
\newcommand{\prob}{\mathbb{P}}

\newcommand{\E}{\mathbb{E}}
\newcommand{\ucb}{\texttt{UCB}}

\usepackage{bm}

\usepackage{amsmath,amssymb}
\usepackage{array}

\usepackage[textsize=tiny]{todonotes}

\newcommand{\safe}{\beta}
\newcommand{\safebudg}{\underline{\rho}}

\newcommand{\totv}{V}
\newcommand{\totc}{D}

\usepackage{algorithm,algpseudocode}

\makeatletter
\newcounter{phase}[algorithm]
\newlength{\phaserulewidth}
\newcommand{\setphaserulewidth}{\setlength{\phaserulewidth}}

\makeatother

\setphaserulewidth{.7pt}

\usepackage{amsmath,amssymb,amsfonts}
\usepackage{enumitem}
\usepackage{bm}
\usepackage{thmtools,thm-restate}
\usepackage{cleveref}
\usepackage{subcaption}
\usepackage{graphicx}

\newenvironment{itquote}
  {\begin{quote}\itshape}
  {\end{quote}\ignorespacesafterend}

\def\dualub{C}

\def\ucbtot{R}

\def\cont{\bm{c}}
\def\discerr{\delta}
\def\contset{\mathcal{C}}
\def\budgset{\mathcal{A}}
\def\budg{B}
\def\roi{R}

\def\slope{\sigma}
\def\adj{\mathcal{U}}

\newcommand{\cond}[1]{\left. \right|}

\newtheorem{theorem}{Theorem}[section]
\newtheorem{corollary}[theorem]{Corollary}
\newtheorem{lemma}[theorem]{Lemma}
\newtheorem{remark}{Remark}[section]
\newtheorem{proposition}[theorem]{Proposition}
\newtheorem{definition}{Definition}[section]
\newtheorem{example}{Example}[section]
\newtheorem{assumption}{Assumption}[section]
\newtheorem{claim}{Claim}[section]

\begin{document}

\TITLE{Multi-channel Autobidding with Budget and ROI Constraints}
\ARTICLEAUTHORS{%
\AUTHOR{Yuan Deng}
\AFF{Google Research, \EMAIL{dengyuan@google.com} \URL{}}
\AUTHOR{Negin Golrezaei}
\AFF{Sloan School of Management, Massachusetts Institute of Technology, \EMAIL{golrezae@mit.edu} \URL{}}
\AUTHOR{Patrick Jaillet}
\AFF{Department of Electrical Engineering and Computer Science, Massachusetts Institute of Technology, \EMAIL{jaillet@mit.edu} \URL{}}
\AUTHOR{Jason Cheuk Nam Liang}
\AFF{Operations Research Center, Massachusetts Institute of Technology, \EMAIL{jcnliang@mit.edu} \URL{}}
\AUTHOR{Vahab Mirrokni}
\AFF{Google Research, \EMAIL{mirrokni@google.com}\URL{}}
} 

\ABSTRACT{
In digital online advertising, advertisers procure ad impressions simultaneously on multiple  platforms, or so-called \textit{channels}, such as Google Ads, Meta Ads Manager, etc. We study how an advertiser maximizes their total conversion (e.g. ad clicks) while satisfying aggregate \textit{return-on-investment (ROI)} and budget constraints across all channels. In practice, an advertiser does not have control over, and thus cannot globally optimize, which individual ad auctions she participates in for each channel, and instead authorizes a channel to procure impressions on her behalf: the advertiser can only utilize two levers on each channel, namely setting a per-channel budget and per-channel target ROI. In this work, we first analyze the effectiveness of each of these levers for solving the advertiser's global multi-channel problem. We show that when an advertiser only optimizes over per-channel ROIs, her total conversion  can be arbitrarily worse than what she could have obtained in the global problem. Further, we show that the advertiser can achieve the global optimal  conversion when she only optimizes over per-channel budgets. In light of this finding, under a bandit feedback setting that mimics real-world scenarios where advertisers have limited information on ad auctions in each channels and how channels procure ads, we present an efficient learning algorithm that produces per-channel budgets whose resulting conversion approximates that of the global optimal problem.   
Finally, we conduct numerical studies to demonstrate that our proposed algorithm accurately approximates optimal per-channel budgets in practical setups. }
\maketitle

\section{Introduction}

In today's online advertisers world, advertisers (including but not limited to small businesses,  marketing practitioners, non-profits, etc) have been embracing an expanding array of advertising platforms such as search engines, social media platforms, web publisher display etc. which present a plenitude of channels for advertisers to procure ad impressions and obtain traffic. In this growing multi-channel environment, the booming online advertising activities have fueled extensive research and technological advancements in \textit{attribution analytics} to answer questions like which channels are more effective in targeting certain users? Or, which channels produce more user conversion (e.g. ad clicks)  or  \textit{return-on-investment} (ROI) with the same amount of investments? (see \cite{kannan2016path} for a comprehensive survey on attribution analytics). Yet, this area of research has largely left out a crucial phase in the workflow of advertisers' creation of a digital ad campaign, namely how advertisers interact with advertising channels, which is the physical starting point of a campaign. 

\begin{figure}[h]
\includegraphics[width=\textwidth]{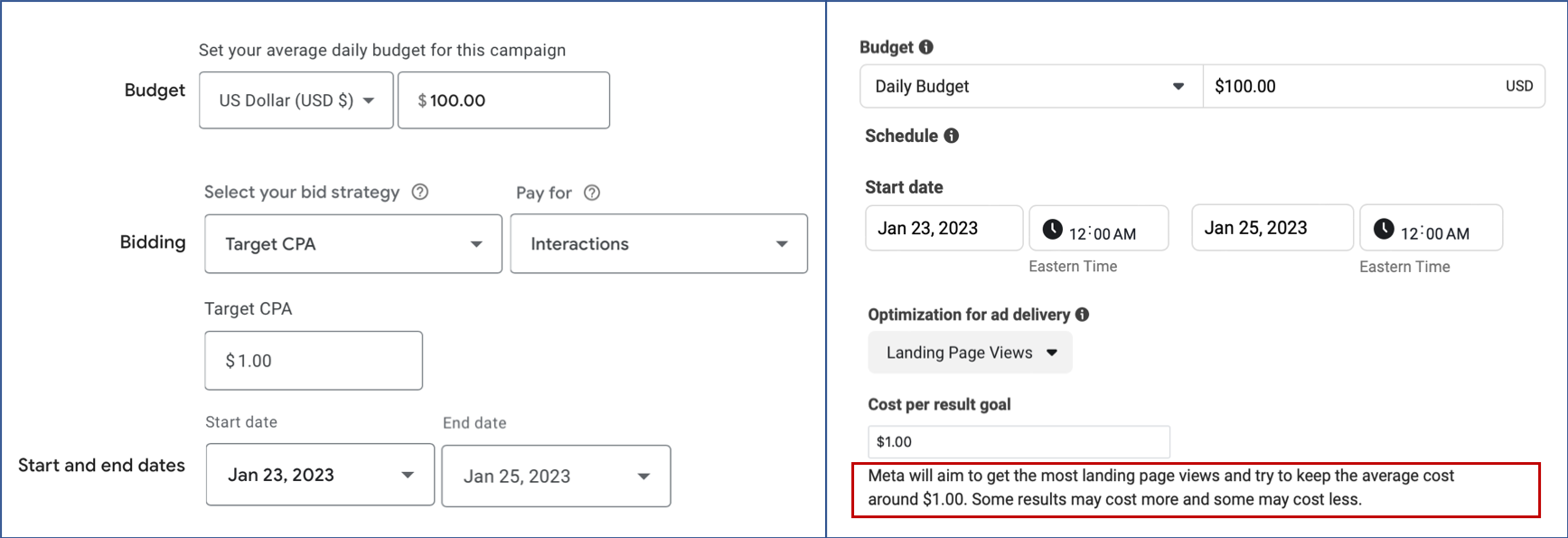}
\caption{Interfaces on Google Ads (left) and Meta Ads Manager (right) for creating advertising campaigns that allow advertisers to set budgets, target ROIs, and campaign duration. CPA, or cost per action on  Google Ads, as well as cost per result goal on Meta Ads Manager, is effectively the inverse value for an advertiser's per-channel target ROI. Meta Ads Manager specifically highlights that the impression procurement methodology via autobidding maximizes total conversion while respecting advertisers' per-channel target ROI (see red box highlighted), providing evidence that supports the \textsc{GL-OPT} and \textsc{CH-OPT} models in Eqs. \eqref{eq:glopt} and  \eqref{eq:chopt}, respectively.}
\label{fig:manager}
\end{figure}

To illustrate the significance of advertiser-channel interactions, consider for example a small business who is relatively well-informed through attribution research that Google Ads and Meta ads are the two most effective channels for its products. The business instantiates its ad campaigns through interacting with the platforms' ad management interfaces (see Figure \ref{fig:manager}), on which the business utilizes levers such as specifying budget and a target ROI\footnote{Target ROI is the numerical inverse of  CPA or cost per action on Google Ads, and cost per result goal in Meta Ads.} to control campaigns. Channels then input these specified parameters into their \textit{autobidding} procedures, where they procure impressions on the advertiser's behalf through automated blackbox algorithms. Eventually, channels report performance metrics such as expenditure and conversion back to the advertiser once the campaign ends. Therefore, one of the most important decisions advertisers need to make involves how to optimize over these levers provided by channels. Unfortunately, this has rarely been addressed in attribution analytics and relevant literature. Hence, this works contributes to filling this vacancy by addressing two themes of practical significance:
\begin{itquote}
How effective are these channel levers for advertisers to achieve their conversion goals? And how should advertisers optimize decisions for such levers?
\end{itquote}
To answer these questions, we study a setting where an advertiser simultaneously procures ads on multiple channels, each of which consists of multiple ad auctions that sell ad impressions. The advertiser's \emph{global optimization problem} is to maximize total conversion over all channels, while respecting a global budget constraint that limits total spend, and a global ROI constraint that ensures total conversion is at least the target ROI times total spend. However,  channels operate as independent entities and conduct autobidding procurement on behalf of advertisers, thereby there are no realistic means for an advertiser to implement the global optimization problem via optimizing over individual auctions. Instead, advertisers can only use two levers, namely a per-channel ROI and per-channel budget,  
to influence how channels should autobid for impressions. Our goal is to understand how effective  these levers are by comparing the total conversion via optimizing levers with the globally optimal conversion, and also present methodologies to help advertisers optimize over the usage of these levers. We summarize our contributions as followed:\\
\subsection{Main contributions}
\paragraph{1. Modelling ad procurement through per-channel ROI and budget levers.} In Section \ref{sec:model} we develop a novel model for online advertisers to optimize over the per-channel ROI and budget levers to maximize total conversion over channels while respecting a global ROI and budget constraint. This multi-channel optimization model closely imitates real-world practices (see Figure \ref{fig:manager} for evidence), and to the best of our knowledge is  the first of its kind to characterize advertisers' interactions with channels to run ad campaigns. 
\paragraph{2. Solely optimizing per-channel budgets  are sufficient to maximize conversion.} In Theorem \ref{thm:roifail} of Section \ref{sec:futil}, we show that solely optimizing for per-channel ROIs is inadequate to optimize conversion across all channels, possibly resulting in arbitrary worse total conversions compared to the hypothetical global optimal where advertisers can optimize over individual auctions. In contrast, in Theorem \ref{thm:budgonlysuff} and Corollary \ref{cor:roiredund} we show that solely optimizing for per-channel budgets allows an advertiser to achieve the global optimal.
\paragraph{3. Algorithm to optimize per-channel budget levers.} Under a realistic bandit feedback structure where advertisers can only observe the total conversion and spend in each channel after making a per-channel budget decision, in Section \ref{sec:online}, we develop an algorithm that augments stochastic gradient descent (SGD) with the upper-confidence bound (UCB) algorithm, and eventually outputs within $T$ iterations a per-channel budget profile with which advertisers can achieve $\mathcal{O}(T^{-1/3})$ approximation accuracy in total conversion to that of the optimal per-channel budget profile.  Our algorithm relates to constrained convex optimization with uncertain constraints and bandit feedback under a ``one point estimation'' regime, and to the best of our knowledge, our proposed algorithm is the first to handle such a setting; see more discussions in Section \ref{subsec:litrev} and Remark \ref{rmk:onepoint} of Section \ref{sec:online}. 

\paragraph{4. Extensions to general advertiser objectives and mutli-impression auctions.} In Sections \ref{sec:multi} and \ref{sec:extend}, we shed light on the applicability of our results in Sections \ref{sec:futil} and \ref{sec:online} to more general settings  when auctions correspond to the sale of multiple auctions, or when advertisers aim to optimize a private cost model instead of conversion.

{
\paragraph{5. Numerical studies.} In Section \ref{sec:num}, we conduct numerical studies to demonstrate that our proposed
algorithm accurately approximates optimal per-channel budgets even with a moderately small number of
data points, and that its performance gracefully deteriorates when channels do not optimally procure ads on
advertisers’ behalf.
}

\subsection{Related works.}
\label{subsec:litrev}
Generally speaking, our work focuses on advertisers' impression procurement process or the interactions between advertisers and impression sellers, which has been addressed in a vast amount of literature in mechanism design and online learning; see e.g. \cite{braverman2018selling,deng2019prior,golrezaei2019dynamic,golrezaei2019incentive,balseiro2019black, golrezaei2021bidding} to name a few. Here, we review literature that relate to key themes of this work, namely autobidding, budget and ROI management, and constrained optimization with bandit feedback. 
\paragraph{Autobidding. } There has been a rich line of research that model the autobidding setup as well as budget and ROI management strategies. The autobidding model has been formally developed in \cite{aggarwal2019autobidding}, and has been analyzed through the lens of welfare efficiency or price of anarchy in \cite{deng2021towards,balseiro2021robust,deng2022efficiency,mehta2022auction}, as well as individual advertiser welfare in \cite{deng2022fairness}. The autobidding model has also been compared to classic quasi-linear utility models in \cite{balseiro2021landscape}. The autobidding model considered in these papers assume advertisers can directly optimize over individual auctions, whereas in this work we address a more realistic setting that mimics practice where advertisers can only use levers provided by channels, and let channels procure ads on their behalf. Finally, a recent paper \cite{alimohammadi2023incentive} studies whether advertisers have incentive to misreport their target ROIs or budgets to a single autobidding platforms, whereas in this paper we optimize over per-channel budget decisions over multiple channels.

\paragraph{Budget and ROI management.} 
Budget and ROI management strategies have been widely studied in the context of mechanism design and online learning. \cite{balseiro2017budget} studies the ``system equilibria'' of a range of budget management strategies in terms of the platforms’ profits and advertisers’ utility; \cite{balseiro2019learning,balseiro2022best} study online bidding algorithms (called pacing) that help advertisers achieve high utility in repeated second-price auctions while maintaining a budget constraint, whereas \cite{feng2022online} studies similar algorithms but considers respecting a long term ROI constraint in addition to a fixed budget. On the other hand, there has been a recent line of work that study the setting where multiple budget or ROI constrained bidders run pacing-type algorithms, and analyze time-average welfare guarantees among all bidders \cite{gaitonde2022budget,lucier2023autobidders}.
All of these works on budget and ROI management focus on bidding strategies in a single repeated auction where advertisers' decisions are bid values submitted directly to the auctions. In contrast, this work focuses on the setting where advertisers procure ads from multiple auctions through channels, and make decisions on how to adjust the per-channel ROI and budget levers while leaving the bidding to channels' blackbox algorithms.

\paragraph{Online optimization. }
Section \ref{sec:online} where we develop an  algorithm to optimize over per-channel target ROI and budgets relates to the area of convex constrained optimization with bandit feedback (also referred to as zero-order or gradient-less feedback) since in light of Lemma \ref{lem:Vstruct} in Section \ref{sec:online} our problem of interest is also constrained and convex. First, there has been a plenitude of algorithms developed for deterministic constrained convex optimization under a bandit feedback structures where function evaluations for the objective and constraints are non-stochastic. Such algorithms include filter methods
\cite{audet2004pattern,pourmohamad2020statistical}, barrier-type methods \cite{fasano2014linesearch, dzahini2022constrained}, as well as Nelder-Mead type algorithms \cite{bHurmen2006grid,audet2018mesh}; see \cite{nguyen2022stochastic} and references therein for a comprehensive survey. In contrast to these works, our optimization algorithm developed in Section \ref{sec:online} handles 
noisy bandit feedback.

Regarding works that also address stochastic settings,  \cite{flaxman2004online} presents online optimization algorithms under the \textit{known constraint} regime, which assumes the optimizer can evaluate whether all constraints are satisfied, i.e. constraints are analytically available. Further, the algorithm achieves a $\mathcal{O}(T^{-1/4})$ accuracy. In this work, our setting is more complex as the optimizer (i.e. the advertiser) cannot tell whether the ROI constrained is satisfied (due to unknown value and cost distributions in each channels' auctions). Yet our proposed algorithm can still achieve a more superior $\mathcal{O}(T^{-1/3})$ accuracy. 

Most relevant to this paper is the very recent works \cite{usmanova2019safe,nguyen2022stochastic}, which considers a similar setting to ours that optimizes for a constrained optimization problem where the objective and constraints are only available through noisy function value evaluations (i.e. unknown constraints). \cite{usmanova2019safe} focuses on a special (unknown) linear constraint setting, and  \cite{nguyen2022stochastic} extends to general convex constraints. Although \cite{usmanova2019safe} and \cite{nguyen2022stochastic} achieve  $\mathcal{O}(T^{-1})$ and $\mathcal{O}(T^{-1/2})$ approximation accuracy to the optimal solution which contrasts our $\mathcal{O}(T^{-1/3})$ accuracy, these works imposes several assumptions that are stronger than the ones that we consider. First, the objective and constraint functions are strongly smooth (i.e. the gradients are Lipschitz continuous) and \cite{nguyen2022stochastic} further assume strong convexity. But in our work,  our objectives and constraints are piece-wise linear and do not satisfy such salient properties.  Second, and most importantly, both works consider a setting with ``two point estimations'' that allows the optimizer to access the objective and constraint function values twice in each iteration, enabling more efficient estimations. This work, however, lies in the one-point setting where we can only access function values once per iteration. Finally, we remark that the optimal accuracy/oracle complexity for the one-point setting for constrained (non-smooth) convex optimization with bandit feedback and unknown constraints remains an open question; see  Remark \ref{rmk:onepoint} in Section \ref{sec:online} for more details. We refer readers to Table 4.1 in \cite{larson2019derivative} for a survey on best known bounds under different one-point bandit feedback settings.

\section{Preliminaries}
\label{sec:model}
\textbf{Advertisers'  global optimization problem.} Consider an advertiser running a digital ad campaign to procure ad impressions on $M \in \N$ platforms such as Google Ads, Meta Ads Manager etc., each of which we call a \textit{channel}. 
Each channel $j$ consists of $m_{j}\in \N$ parallel ad auctions, each of which corresponds to the sale of an ad impression.\footnote{Ad auctions for each channel may be run by the channel itself or other external ad inventory suppliers such as web publishers.} An ad auction $n\in [m_{j}]$ is associated with a value $v_{j,n}\geq 0$ that represents the expected conversion (e.g. number of clicks) of the impression on sale, and a cost $d_{j,n}\geq 0$ that is required for the purchase of the impression. For example,  the cost in a single slot second-price auction is the highest competing bid of competitors in the market, and in a posted price auction the cost is simply the posted price by the seller of the impression. Writing $\bm{v}_{j} = (v_{j,n})_{n\in[m_{j}]}$ and $\bm{d}_{j} = (d_{j,n})_{n\in[m_{j}]}$, we assume that $\bm{z}_{j}:= (\bm{v}_{j},\bm{d}_{j})$ is sampled from some discrete distribution $\bm{p}_{j}$ supported on some finite set $F_{j}\subseteq \R_{+}^{m_{j}} \times \R_{+}^{m_{j}}$.

The advertiser's goal is to maximize total conversion  of procured ad impressions, while subject to a \textit{ return-on-investment (ROI)} constraint that states total conversion across all channels is no less than $\gamma$ times total spend for some pre-specified target ROI $0 < \gamma < \infty$, as well as a budget constraint that states total spend over all channels is no greater than the total budget $\rho \geq 0$. Mathematically, the advertiser's \emph{global optimization problem} across all $M$ channels  can be written as:
\begin{align}
\label{eq:glopt}
\begin{aligned}
\textsc{GL-OPT} ~=~\max_{\bm{x}_{1},\ldots, \bm{x}_{M}} ~&~ \sum_{j\in [M]}\E\left[\bm{v}_{j}^{\top} \bm{x}_{j}\right] \\
     \text{s.t. } & \sum_{j\in [M]}\E\left[\bm{v}_{j}^{\top} \bm{x}_{j}\right]~\geq~ \gamma \sum_{j\in [M]} \E\left[\bm{d}_{j}^{\top} \bm{x}_{j}\right]\\
     & \sum_{j\in [M]} \E\left[\bm{d}_{j}^{\top} \bm{x}_{j}\right] ~\leq~\rho \\
     &\bm{x}_{j}\in [0,1]^{m_{j}} \quad j\in [M]\,.
    \end{aligned}
\end{align} 
Here, the decision variable $\bm{x}_{j} \in [0,1]^{m_{j}}$ is a vector where
$x_{j,n}$ denotes whether impression in auction $n$ for channel $j$ is procured. We remark that $\bm{x}$ depends on the realization of $\bm{z} = (\bm{v}_{j},\bm{d}_{j})_{j \in [M]}$ and is also random. We note that the ROI and budget constraints are taken in expectation because an advertiser procures impressions from a very large number of auctions (since the number of auctions in each platform is typically very large) and thus the advertiser only demands to satisfy constraints in an average sense. We note that \textsc{GL-OPT}  is a widely adopted formulation for autobidding practices in modern online advertising, which represents advertisers' conversion maximizing behavior while respecting certain financial targets for ROIs and budgets; see e.g. \cite{aggarwal2019autobidding,balseiro2021robust,deng2021towards,deng2022efficiency}. In Section \ref{subsec:genutil} we discuss more general advertiser objectives, e.g. maximizing quasi-linear utility.

 Our overarching goal of this work is to develop methodologies that enable an advertiser to achieve total campaign conversion that match \textsc{GL-OPT} while respecting her global ROI $\gamma$ and budget $\rho$. However, directly optimizing \textsc{GL-OPT} may not be plausible as we discuss in the following.

\textbf{Advertisers' levers to solve their global problems.} To solve the global optimization problem \textsc{GL-OPT}, ideally advertisers would like to optimize over individual auctions across all channels. However, in reality channels operate as independent entities, and typically do not provide means for general advertisers to participate in specific individual auctions at their discretion. Instead, channels provide advertisers with specific \emph{levers} to express their ad campaign goals on spend and conversion. In this work,  we focus on two of the  most widely used levers, namely the per-channel ROI target and per-channel budget (see illustration in Fig. \ref{fig:manager}). After an advertiser inputs these parameters to a channel, the channel then procures on behalf of the advertiser through autonomous programs (we call this programmatic process \textit{autobidding}) to help advertiser achieve procurement results that match with the inputs. We will elaborate on this process later.

Formally, we consider the setting where for each channel $j\in [M]$,  an advertiser is allowed to input a per-channel target ROI $0\leq \gamma_{j} < \infty$, and a per-channel budget $\rho_{j} \in [0,\rho]$ where we recall $\rho >0$ is the total advertiser budget for a certain campaign. Then, the channel uses these inputs in its autobidding algorithm to procure ads, and returns the total conversion $\totv_{j}(\gamma_{j},\rho_{j};\bm{z}_{j})\geq 0$ , as well as total spend $\totc_{j}(\gamma_{j},\rho_{j};\bm{z}_{j})\geq 0$ to the advertiser, where we recall $\bm{z}_{j} =  (\bm{v}_{j},\bm{d}_{j}) \in \R^{m_{j}}\times  \R^{m_{j}}$ is the vector of value-cost pairs in channel $j$ sampled from discrete support $F_{j}$ according to distribution $\bm{p}_{j}$; $\totv_{j}$ and $\totc_{j}$  will be further specified later.

As the advertiser has the freedom of choice to input either per-channel target  ROI's, budgets, or both, we consider three options for the advertiser: 1. input only a per-channel target ROI for each channel; 2. input only a per-channel budget for each channel; 3. input both per-channel target ROI and budgets for each channel. Such options correspond to the following decision sets for $(\gamma_{j},\rho_{j})_{j\in [M]}$:
\begin{align}
\begin{aligned}
\label{eq:option}
    & \textbf{Per-channel budget only option: } \mathcal{I}_{\budg} = \{(\gamma_{j},\rho_{j})_{j\in [M]}\in \R_{+}^{2\times M}: \gamma_{j} = 0, \rho_{j}\in [0,\rho]\text{ for } \forall j\}.\\
    & \textbf{Per-channel target ROI only option: } \mathcal{I}_{\roi} = \{(\gamma_{j},\rho_{j})_{j\in [M]} \in \R_{+}^{2\times M}: \gamma_{j}\geq 0, \rho_{j} = \infty \text{ for } \forall j\}. \\
    & \textbf{General option: }\mathcal{I}_{G} = \{(\gamma_{j},\rho_{j})_{j\in [M]}: \gamma_{j}\geq 0,\rho_{j}\in [0,\rho]\text{ for } \forall j\}.
\end{aligned}
\end{align}
The advertiser's goal in practice is to maximize their total conversion of procured ad impressions through optimizing over per-channel budgets and target ROIs, while subject to the global ROI and budget constraint similar to those in $\textsc{GL-OPT}$. Mathematically, for any option $\mathcal{I} \in\{\mathcal{I}_{\budg} ,\mathcal{I}_{\roi} ,\mathcal{I}_{G}\}$, the advertiser's optimization problem through channels can be written as
\begin{align}
\label{eq:chopt}
\begin{aligned}
   \textsc{CH-OPT}(\mathcal{I})=  \max_{(\gamma_{j},\rho_{j})_{j\in [M]}\in \mathcal{I}} ~&~ \sum_{j \in M} \E\left[\totv_{j}(\gamma_{j},\rho_{j};\bm{z}_{j})\right]\\
    \text{s.t. } &  \sum_{j \in M} \E\left[\totv_{j}(\gamma_{j},\rho_{j};\bm{z}_{j})\right] \geq \gamma \sum_{j \in M} \E\left[\totc_{j}(\gamma_{j},\rho_{j};\bm{z}_{j})\right]\\
   &   \sum_{j\in [M]}\E\left[\totc_{j}(\gamma_{j},\rho_{j};\bm{z}_{j})\right]\leq \rho\,,
    \end{aligned}
\end{align}
where the expectation is taken w.r.t. randomness in $\bm{z}_{j}$. We remark that for any channel $j\in [M]$, the number of auctions $m_{j}$ as well as the distribution $\bm{p}_{j}$ are fixed and not a function of  the input parameters  $\gamma_{j},\rho_{j}$.

The functions $(\totv_{j}, \totc_{j})$ that map 
per-channel target ROI and budgets $\gamma_{j},\rho_{j}$ to the total conversion and expenditure are specified by various factors including but not limited to channel $j$'s autobidding algorithms deployed to procure ads on advertisers' behalf as well as the auctions mechanisms that sell impressions. In this work, we study a general setup that closely mimics industry practices. We assume that on the behalf of the advertiser,  each channel aims to optimize their conversion over all $m_j$ auctions while respecting the advertiser's input (i.e., per-channel target ROI and budgets).  (See e.g. Meta Ads Manager in Figure \ref{fig:manager} specifically highlights the channel's autobidding procurement methodology provides evidence to support the aforementioned setup). Hence, each channel $j$'s optimization problem can be written as
\begin{align}
\begin{aligned}
\label{eq:budperchannelsol}
    \bm{x}_{j}^{*}(\gamma_{j},\rho_{j};\bm{z}_{j}) ~=~ \arg\max_{\bm{x}\in [0,1]^{m_{j}}} ~&~ \bm{v}_{j}^{\top} \bm{x}\quad 
     \text{s.t. }\quad   \bm{v}_{j}^{\top} \bm{x}\geq \gamma_{j}\bm{d}_{j}^{\top}\bm{x} ,\quad  \bm{d}_{j}^{\top} \bm{x}\leq \rho_{j}\,,
    \end{aligned}
\end{align} 
where $\bm{x} =(x_n)_{n\in[m_j]}\in [0,1]^{m_{j}}$ denotes the vector of probabilities to win each of the parallel auctions, i.e. $x_{n}\in [0,1]$ is the probability to win auction $n\in [m_{j}]$ in channel $j$. In light of this representation, the corresponding conversion and spend functions are given by
\begin{align}
\begin{aligned}
\label{eq:budperchannel}
    & \totv_{j}(\gamma_{j},\rho_{j};\bm{z}_{j}) = \bm{v}_{j}^{\top} \bm{x}_{j}^{*}(\gamma_{j},\rho_{j};\bm{z}_{j})\quad \text{ and }\quad \totv_{j}(\gamma_{j},\rho_{j}) = \E[\totv_{j}(\gamma_{j},\rho_{j};\bm{z}_{j})] 
    \\ 
    & 
    \totc_{j}(\gamma_{j},\rho_{j};\bm{z}_{j}) = \bm{d}_{j}^{\top} \bm{x}_{j}^{*}(\gamma_{j},\rho_{j};\bm{z}_{j})  \quad \text{ and }\quad  \totc_{j}(\gamma_{j},\rho_{j}) = \E[ \totc_{j}(\gamma_{j},\rho_{j};\bm{z}_{j})]
    \,.
\end{aligned}
\end{align}
Here, the expectation is taken w.r.t. randomness in $\bm{z}_{j} = (\bm{v}_{j},\bm{d}_{j}) \in \R_{+}^{m_{j}} \times\R_{+}^{m_{j}}$. We assume that for any $(\gamma_{j},\rho_{j})$ and realization $\bm{z}_{j}$, $ \totv_{j}(\gamma_{j},\rho_{j};\bm{z}_{j})$ is bounded above by some absolute constant $\Bar{\totv} \in (0,\infty)$ almost surely. We remark that Eq.\eqref{eq:budperchannel} assumes channels are able to achieve optimal procurement performance. Later in Section \ref{sec:num}, we conduct numerical studies to address
setups where channels does not optimally solve for Eq.\eqref{eq:budperchannelsol}, and in 
 Section \ref{subsec:nonoptautobid}, we will briefly discuss theoretical approaches to handle non-optimal autobidding.

\textbf{Key focuses and organization of this work.} In this paper, we address two key topics:
\begin{enumerate}
    \item How effective are the per-channel ROI and budget levers to help advertisers achieve the globally optimal conversion \textsc{GL-OPT} while respecting the global ROI and budget constraints? In particular, for each of the advertiser options $\mathcal{I} \in\{\mathcal{I}_{\budg} ,\mathcal{I}_{\roi} ,\mathcal{I}_{G}\}$ defined in Eq. \eqref{eq:option}, what is the discrepancy between $\textsc{CH-OPT}(\mathcal{I})$, i.e. the optimal conversion an advertiser can achieve in practice, versus the optimal $\textsc{GL-OPT}$? 
    \item Since in reality advertisers can only utilize the two per-channel levers offered by channels, how can advertisers optimize per-channel target ROIs and budgets to solve for $\textsc{CH-OPT}(\mathcal{I})$?
\end{enumerate}
In Section \ref{sec:futil}, we address the first question to determine the gap between $\textsc{CH-OPT}(\mathcal{I})$ and $\textsc{GL-OPT}$ for different advertiser options. In Section \ref{sec:online}, we develop an efficient algorithm to solve for per-channel levers that optimize $\textsc{CH-OPT}(\mathcal{I})$.

\section{On the efficacy of the per-channel target ROIs and budgets as levers in solving the global problem}
\label{sec:futil}

In this section,
 we examine the effectiveness of the per-channel target ROI and per-channel budget levers in achieving the global optimal  \textsc{GL-OPT}. In particular,  we study if the 
 optimal solution to the channel problem \textsc{CH-OPT}($\mathcal I$) defined in Eq. \eqref{eq:chopt} for $\mathcal I \in \{\mathcal{I}_{\budg},\mathcal{I}_{\roi},\mathcal{I}_{G}\}$ is equal to the  global optimal \textsc{GL-OPT}.  As a summary of our results, we show that the per-channel budget only option, and the general option achieves \textsc{GL-OPT}, but the per-channel ROI only option can yield conversion arbitrarily worse than \textsc{GL-OPT} for certain instance, even when there is no global budget constraint (i.e., $\rho = \infty$). This implies that the per-channel ROI lever is inadequate to help advertisers achieve the globally optimal conversion, whereas the per-channel budget lever is effective to attain optimal conversion even when the advertiser solely uses this lever.

Our first result in this section is the following Lemma \ref{lem:glbest} which shows that \textsc{GL-OPT} serves as a theoretical upper bound
for an advertiser's conversion through optimizing $\textsc{CH-OPT}(\mathcal{I})$ with any option $\mathcal{I}$. 
\begin{lemma}[\textsc{GL-OPT} is the theoretical upper bound for conversion]
\label{lem:glbest}
For any option $\mathcal{I} \in \{\mathcal{I}_{\budg},\mathcal{I}_{\roi},\mathcal{I}_{G}\}$ defined in Eq. \eqref{eq:option}, we have $\textsc{GL-OPT}\geq  \textsc{CH-OPT}(\mathcal{I})$, where we recall the definitions of  \textsc{GL-OPT} and  \textsc{CH-OPT} in Eq. \eqref{eq:glopt} and \eqref{eq:chopt}, respectively.  
\end{lemma}
The proof of Lemma \ref{lem:glbest} is deferred to Appendix \ref{pf:lem:glbest}. In light of the theoretical upper bound \textsc{GL-OPT}, we are now interested in the gap between \textsc{GL-OPT} and  $\textsc{CH-OPT}(\mathcal{I})$ for option
$\mathcal{I} \in \{\mathcal{I}_{\budg},\mathcal{I}_{\roi},\mathcal{I}_{G}\}$. In the following Theorem \ref{thm:roifail}, we show that there exists a problem instance under which the ratio $\frac{\textsc{CH-OPT}(\mathcal{I}_{\roi})}{\textsc{GL-OPT}}$ nears 0, implying the per-channel ROIs alone 
fail to help advertisers optimize conversion. The proof can be found in Appendix \ref{pf:thm:roifail}.
\begin{theorem}[Per-channel ROI only option fails to optimize conversion]
\label{thm:roifail}
Consider an advertiser with a  (global) target  ROI of $\gamma = 1$ procuring impressions from $M=2$ channels, where channel 1 consists of a single auction and channel 2 consists of two auctions. The advertiser has unlimited budget $\rho=\infty$,
and chooses the per-channel target ROI only option $\mathcal{I}_{\roi}$ defined in Eq. \eqref{eq:option}. Assume there is only one realization of value-cost pairs $\bm{z} = (\bm{v}_{j},\bm{d}_{j})_{j\in [M]}$ (i.e. the support $F = F_{1}\times F_{2}$ is a singleton), and the realization is presented in the following table, where $ X>0$ is some arbitrary parameter. Then, for this problem instance we have $\lim_{X\to \infty}\frac{\textsc{CH-OPT}(\mathcal{I}_{\roi})}{\textsc{GL-OPT}} = 0$.
\vspace{0.5cm}
\renewcommand{\arraystretch}{1.5}
\begin{center}
\begin{tabular}{|c | c| c c | } 
\hline
  & Channel 1 &    \multicolumn{2}{|c|}{Channel 2}\\
 \hline
  & Auction 1 & Auction 2   & Auction 3\\
 \hline\hline
 Value $v_{j,n}$ & 1 & $X$  & $2X$\\
 \hline
 Spend $d_{j,n}$ & 0  & $1+X$ & $2(1+X)$ \\
 \hline
\end{tabular}
\end{center}
\vspace{0.5cm}
\end{theorem}



In contrast to the per-channel ROI only option, the budget only option in fact allows an advertiser's conversion to reach the theoretical upper bound $\textsc{GL-OPT}$ through solely optimizing for per-channel budgets. This is formalized in the following theorem whose proof we present in Appendix \ref{pf:thm:budgonlysuff}.
\begin{theorem}[Per-channel budget only option suffices to achieve optimal conversion]
\label{thm:budgonlysuff}
 For the budget only option $\mathcal{I}_{\budg}$ defined in Eq.\eqref{eq:option}, we have    $  \textsc{GL-OPT}= \textsc{CH-OPT}(\mathcal{I}_{\budg})$ for any global target ROI $\gamma > 0$ and total budget $\rho > 0$, even for $\rho=  \infty$.
\end{theorem}
As an immediate extension of Theorem \ref{thm:budgonlysuff}, 
the following Corollary \ref{cor:roiredund} states  per-channel ROIs in fact become redundant once advertisers optimize for per-channel budgets.
\begin{corollary}[Redundancy of per-channel ROIs]
\label{cor:roiredund}
For the general option $\mathcal{I}_{G}$ defined in Eq.\eqref{eq:option} where an advertiser sets both per-channel ROI and budgets, we have  $\textsc{GL-OPT}= \textsc{CH-OPT}(\mathcal{I}_{G})$ for any aggregate ROI $\gamma > 0$ and total budget $\rho > 0$, even for $\rho= \infty$. Further, there exists and optimal solution $(\gamma_{j},\rho_{j})_{j\in [M]}$ to $\textsc{CH-OPT}(\mathcal{I}_{G})$, s.t. $\gamma_{j} = 0$ for all $j \in [M]$.
\end{corollary}

In light of the redundancy of per-channel ROIs  as illustrated in  Corollary \ref{cor:roiredund}, in the rest of the paper we will fix $\gamma_{j} = 0$ for any channel $j \in [M]$, and omit $\gamma_{j}$ in all relevant notations; e.g. we will write $ \totc_{j}(\rho_{j};\bm{z}_{j})$ and $ \totc_{j}(\rho_{j})$, instead of $ \totc_{j}(\gamma_{j},\rho_{j};\bm{z}_{j})$ and $ \totc_{j}(\gamma_{j},\rho_{j})$. Equivalently, we will only consider the per-channel budget only option $\mathcal{I}_{\budg}$. 

\section{Optimization algorithm for per-channel budgets under bandit feedback}
\label{sec:online}
In this section, we develop an efficient algorithm to solve for per-channel budgets that optimize
$\textsc{CH-OPT}(\mathcal{I}_{\budg})$ defined in Eq. \eqref{eq:chopt}, which achieves the theoretical optimal conversion, namely $\textsc{GL-OPT}$, as illustrated in Theorem \ref{thm:budgonlysuff}. In particular, we consider algorithms that run over $T > 0$ periods, where each period for example corresponds to the duration of 1 hour or 1 day. At the end of $T$ periods, the algorithm  produces some per-channel budget profile $(\rho_{j})_{j\in [M]}\in [0,\rho]^{M}$ that approximates $\textsc{CH-OPT}(\mathcal{I}_{\budg})$,  and satisfies aggregate ROI and budget constraints, namely 
\begin{align*}
\begin{aligned}
  \textstyle \sum_{j \in M} \totv_{j}(\rho_{j}) \geq \gamma \sum_{j \in M} \totc_{j}(\rho_{j}),~\sum_{j\in [M]}\totc_{j}(\rho_{j})\leq \rho\,,
 \end{aligned}
\end{align*}
where we recall $(\totv_{j}(\rho_{j}),\totc_{j}(\rho_{j}))$ are defined in Eq. \eqref{eq:budperchannel}.

The algorithm proceeds as follows: at the beginning of period $t\in [T]$, the advertiser sets per-channel budgets $(\rho_{j,t})_{j\in [M]}$, while simultaneously values and costs $\bm{z}_{t}=(\bm{z}_{j,t})_{j\in [M]}=(\bm{v}_{j,t}, \bm{d}_{j,t})_{j\in[M]} $, where 
$(\bm{v}_{j,t}, \bm{d}_{j,t})\in \R_{+}^{m_{j}}\times \R_{+}^{m_{j}}$ are sampled (independently in each period) from  finite support $F = F_{1}\times \dots F_{M}$ according to discrete distributions $(\bm{p}_{j})_{j\in[M]}$. Each channel $j$ then takes as input $\rho_{j,t} \in [0,\rho]$ and procures ads on behalf of the advertiser, and reports the total realized conversion $\totv_{j}(\rho_{j,t};\bm{z}_{j,t})$ as well as total spend $\totc_{j}(\rho_{j,t};\bm{z}_{j,t}) $ to the advertiser (see definitions in Eq. \eqref{eq:budperchannel}). For simplicity we  assume any realization $\bm{z}_{j} = (\bm{v}_{j},\bm{d}_{j}) \in F_{j}$ admits an ordering $\frac{v_{j,1}}{d_{j,1}} > \dots  > \frac{v_{j,m_{j}}}{d_{j,m_{j}}}$ for all channels $j \in [M]$.

\textbf{Bandit feedback:} We highlight that the advertiser receives \textit{bandit feedback} from channels, i.e. the advertiser only observes the numerical values $\totv_{j}(\rho_{j,t}; \bm{z}_{j,t})$ and $\totc_{j}(\rho_{j,t}; \bm{z}_{j,t})$, but does not get to observe 
$\totv_{j}(\rho_{j}'; \bm{z}_{j}')$ and $\totc_{j}(\rho_{j}'; \bm{z}_{j}')$ evaluated at any other per-channel budget $\rho_{j}'\neq \rho_{j,t}$ and realized value-cost pairs $\bm{z}_{j}'\neq \bm{z}_{j,t}$. 


We also make two mild assumptions: In Assumption \ref{ass:budgetbind}, we assume that any channel will deplete input per-channel budgets. This is a natural assumption that mimics practical scenarios, e.g. marketing for small businesses that  have moderate-sized budgets. In Assumption \ref{ass:feas}, we assume for any realization of value-cost pairs  $\bm{z}_{j} = (\bm{v}_{j},\bm{d}_{j})$  in a channel $j\in [M]$, there always exists an auction $n\in [m_{j}]$ in this channel  whose value-to-cost ratio is at least $\gamma$, i.e. $v_{j,n} >  \gamma d_{j,n}$.
\begin{assumption}[Moderate budgets]
\label{ass:budgetbind}
Assume $\rho < \infty$, and for any channel $j \in [M]$, value-cost realization $\bm{z}_{j} = (\bm{v}_{j}, \bm{d}_{j}) \in F_{j}$, and per-channel budget $\rho_{j} \in [0,\rho]$, the optimal solution $\bm{x}_{j}^{*}(\rho_{j};\bm{z}_{j})$ defined in Eq. \eqref{eq:budperchannelsol} is budget binding, i.e. $\totc_{j}(\rho_{j};\bm{z}_{j}) = \bm{d}_{j}^{\top}\bm{x}_{j}^{*}(\rho_{j};\bm{z}_{j}) = \rho_{j}$.
\end{assumption}
\begin{assumption}[Strictly feasible global ROI constraints]
\label{ass:feas}
Fix any channel $j\in [M]$ and any realization of value-cost pairs  $\bm{z}_{j} =(\bm{v}_{j},\bm{d}_{j}) \in F_{j}$. Then, the channel's optimization problem in Eq. \eqref{eq:budperchannelsol} is strictly feasible, i.e. the set $\left\{\bm{x}_{j} \in [0,1]^{m_{j}}: \bm{v}_{j}^{\top} \bm{x}_{j} > \gamma\bm{d}_{j}^{\top}\bm{x}_{j}\right\}$ is nonempty.
\end{assumption}

\subsection{Optimize per-channel budgets with SGD-UCB}
\label{subsec:algo}
Here, we describe our algorithm to solve for optimal per-channel budgets w.r.t.  $\textsc{CH-OPT}(\mathcal{I}_{\budg})$. Similar to most algorithms for constrained optimization, we take a dual stochastic gradient descent (SGD) approach; see a comprehensive survey on dual descent methods in \cite{bertsekas2014constrained}. First, we consider the Lagrangian functions w.r.t. $\textsc{CH-OPT}(\mathcal{I}_{\budg})$ where we let  $\cont = (\lambda, \mu) \in \R_{+}^{2}$ be the dual variables corresponding to the ROI and budget constraints, respectively:
\begin{align}
    \label{eq:Lagrangian}
    \begin{aligned}
   & \mathcal{L}_{j}(\rho_{j}, \cont; \bm{z}_{j}) =  (1+\lambda)\totv_{j}(\rho_{j};\bm{z}_{j}) - (\lambda \gamma  + \mu) \rho_{j} \\
   &  \mathcal{L}_{j}(\rho_{j}, \cont) = \E\left[  \mathcal{L}_{j}(\rho_{j}, \cont; \bm{z}_{j})\right]\,.
    \end{aligned}
\end{align}
Then, in each period $t\in [T]$ given dual  variables $\cont_{t} = (\lambda_{t},\gamma_{t})$, SGD decides on a primal decision, i.e. per-channel budget $(\rho_{j,t})_{j\in[M]}$ by optimizing the following 
\begin{align}
\label{eq:SGDprimal}
     \textstyle \rho_{j,t} ~=~ \arg\max_{\rho_{j}\in [0,\rho]}\mathcal{L}_{j}(\rho_{j}, \cont_{t};\bm{z}_{j,t})\,.
\end{align}
Having observed the realized values $\left(\totv_{j}(\rho_{j,t};\bm{z}_{j,t})\right)_{j\in[M]}$ (note that spend is $(\rho_{j,t})_{j\in [M]}$ under Assumption \ref{ass:budgetbind}),  we calculate the current period violation in budget and ROI constraints, namely  $g_{1,t}:= \sum_{j \in M} \left(\totv_{j}(\rho_{j,t};\bm{z}_{j,t}) -  \gamma \rho_{j,t}\right)$ and $g_{2,t} = \rho - \sum_{j\in [M]} \rho_{j,t}$. Next, we update dual variables via $ 
\Pi_{[0,\dualub_{F}]}\left(\lambda_{t} - \eta g_{1,t}\right) \text{ and } \mu_{t+1}  = \Pi_{[0,\dualub_{F}]}\left(\mu_{t} - \eta g_{2,t}\right)$, where $\Pi$ is the projection operator, $\eta$ is some pre-specified step size, and $\dualub_{F}$ is some dual variable upper bound specified in Eq. \eqref{def:daulvarub}.\footnote{One can also employ more general mirror descent dual variable updates; see e.g. \cite{balseiro2022best}.}

 However, we cannot realistically find the primal decisions by solving Eq. \eqref{eq:SGDprimal} since the function $\mathcal{L}_{j}(\cdot, \cont_{t};\bm{z}_{j,t})$ is unknown due to the  bandit feedback  structure. Therefore, we provide a modification to SGD to handle this issue. We briefly note that although bandit feedback prevents naively applying SGD to our problem, this may not be the case in other online advertising scenarios that involve relevant learning tasks, underlining the challenges of our problem; see following Remark \ref{rmk:banditfeedback} for details.
\begin{remark}
\label{rmk:banditfeedback}
Our problem of interest under bandit feedback is more difficult than similar problems in related works that study online bidding strategies under budget and ROI constraints; see e.g. \cite{balseiro2017budget,balseiro2022best,feng2022online}. To illustrate, consider for instance  \cite{balseiro2017budget} in which a budget constrained advertiser's primal decision at period $t$ is to submit a bid value  $b_{t}$ after observing her value $v_{t}$. The advertiser competes with some unknown  highest competing bid $d_{t}$ in the market, and after submitting bid $b_{t}$, does not observe $d_{t}$ if she does not win the competition, which involves a semi-bandit feedback structure. Nevertheless, the corresponding Lagrangian under SGD takes the special form $\mathcal{L}_{j}(b, \mu_{t};\bm{z}_{t}) =\left(v_{t} - (1+\mu_{t})d_{t}\right)\I\{b_{t} \geq d_{t}\}$ where $\mu_{t}$ is the dual variable w.r.t. the budget constraint.  This simply allows an advertiser to optimize for her primal decision by bidding $\arg\max_{b\geq 0} \mathcal{L}_{j}(b, \cont_{t};\bm{z}_{t}) = \frac{v_{t}}{1+\mu_{t}}$. So even though \cite{balseiro2017budget,balseiro2022best,feng2022online} study dual SGD under bandit feedback, the special structures of their problem instances permits SGD to effectively optimize for primal decisions in each period, as opposed to Eq. \eqref{eq:SGDprimal} in our setting which can not be  solved.
\end{remark}

To handle bandit feedback, we take a natural approach to augment SGD with the celebrated \textit{upper-confidence bound (UCB)} algorithm; see intro to UCB and multi-arm bandits in \cite{slivkins2019introduction}.
In particular, we first discretize our per-channel budget decision set $[0,\rho]$ into granular ``arms'' separated by distance $\discerr > 0$:
\begin{align}
\label{eq:budgetdisc}
   \textstyle \budgset(\discerr) = \{a_{k}\}_{k \in [K]} \text{ where } a_{k} = (k-1)\discerr\,.
\end{align}
for $K := \lceil{\rho}/{\discerr}\rceil + 1$. In the following we will use the terms ``per-channel budget'' and ``arm'' interchangeably. In the spirit of UCB, in each period $t$ we maintain some estimate $\left(\overline{\totv}_{j}(a_{k})\right)_{j\in[M]}$ of the conversions $\left(\totv_{j}(a_{k})\right)_{j\in [M]}$ as well as an upper confidence bound $\ucb_{j,t}(a_{k})$ for each arm $a_{k}$ using historical payoffs from periods in which arm $a_{k}$ is pulled. Finally, we update primal decisions for each channel $j\in[M]$ using the ``best arm'' $\rho_{j,t} = \arg\max_{a_{k}\in \budgset(\discerr)}~(1+\lambda_{t})\left(\overline{\totv}_{j,t}(a_{k}) +  \ucb_{j,t}(a_{k})\right) - \left(\lambda_{t} \gamma + \mu_{t} \right)a_{k}$. 

Finally, to ensure aggregate ROI and budget constraint satisfaction, we maintain variables that check ROI and budget balances, namely $S_{1,t}$ and $S_{2,t}$, to record the cumulative ROI and spend across all channels up until period $t$. When the ROI balance check  $S_{1,t}$ is too negative, or the budget balance check is too large, we ``stop'' the algorithm, and naively set some pre-defined small per-channel budget $\safebudg\in (0,\rho)$ (later chosen in Theorem \ref{thm:regret}) during all periods after the ``stopping time'' denoted as $\tau_{A}$. We remark that similar approaches to ensure constraint satisfaction has been introduced in e.g. \cite{balseiro2022best,feng2022online}.

We summarize our algorithm, called SGD-UCB, in Algorithm \ref{alg:omducb}.\footnote{There has been very recent works that combine SGD with 
adversarial bandit type algorithms such as EXP3 \cite{castiglioni2022unifying,castiglioni2023online}, or with
Thompson sampling which is another well-known algorithm for stochastic bandit problems (e.g. \cite{ding2021efficient}), and works that employ SGD in bandit problems (e.g. \cite{han2021generalized}). Yet to the best of our knowledge, our approach to integrate SGD with UCB is novel.}

\begin{algorithm}[!h]
   \caption{SGD-UCB}\label{alg:omducb}
   \footnotesize
\begin{algorithmic}[1]
   \State {\bfseries Input:} Budget discretization set of arms $\budgset(\discerr)$ defined in Eq.\eqref{eq:budgetdisc}. Step size $\eta > 0$. Initialize $ N_{j,1}(a_{k}) = \overline{\totv}_{j,1}(a_{k})  = 0$ for all $j \in [M]$ and $k\in [K]$, and dual variables $\lambda_{1} = \mu_{1}= 0$. Set $\safebudg\in (0,\rho/M)$, $\safe > 0$ and dual variable upper bound 
   \begin{align}
   \label{def:daulvarub}
       \textstyle \dualub_{F} = M \overline{\totv}  \max\Big\{\frac{1}{\safe \safebudg},\frac{1}{\rho - M\safebudg}\Big\}
   \end{align}
   where $\overline{\totv} \geq \max_{j\in[M]}\max_{\rho_{j}\in [0,\rho]}\max_{\bm{z}_{j}\in F_{j}}V_{j}(\rho_{j},\bm{z}_{j})$ is the conversion upper bound.
   \State Set initial constraint balance checks: $S_{1,t} = S_{2,t}=0$ for $t=1$, and start period counter $t= 1$.
   \While{$t\leq T$ and $S_{1,t} - \gamma M\rho +\safe \safebudg (T-t)  \geq 0$ and $S_{2,t} + M\rho + M\safebudg (T-t) \leq \rho T$} 
   \State \textbf{Set per-channel budget.} For each channel $j\in [M]$: If $t \leq K$, set $\rho_{j,t} = a_{t}$. Else if $t > K$, set $\rho_{j,t} =$
      \begin{align*}
        \begin{aligned}
         \textstyle \arg\max_{a_{k}\in \budgset(\discerr)}~\overline{\totv}_{j,t}(a_{k}) +  \ucb_{j,t}(a_{k})- \frac{\left(\lambda_{t} \gamma + \mu_{t} \right)a_{k}}{1+\lambda_{t}},
         \end{aligned}
          \end{align*}
          where $ \ucb_{j,t}(a_{k}) = \sqrt{\frac{2\log(T)}{N_{j,t}(a_{k})}} $.
    \State  Observe realized conversion $\left\{\totv_{j}(\rho_{j,t};\bm{z}_{j,t}) \right\}_{j \in [M]}$, and update for each arm $k \in [K]$ and channel $j \in [M]$
    \begin{align*}
    \begin{aligned}
        &  \textstyle N_{j,t+1}(a_{k}) =  N_{j,t}(a_{k}) + \I\{\rho_{j,t} = a_{k}\}, \quad \overline{\totv}_{j,t+1}(a_{k}) = \frac{N_{j,t}(a_{k}) \overline{\totv}_{j,t}(a_{k}) + \totv_{j}(\rho_{j,t};\bm{z}_{j,t})\I\{\rho_{j,t} = a_{k}\}}{N_{j,t+1}(a_{k})} 
        \end{aligned}
    \end{align*}
    \State \textbf{Update dual variables.} Calculate $g_{1,t}= \sum_{j \in M} \left(\totv_{j}(\rho_{j,t};\bm{z}_{j,t}) -  \gamma \rho_{j,t}\right)$ and $g_{2,t} = \rho - \sum_{j\in [M]} \rho_{j,t}$. Then, set
        \begin{align}
        \label{eq:dualupdate}
        \begin{aligned}
     & \textstyle \lambda_{t+1} = \Pi_{[0,\dualub_{F}]}\left(\lambda_{t} - \eta g_{1,t}\right) ~\text{ and }~ \mu_{t+1}  = \Pi_{[0,\dualub_{F}]}\left(\mu_{t} - \eta g_{2,t}\right)\,.
        \end{aligned}
        \end{align}
    \State Update balance check:  $S_{1,t+1} = S_{1,t} + g_{1,t}$  and $S_{2,t+1} = S_{2,t} + \sum_{j\in[M]}\rho_{j,t}$. 
    \State Increment period counter $t \leftarrow t+1$.
	\EndWhile
 \State Record $\tau_{A} = t-1$ and for all $t = \tau_{A}+1 \ldots T$ set $\rho_{j,t} = \underline{\rho}$ for all $j\in [M]$.
 \State Output $\overline{\bm{\rho}}_{T} = \left(\frac{1}{T}\sum_{t\in[T]} \rho_{j,t}\right)_{j\in [M]}$.
\end{algorithmic}
\end{algorithm}
\vspace{-0.3cm}

\subsection{Analyzing the SGD-UCB algorithm}
\label{subsec:algoanalysis}

In this subsection, we analyze the performance of SGD-UCB in Algorithm \ref{alg:omducb}, and present accuracy guarantees on the final output $\overline{\bm{\rho}}_{T} = \left(\frac{1}{T}\sum_{t\in[T]} \rho_{j,t}\right)_{j\in [M]}$. The backbone of our analysis strategy is to show the cumulative loss over $T$ periods, namely $T\cdot  \textsc{GL-OPT} - \E\left[\sum_{t\in [T]}\sum_{j\in[M]}\totv_{j}(\rho_{j,t})\right]$ consists of three main parts, namely the ``stopping error''  due to some condition for the while loop being violated and  naively setting a small per-channel budget $\safebudg$ after the ``stopping time'' $\tau_{A}$ (see step 10); the error induced by UCB in our algorithm; and the error due to SGD (or what is typically viewed as the deviations from  complementary slackness); see following Proposition \ref{prop:ubbenchlag}. Then we further bound each part.
\begin{proposition}[Regret decomposition]
\label{prop:ubbenchlag}
For any channel $ j\in [M]$ define $\rho_{j}^{*}(t) = \arg\max_{\rho_{j\in[M]}}\mathcal{L}_{j}(\rho_{j};\cont_{t})$ to be the optimal per-channel budget w.r.t. dual variables $\cont_{t}=(\lambda_{t},\mu_{t})_{t\in [T]}$. Then $T \cdot\textsc{GL-OPT} - \sum_{t\in [T]}\sum_{j\in[M]}\totv_{j}(\rho_{j,t})$ is bounded by
 \begin{align*}
 \begin{aligned}
      &   \underbrace{M\overline{\totv}(T-\tau_{A})}_{\text{Stopping error}} + \underbrace{ \sum_{t\in[\tau_{A}]}(\lambda_{t}g_{1,t} + \mu_{t}g_{2,t})}_{\text{SGD complementary slackness deviations}} + 
     \sum_{j\in [M]}\underbrace{\sum_{t\in[\tau_{A}]} 
     \mathcal{L}_{j}(\rho_{j}^{*}(t), \lambda_{t}, \mu_{t})  - \mathcal{L}_{j}(\rho_{j,t}, \lambda_{t}, \mu_{t})}_{\text{UCB error}} \,.
      \end{aligned}
 \end{align*}
where $\tau_{A}\in [T]$ is defined in step 10 of Algorithm \ref{alg:omducb}.
\end{proposition}
 Recall the definitions of $g_{1,t}$ and $g_{2,t}$ in step 5 of Algorithm \ref{alg:omducb}, and the fact that the conversion $ \totv_{j}(\rho_{j};\bm{z}_{j})$ is bounded above by absolute constant $\overline{\totv} \in (0,\infty)$ almost surely.

We bound the stopping error together with SGD complementary slackness violations in the following Lemma \ref{lem:boundcs}, which follows standard analyses for SGD; see proof in Appendix \ref{pf:lem:boundcs}. 
\begin{lemma}[Bounding stopping error and complementary slackness deviations]
\label{lem:boundcs}
Assume Assumptions \ref{ass:budgetbind} and \ref{ass:feas} hold. Recall $\eta > 0$ is the step size. Then we have $M\overline{\totv}(T-\tau_{A})  + \sum_{t\in [\tau_{A}]} (\lambda_{t}g_{1,t} + \mu_{t}g_{2,t}) ~\leq~ \mathcal{O}\left( \eta  T + \frac{1}{\eta}\right)$.
\end{lemma}
\textbf{Challenges in bounding UCB error due to adversarial contexts and continuum-arm dicretization.}
Bounding our UCB error is much more challenging than doing so in classic stochastic multi-arm bandit settings: first, our setup involves discretizing a continuum of arms i.e. our discretization in  Eq.\eqref{eq:budgetdisc} for $[0,\rho]$; second, and more importantly, the dual variables $\{\cont_{t}\}_{t\in[T]}$ are effectively \textit{adversarial contexts} since they are updated via SGD instead of being stochastically sampled from some nice distribution, and correspondingly the Lagrangian function $\mathcal{L}_{j}(a_{k},\cont_{t};\bm{z}_{t})$ can be viewed as a reward function that maps any arm-context pair $(a_{k},\cont_{t})$ to (stochastic) payoffs. Both continuum-arms and adversarial contexts have been notorious in making reward function estimations highly inefficient; see e.g. discussions in \cite{agrawal1995continuum,agarwal2014taming}. We further elaborate on specific challenges that adversarial contexts bring about:\\
\textbf{1. Boundedness of rewards.} In classic stochastic multi-arm bandtis and UCB, losses in total rewards grow linearly with the magnitude of rewards. In our setting, the reward function, i.e. the Lagrangian function $\mathcal{L}_{j}(a_{k},\cont_{t};\bm{z}_{t})$, scales linearly with the magnitude of contexts (see Eq. \eqref{eq:Lagrangian}, so large contexts (i.e. large dual variables) may lead to large losses. \\
\textbf{2. Context-dependent exploration-exploitation tradeoffs.} The typical trade-off for arm exploration and exploitation in our setting depends on the particular values of the contexts (i.e. the dual variables), which means there may exist ``bad'' contexts that lead to poor tradeoffs that require significantly more explorations to achieve accurate estimates of arm rewards than  other ``good'' contexts. We elaborate more in Lemma \ref{lem:ucbregret} and discussions thereof.

\begin{figure}[ht]
\centerline{\includegraphics[width=0.65\columnwidth]{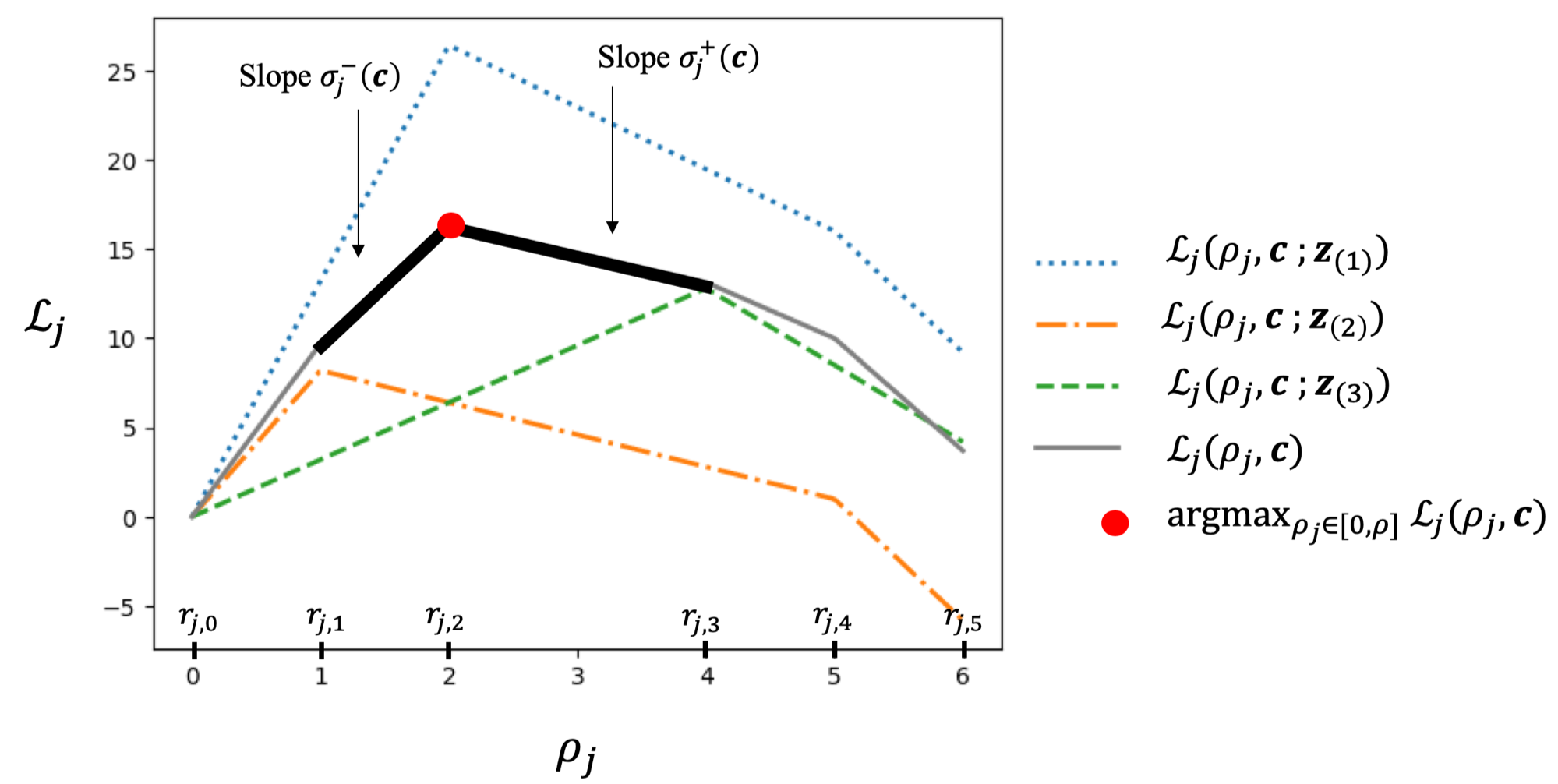}}
\caption{Illustration of Lagrangian functions defined in Eq. \eqref{eq:Lagrangian} with $M_{j} = 2$ auctions in channel $j$, and support $F_{j}$ that contains 3 elements,  $\bm{z}_{(1)} = ((8,2), (2,3))$, $\bm{z}_{(2)} = ((3,4), (1,4))$, $\bm{z}_{(3)} = ((8,1), (4,2))$, and context $\cont = (\lambda,\mu) = (4,2)$. Under Lemma \ref{lem:Vstruct}, $S_{j} = 5$, where the ``turning points'' $r_{j,0} \dots r_{j,S_{j}}$ are indicated on the x-axis, and the optimal budget w.r.t. $\cont$ is $\arg\max_{\rho_{j}\in [0,\rho]}\mathcal{L}_{j}(\rho_{j};\cont) = r_{j,2}$. The adjacent slopes in Eq. \eqref{eq:adj} are $\slope_{j}^{-}(\cont) = \slope_{j,2}(\cont)$, and  $\slope_{j}^{+}(\cont) = \slope_{j,3}(\cont)$.}
\vspace{-0.3cm}
\label{fig:lagr}
\end{figure}

We first handle continuum arm discretization via showing the specific form of conversion functions $\totv(\rho_{j};\bm{z})$ in Eq. \eqref{eq:budperchannelsol} induces salient structures for the Lagrangian function, namely it is continuous, piecewise linear, concave, and unimodal\footnote{A function $f:\R\to \R$ is unimodal if $\exists y^{*}$ such that $f(y)$ strictly increases when $y \leq y^{*}$ and  strictly decreases when $y \geq y^{*}$.}; we present the proof in Appendix \ref{pf:lem:Vstruct}
\begin{lemma}[Structural properties]
\label{lem:Vstruct}
For any channel $j\in [M]$: 
\begin{itemize}
\item
The conversion function $\totv_{j}(\rho_{j})$ is continuous, piece-wise linear, strictly increasing, and concave. In particular,  $\totv_{j}(\rho_{j})$ takes the form
\begin{align*}
         \textstyle \totv_{j}(\rho_{j}) = \sum_{n\in [S_{j}]} \left(s_{j,n} \rho_{j} + b_{j,n}\right)\I\{\rho_{j}\in [r_{j,n-1},r_{j,n}]\}\,,
\end{align*}
where $S_{j} \in \N$ and $\{(s_{j,n},b_{j,n},r_{j,n})\}_{n\in [S_{j}]}$ only depend on the support $F_{j}$ and distribution $\bm{p}_{j}$ from which value and costs are sampled. These parameters satisfy
$s_{j,1} >  \dots > s_{j,S_{j}} > 0$ and $0=r_{j,0} < r_{j,1} < \dots < r_{j,S_{j}} =\rho$, as well as $b_{j,n} \geq 0$ s.t. $s_{j,n} r_{j,n} + b_{j,n} = s_{j,n+1} r_{j,n} + b_{j,n+1}$ for all $n \in [S_{j}-1]$, with $b_{j,1} = 0$.  This implies $\totv_{j}(\rho_{j}) $ is continous  in $\rho_j$.
\item For any dual variables $\cont = (\lambda,\mu) \in \R_{+}^{2}$,  $ \mathcal{L}_{j}(\rho_{j},\cont)$ defined in Eq. \eqref{eq:Lagrangian} is continuous, piece-wise linear, concave, and unimodal in $\rho_{j}$. In particular,
\begin{align*}
        \textstyle \mathcal{L}_{j}(\rho_{j}, \cont) 
        =  \sum_{n=1}^{S_{j}} \left( \slope_{j,n}(\cont) \rho_{j} + b_{j,n}'\right)\I\{\rho_{j}\in [r_{j,n-1},r_{j,n}]\}
\end{align*}
where the slope $\slope_{j,n}(\cont) = (1+\lambda) s_{j,n} -(\mu + \gamma \lambda)$ and $b_{j,n}' = (1+\lambda)b_{j,n}$.  This implies $\arg\max_{\rho_{j}\geq 0} \mathcal{L}_{j}(\rho_{j}, \cont) = \max\{r_{j,n}:n=0,1\dots, S_{j}~,\slope_{j,n}(\cont) \geq 0\}$.
\end{itemize}
\end{lemma}
In fact,  for any realized value-cost pairs $\bm{z}$, the ``realization 
versions'' of the conversion and Lagrangians functions, namely $\totv_{j}(\rho_{j};\bm{z}) $ and $\mathcal{L}_{j}(\rho_{j}, \cont;\bm{z})$,  also satisfy the same properties as those of $\totv_{j}(\rho_{j})$ and $\mathcal{L}_{j}(\rho_{j}, \cont)$. We provide a visual illustration for these  properties in Figure \ref{fig:lagr}. 

We now handle the reward boundedness issue for the Lagrangian functions that arise from adversarial contexts: in Lemma \ref{lem:dualvarub} (proof in Appendix \ref{pf:lem:dualvarub}), we show 
the Lagrangian functions are bounded by some absolute constants:

\begin{lemma}[Bounding Lagrangian functions]
\label{lem:dualvarub}
For any $t\in [T]$, $j\in [M]$ and $\rho_{j} \in [0,\rho]$ we have $-\left(1+\gamma \right)\rho\dualub_{F}
   ~\leq~ \mathcal{L}_{j}(\rho_{j},\lambda_{t},\mu_{t})~\leq~(1+\dualub_{F})\overline{\totv}$, where the dual variables $(\lambda_{t},\mu_{t})_{t\in [T]}$ are generated from Algorithm \ref{alg:omducb}.
\end{lemma}
We now address the context-dependent exploration-exploitation tradeoff. To illustrate (e.g. Figure \ref{fig:lagr}), define the slopes that are adjacent to the optimal per-channel budget w.r.t. $\cont = (\lambda,\mu)$ as followed: assume the $n$th ``turning point'' $r_{j,n} =\arg\max_{\rho_{j}\in[0,\rho]} \mathcal{L}_{j}(\rho_{j},\cont)$, then
\begin{align}
\label{eq:adj}
\begin{aligned}
     \textstyle \slope_{j}^{-}(\cont) = \slope_{j,n}(\cont) 
    ~\text{ and }~ \slope_{j}^{+}(\cont) = \slope_{j,n+1}(\cont)
  \end{aligned}  
\end{align}
Similar to standard exploration-exploitation tradeoffs in bandits, the flatter the slope (e.g. $\slope_{j}^{-}(\cont)$ is close to 0), the more pulls required to accurately estimate rewards for sub-optimal arms on the slope, but the lower the loss in conversion for pulling sub-optimal arms. Our setting is challenging because the magnitude of this tradeoff, or equivalently adjacent slopes $\slope_{j}^{-}(\cont)$ and $\slope_{j}^{+}(\cont)$, depend on the adversarial contexts. In Lemma  \ref{lem:ucbregret} we bound the UCB error by handling this context-dependent tradeoff through separately analyzing periods when the adjacent slopes $\slope_{j}^{-}(\cont)$ and $\slope_{j}^{+}(\cont)$ are less or greater than some parameter $\underline{\slope} > 0$ chosen later, and characterize the context-dependent tradeoff using $\underline{\slope}$.
\begin{lemma}[Bounding UCB error]
\label{lem:ucbregret}
Assume the discretization width $\discerr$ satisfies $\discerr < \underline{r}_{j}:= \min_{n \in [S_{j}]}r_{j,n} - r_{j,n-1}$, where $S_{j}$ and $\{r_{j,n}\}_{n=0}^{S_{j}}$ are defined in Lemma \ref{lem:Vstruct}. Then the UCB error in Proposition \ref{prop:ubbenchlag} is upper bounded by $\mathcal{O}\left(\discerr T + \underline{\slope} T + \frac{1}{\underline{\slope}\discerr}\right)$, 
where $\underline{\slope} > 0$ is any positive number.
\end{lemma}
See Appendix \ref{pf:lem:ucbregret} for the proof. Finally, we put together Proposition \ref{prop:ubbenchlag}, Lemma \ref{lem:boundcs} and \ref{lem:ucbregret}, and obtain the main result Theorem \ref{thm:regret} whose proof we detail in Appendix \ref{pf:thm:regret}
\begin{theorem}[Putting everything together]
\label{thm:regret}
Assume assumptions  \ref{ass:budgetbind} and \ref{ass:feas} hold. Take step size $\eta = \Theta(1/\sqrt{T})$, discretization width $\discerr = \Theta(T^{-1/3}) $ and  $\safe = \safebudg =  \frac{1}{\log(T)}$ in Algorithm \ref{alg:omducb}, as well as  $\underline{\slope} = \Theta(T^{-1/3})$ in Lemma \ref{lem:ucbregret}. Then, for large enough $T$ we have $ T\cdot  \textsc{GL-OPT} - \E\left[\sum_{t\in [T]}\sum_{j\in[M]}\totv_{j}(\rho_{j,t})\right] \leq \mathcal{O}(T^{2/3})$.
Further, recalling $\overline{\bm{\rho}}_{T}$ is the final output of Algorithm \ref{alg:omducb}, we have 
\begin{align*}
     \textstyle \textsc{GL-OPT} - \sum_{j \in [M]}\E\left[\totv_{j}(\overline{\rho}_{j,T})\right] \leq \mathcal{O}(T^{-1/3}) \text{ and}
\end{align*}
constraint satisfaction: $\rho - \sum_{j \in [M]}\E[\overline{\rho}_{j,T}]\geq 0$, as well as $\sum_{j \in [M]} \E[\totv_{j}(\overline{\rho}_{j,T}) -\gamma\overline{\rho}_{j,T}] \geq 0 $.
\end{theorem}
We make an important remark that distinguishes our result in Theorem \ref{thm:regret} with related literature on convex optimization:

\begin{remark}
\label{rmk:onepoint}
In light of Lemma \ref{lem:Vstruct}, the 
advertiser's optimization problem $\textsc{CH-OPT}(\mathcal{I}_{\budg})$ in Eq. \eqref{eq:chopt} effectively becomes a convex problem (see Proposition \ref{prop:convexprob} in Appendix \ref{app:onlineadd}). Hence it may be tempting for one to directly employ off-the-shelf convex optimization algorithms. However, our problem involves stochastic bandit feedback, and more importantly, \textit{uncertain constraints}, meaning that we cannot analytically determine whether a primal decision satisfies the constraints of the problem. For example, in $\textsc{CH-OPT}(\mathcal{I}_{\budg})$, for some primal decision $(\rho_{j})_{j\in[M]}$, we cannot determine whether the ROI constraint  $\sum_{j \in M} \E\left[\totv_{j}(\gamma_{j},\rho_{j};\bm{z}_{j}) - \gamma \totc_{j}(\gamma_{j},\rho_{j};\bm{z}_{j})\right]\geq 0$ holds because the distribution $(\bm{p}_{j})_{j\in[M]}$ from which $\bm{z}$ is sampled is unknown. To the best of our knowledge, there are only two recent works that handle a similar stochastic bandit feedback, and uncertain constraint setting, namely \cite{usmanova2019safe} and \cite{nguyen2022stochastic}. Nevertheless, our setting is more challenging because these works consider a ``two-point estimation'' regime where one can make function evaluations to the objective and constraints twice each period, whereas our setting involves ``one-point estimation'' such that we can only make function calls once per period. We note the optimal oracle complexities for unknown constraint convex optimization with one-point bandit feedback, remains an open problem.\footnote{See Table 4.1 in \cite{larson2019derivative} for best known  complexity bounds for one-point bandit feedback setups.}
\end{remark}

\section{Generalizing to autobidding in multi-item  auctions}
\label{sec:multi}
In previous sections, we assumed that
each channel consists of multiple auctions, each of which is associated with the sale of a single ad  impression (see Eq. \eqref{eq:budperchannelsol} and discussions thereof). Yet, in practice, there are many scenarios in which ad platforms sell multiple impressions in each auction (see e.g. \cite{varian2007position,edelman2007internet}). Thereby in this section, we extend all our results for the single-item auction setting in previous sections to the multi-item auction setup. In Section \ref{subsec:multisetup}, we formally describe the multi-item setup; in Section \ref{subsec:multifutil} we show that in the multi-item setting, the per-budget ROI lever is again redundant (similar to what is shown in Theorem \ref{thm:budgonlysuff} and Corollary \ref{cor:roiredund}), and an advertiser can solely optimize over per-channel budgets to achieve the global optimal conversion; in Section \ref{subsec:multionline}, we show
 our proposed UCB-SGD algorithm is directly applicable to the multi-item auction setup for a broad class of auctions, and similar to Theorem \ref{thm:regret}, our algorithm produces accurate lever estimates with which the advertiser can approximate the globally optimal lever decisions. 

\subsection{Multi-item autobidding setup} 
\label{subsec:multisetup}

We first formalize our multi-item setup as followed. For each auction $n\in [m_{j}]$ of channel $j\in [M]$, assume $L_{j,n} \in \N$ impressions are sold, and channel $j$ is only allowed to procure at most 1 impression in auction $n$ on the advertiser's behalf. The value acquired and cost incurred by the advertiser when procuring impression $\ell \in [L_{j,n}]$ are $v_{j,n}(\ell)$ and $d_{j,n}(\ell)$, respectively. With a slight abuse of notation from previous sections, we write
 $\bm{v}_{j,n} = (v_{j,n}(1),\ldots v_{j,n}(L_{j,n})) \in \R_{+}^{L_{j,n}}$ as the $L_{j,n}$-dimensional vector that includes all impression values of auction $n$ in channel $j$, and further write  $\bm{v}_{j} = (\bm{v}_{j,1} \ldots \bm{v}_{j,m_{j}}) \in \R_{+}^{\sum_{n \in [m_{j}]}L_{j,n}}$ 
as the vector that concatenates all value vectors across auctions in channel $j$. 
We also define $\bm{d}_{j,n}\in \R_{+}^{L_{j,n}}$ and $\bm{d}_{j} \in \R_{+}^{\sum_{n \in [m_{j}]}L_{j,n}}$ accordingly for costs. Similar to Section \ref{sec:model}, we assume $\bm{z}_{j} = (\bm{v}_{j},\bm{d}_{j})$ is sampled from finite support $F_{j}$ according to discrete distribution $\bm{p}_{j}$ for any channel $j\in [M]$, and without loss of generality,  we assume that for any element $\bm{z}_{j} \in F_{j}$,  the value and costs for individual impressions in any auction $n \in [m_{j}]$ satisfy 
$v_{j,n}(1) > \ldots > v_{j,n}(L_{j,n}) > 0$  and $d_{j,n}(1) > \ldots > d_{j,n}(L_{j,n}) > 0$.

Under the above multi-item setup, an advertiser's global optimization problem (analogous to \textsc{GL-OPT}
in Eq. \eqref{eq:glopt} for the single-item auction setup in previous sections), can be written as the following problem called $\textsc{GL-OPT}^{+}$:
\begin{align}
\label{eq:gloptmulti}
\begin{aligned}
\textsc{GL-OPT}^{+}
~=~ & \\
\max_{ \left(\bm{x}_{j} = (\bm{x}_{j,1}\ldots, \bm{x}_{j,m_{j}})\right)_{j\in [M]}} ~&~ \sum_{j\in [M]}\E\left[\bm{v}_{j}^{\top} \bm{x}_{j}\right] \\
     \text{s.t. } & \sum_{j\in [M]}\E\left[\bm{v}_{j}^{\top} \bm{x}_{j}\right]~\geq~ \gamma \sum_{j\in [M]} \E\left[\bm{d}_{j}^{\top} \bm{x}_{j}\right]\\
     & \sum_{j\in [M]} \E\left[\bm{d}_{j}^{\top} \bm{x}_{j}\right] ~\leq~\rho \\
     & \bm{x}_{j,n}\in [0,1]^{\sum_{n \in [m_{j}]}L_{j,n}} \text{ and } \sum_{\ell \in [L_{j,n}]}x_{j,n}(\ell) \leq 1,  \quad \forall j\in [M],~ n\in [m_{j}]\,.
    \end{aligned}
\end{align} 
Here
$\bm{x}_{j,n} = (x_{j,n}(\ell))_{\ell \in [L_{j,n}]}$ denotes the indicator vector for procuring impressions $\ell \in L_{j,n}$ in auction $n \in [m_{j}]$ of channel $j\in [M]$. Compared to \textsc{GL-OPT}, the key difference for $\textsc{GL-OPT}^{+}$ is that we introduced additional constraints which states ``at most 1 impression is procured in every multi-item auction''.

On the other hand, analogous to a channel's autobidding problem for the single-item auction setup in previous sections (Eq. \eqref{eq:budperchannelsol}), in the multi-item setting each channel $j$'s autobidding problem can be written as
\begin{align}
\begin{aligned}
\label{eq:budperchannelsolmulti}
    \bm{x}_{j}^{*,+}(\gamma_{j},\rho_{j};\bm{z}_{j}) ~=~  \arg\max_{\bm{x} = (\bm{x}_{1} \ldots \bm{x}_{m_{j}})} & ~ \bm{v}_{j}^{\top} \bm{x} \\
    \quad 
     \text{s.t. }  \quad  & \bm{v}_{j}^{\top} \bm{x}\geq \gamma_{j}\bm{d}_{j}^{\top}\bm{x} ,\text{ and }  \quad  \bm{d}_{j}^{\top} \bm{x}\leq \rho_{j}\\
    & \bm{x}_{n}\in [0,1]^{L_{j,n}} \text{ and } \sum_{\ell \in [L_{j,n}]}x_{n}(\ell) \leq 1,  \quad \forall  n\in [m_{j}]
    \end{aligned}
\end{align} 
where $\bm{x}_{n} =(x_{n}(\ell))_{\ell\in [L_{j,n}]}\in [0,1]^{m_{j}}$ denotes the (possibly random) vector of indicators to win each impression of auction $n$ in channel $j$. With respect to this per-channel multi-item auction optimization problem in Eq. \eqref{eq:budperchannelsolmulti}, we can further define $\totv_{j}^{+}(\gamma_{j},\rho_{j};\bm{z}_{j})$, $\totc_{j}^{+}(\gamma_{j},\rho_{j};\bm{z}_{j}) $, $\totv_{j}^{+}(\gamma_{j},\rho_{j})$, $\totc_{j}^{+}(\gamma_{j},\rho_{j})$ as in Eq.\eqref{eq:budperchannel}, and $\textsc{CH-OPT}^{+}(\mathcal{I})$ as in Eq.\eqref{eq:chopt} for any advertiser lever option $\mathcal{I}$ in Eq.\eqref{eq:option}.

\subsection{Optimizing per-channel budgets is sufficient to achieve global optimal}
\label{subsec:multifutil}

Our first main result for the multi-item setting is the following Theorem \ref{thm:roiredundmulti} which again shows an advertiser can achieve the global optimal conversion $\textsc{GL-OPT}^{+}$ via solely optimizing over per-channel budgets (analogous to Theorem \ref{thm:budgonlysuff} and Corollary \ref{cor:roiredund}).
\begin{theorem}[Redundancy of per-channel ROIs in multi-slot auctions]
\label{thm:roiredundmulti}
For the per-channel budget option $\mathcal{I}_{\budg}$ and general options $\mathcal{I}_{G}$ defined in Eq.\eqref{eq:option},  we have  $\textsc{GL-OPT}^{+}=\textsc{CH-OPT}^{+}(\mathcal{I}_{\budg})= \textsc{CH-OPT}^{+}(\mathcal{I}_{G})$ for any aggregate ROI $\gamma > 0$ and total budget $\rho > 0$, even for $\rho= \infty$. Further, there exists and optimal solution $(\gamma_{j},\rho_{j})_{j\in [M]}$ to $\textsc{CH-OPT}^{+}(\mathcal{I}_{G})$, s.t. $\gamma_{j} = 0$ for all $j \in [M]$.
\end{theorem}
It is easy to see that the proof of Lemma \ref{lem:glbest}, Theorem \ref{thm:budgonlysuff}, and Corollary \ref{cor:roiredund} w.r.t. the single item setting in Section \ref{sec:futil} can be directly applied to Theorem \ref{thm:roiredundmulti} since we did not rely on specific structures of the solutions to \textsc{GL-OPT} and \textsc{CH-OPT} other than the presence of the respective ROI and budget constraints
(which are still present in 
$\textsc{GL-OPT}^{+}$ and $\textsc{CH-OPT}^{+}$). Thereby we will omit the proof of Theorem \ref{thm:roiredundmulti}.  In light of Theorem \ref{thm:roiredundmulti}, we again conclude that the per-channel ROI lever is redundant, and hence omit per-channel ROI $\gamma_{j}$ when the context is clear.

\subsection{Applying UCB-SGD to the multi-item setting } 
\label{subsec:multionline}
We now turn to our second main focus of the multi-item setting, which is to understand whether our proposed UCB-SGD algorithm can achieve accurate approximations to the optimal per-channel budgets, similar to Theorem \ref{thm:regret} for the single-item setting. A key observation is that the only difference between bounding the error of UCB-SGD in the single and multi-item settings is the structure of the conversion and corresponding Lagrangian functions (see Lemma \ref{lem:Vstruct}), since the only change in the multi-item setting compared to the single-item setting is how a given per-channel budget translates into a certain conversion. 
Therefore, in this section we introduce a broad class of multi-item auction formats that induce the same conversion function structural properties as those illustrated in Lemma \ref{lem:Vstruct}, which will allows us to directly apply the proof for bounding the error of UCB-SGD (Theorem \ref{thm:regret}) to the multi-item setting of interest. 

To begin with, we introduce the following notion of increasing marginal values, which is a characteristic that preserves the 
structural properties for conversion and Lagrangian functions from the single-item setting (in Lemma \ref{lem:Vstruct}), as shown later in Lemma \ref{lem:Vstructmulti}.
\begin{definition}[Multi-item auctions with increasing marginal values]
\label{def:incrmarginal}
 We say an auction $n \in [m_{j}]$ in channel $j\in [M]$ has increasing marginal values if for any realization $\bm{z}_{j} = (\bm{v}_{j},\bm{d}_{j})$, 
we have 
\begin{align*}
    \frac{v_{j,n}(1)-v_{j,n}(2)}{d_{j,n}(1)-d_{j,n}(2)} > \ldots > \frac{v_{j,n}(L_{j,n}-1)-v_{j,n}(L_{j,n})}{d_{j,n}(L_{j,n}-1)-d_{j,n}(L_{j,n})} > \frac{v_{j,n}(L_{j,n})}{d_{j,n}(L_{j,n})} > 0\,,
\end{align*}
where we recall $v_{j,n}(1) > \ldots > v_{j,n}(L_{j,n}) > 0$  and $d_{j,n}(1) > \ldots > d_{j,n}(L_{j,n}) > 0$. 
\end{definition}
Increasing marginal values intuitively means that in some multi-item auction, the marginal value per cost gained increases with procuring impressions of greater values. Many classic position auction formats satisfy increasing marginal gains, such as the Vickrey–Clarke–Groves (VCG) auction; see \cite{varian2007position,edelman2007internet} for more details on position auctions.

\begin{example}[VCG auctions have increasing marginal values]
Let auction $n \in [m_{j}]$ in channel $j \in [M]$ be a VCG auction, where  
for any realization of $(\bm{v}_{j,n},\bm{d}_{j,n}) = (v_{j,n}(\ell), d_{j,n}(\ell))_{\ell\in [L_{j,n}]}$
there exists some $\widetilde{v}_{n,j} > 0$, position discounts $1 \geq \theta_{n,j}(1) >\theta_{n,j}(2) \ldots \theta_{n,j}(L_{n,j}) > 0$, and $L_{n,j}$-highest competing bids from competitors in the market  $\widetilde{d}_{n,j}(1) > \widetilde{d}_{n,j}(2)\ldots > \widetilde{d}_{n,j}(L_{n,j}) > 0$, such that the acquired value for procuring impression $\ell\in L_{n,j}$ is $v_{n,j}(\ell) = \theta_{n,j}(\ell)\cdot \widetilde{v}_{n,j}$, and the corresponding cost  is $d_{j,n}(\ell) = \sum_{\ell'=\ell}^{L_{j,n}}(\theta_{n,j}(\ell')- \theta_{n,j}(\ell'+1))\widetilde{d}_{n,j}(\ell')$ where we denote $\theta_{n,j}(L_{j,n}+1)= 0$.\footnote{Here, the distribution over $(\bm{v}_{j,n},\bm{d}_{j,n})$ can be viewed as the joint distribution over $\widetilde{v}_{n,j}$, $(\theta_{n,j}(\ell))_{\ell\in [L_{j,n}]}$ and $(\widetilde{d}_{n,j}(\ell))_{\ell\in [L_{j,n}]}$.} 
Thereby, under VCG the marginal values are
\begin{align*}
    \frac{v_{j,n}(\ell)-v_{j,n}(\ell+1)}{d_{j,n}(\ell)-d_{j,n}(\ell+1)} =  \frac{\left(\theta_{n,j}(\ell)-\theta_{n,j}(\ell+1)\right)\widetilde{v}_{j,n}}{\left(\theta_{n,j}(\ell)-\theta_{n,j}(\ell+1)\right)\widetilde{d}_{j,n}(\ell)} = \frac{\widetilde{v}_{j,n}}{\widetilde{d}_{j,n}(\ell)}\,,
\end{align*}
which decreases in $\ell$ since $\widetilde{d}_{n,j}(1) > \widetilde{d}_{n,j}(2)\ldots > \widetilde{d}_{n,j}(L_{n,j}) > 0$. Hence  VCG auctions admit increasing marginal values. 
\end{example}
We remark that the generalized second price auction (GSP) does not necessarily have increasing marginal values. Now, if all auctions in a channel have increasing marginal values, then we can show the conversion function $\totv_{j}^{+}(\rho_{j})$ and the corresponding Lagrangian function for multi-item auctions admits the same structural properties as those in Lemma \ref{lem:Vstruct}:
\begin{lemma}[Structural properties for multi-item auctions]
\label{lem:Vstructmulti}
For any channel $j\in [M]$ whose auctions have increasing marginal values (see Definition \ref{def:incrmarginal}), the conversion function $\totv_{j}^{+}(\rho_{j})= \E\left[\bm{v}_{j}^{\top} \bm{x}_{j}^{*,+}(\rho_{j};\bm{z}_{j})\right]$  is continuous, piece-wise linear, strictly increasing, and concave. Here recall $\bm{x}_{j}^{*,+}(\rho_{j};\bm{z}_{j})$ is the optimal solution to the channel's optimization problem in Eq. \eqref{eq:budperchannelsolmulti}. Further, for any dual variables $\cont = (\lambda,\theta) \in \R_{+}^{2}$, the Lagrangian function $\mathcal{L}_{j}^{+}(\rho_{j},\cont):= (1+\lambda)\totv_{j}^{+}(\rho_{j}) - (\theta+\gamma\lambda)\rho_{j}$ is continuous, piece-wise linear, concave, and unimodal in $\rho_{j}$. 
\end{lemma}
See the proof in Appendix \ref{pf:lem:Vstructmulti}. In light of Lemma \ref{lem:Vstructmulti}, we can argue that UCB-SGD produces per-channel budgets that yield the same accuracy as in Theorem \ref{thm:regret} for the single-item setting,

\begin{theorem}[UCB-SGD applied to channel procurement for multi-item auctions]
Assume multi-item auctions in any channel $j\in [M]$ has increasing marginal values (per Definition \ref{def:incrmarginal}), and assume Assumptions \ref{ass:budgetbind} and  \ref{ass:feas} hold for the multi-item setting.\footnote{Assumption \ref{ass:budgetbind} in the multi-item setting again implies the spend in any channel is exactly the input per-channel budget; Assumption \ref{ass:feas} in the multi-item setting states that for any realization of value-cost pairs  $\bm{z} =(\bm{v}_{j},\bm{d}_{j})_{j\in[M]} \in F_{1}\times \dots F_{M}$, the realized version of the ROI constraint in $\textsc{GL-OPT}^{+}$ defined in Eq. \eqref{eq:gloptmulti} is strictly feasible.} Then with the same parameter choices as in Theorem \ref{thm:regret}, and recalling $\overline{\bm{\rho}}_{T} = \left(\frac{1}{T}\sum_{t\in[T]} \rho_{j,t}\right)_{j\in [M]}$ is the vector of time-averaged per-channel budgets produced by UCB-SGD, we have
\begin{align*}
    \textsc{GL-OPT}^{+} - \sum_{j \in [M]}\E\left[\totv_{j}^{+}(\overline{\rho}_{T,j})\right] \leq \mathcal{O}(T^{-1/3})\,,
\end{align*}
as well as  constraint satisfaction
\begin{align*}
     \sum_{j \in [M]}\E\left[\totv_{j}^{+}(\overline{\rho}_{T,j}) -\gamma \overline{\rho}_{T,j}\right] \geq 0 ,~\text{ and }\rho - \sum_{j \in [M]}\E[\overline{\rho}_{T,j}] \geq 0\,,
\end{align*}
where we recall $\textsc{GL-OPT}^{+}$ is defined in Eq. \eqref{eq:gloptmulti}, $\totv_{j}^{+}(\rho_{j})  = \E\left[\bm{v}_{j}^{\top} \bm{x}_{j}^{*,+}(\rho_{j};\bm{z}_{j})\right]$ and $\bm{x}_{j}^{*,+}(\rho_{j};\bm{z}_{j})$ is defined in Eq. \eqref{eq:budperchannelsolmulti}.
\end{theorem}
The proof for this theorem is identical to that of Theorem \ref{thm:regret} given the same structural properties of the conversion and Lagrangian functions in Lemma \ref{lem:Vstructmulti} and Lemma \ref{lem:Vstruct}. Hence we will omit the proof.

\section{Numerical studies}
\label{sec:num}
In this section, we conduct a numerical study using synthetic data to showcase the performance of our proposed SGD-UCB algorithm. In particular, we demonstrate that the final per-channel budget profile output of our proposed algorithm accurately approximates the optimal per-channel budget profile, even when the total number of periods $T$ is moderately small. Further, we show that when channels do not optimally autobid on advertisers' behalf (i.e. channels do not optimally solve for Eq. \eqref{eq:budperchannelsol}), the performance of our proposed algorithm deteriorates gracefully. 

\textbf{Data generation. }
We take the perspective of an advertiser with aggregate budget ratio $\rho = 10$ as well as  aggregate ROI $\gamma=1.3$, and consider $M = 10$ channels, each of which consists of $m_{j} = 100$ auctions for $j\in [M]$. We conduct experiments over $L = 100$ independent trials, where each trial corresponds to a different support $F$ for value-cost pairs, as well as a different distribution $\bm{p} \in \Delta(F)$ from which value-cost pairs are sampled. Here $ \Delta(F)$ is the probability simplex over $F$. In particular, fix a value-cost pair support size of $f = 5000$. For each trial $\ell \in [L]$ and channel $j\in \{1\ldots 5\}$: we sample $f$ values from $\text{Uniform}([0,1]^{m_{j}})$ as well as $f$ costs from $\text{Uniform}([0,1]^{m_{j}})$,
with which we construct the support $F_{j}^{(\ell)} \subset \R^{m_{j}}\times \R^{m_{j}}$ for channel $j$ (note that the support size $|F_{j}^{(\ell)}| = f$); on the other hand, for any channel $j\in \{6\ldots 10\}$, we sample $f$ values from $\text{Uniform}([0,2]^{m_{j}})$ as well as $f$ costs from $\text{Uniform}([0,1]^{m_{j}})$ for channel $j$,
with which we construct the support $F_{j}^{(\ell)} \subset \R^{m_{j}}\times \R^{m_{j}}$.  Correspondingly, for any channel $j\in [M]$ we generate uniformly at random a probability distribution $\bm{p}^{(\ell)} \in \Delta(F^{(\ell)})$ where $F^{(\ell)} = F_{1}^{(\ell)}\times \ldots \times F_{M}^{(\ell)}$. Finally, for trial $\ell \in [L]$, w.r.t. support $F^{(\ell)}$ and distribution $\bm{p}^{(\ell)}$, 
we denote the associated optimal conversion defined in Eq. \eqref{eq:glopt} as $\textsc{GL-OPT}^{(\ell)}$, as well as the expected per-channel conversion defined in Eq. \eqref{eq:budperchannel} as  $V_{j}^{(\ell)}(\rho_{j})$. 

 We remark that it is not difficult to see auctions in channels 6-10  have a higher value-to-cost ratio on average than those of channels 1-5, and therefore we call channel 6-10 the ``lucrative'' channels, and 1-5 the ``non-lucrative'' channels.  Further for any trial, under the hindsight optimal per-channel budget profiles,  lucrative channels consumes approximately $80\%$ of our total budget, whereas non-lucrative channels demand the remaining $20\%$.

\textbf{Non-optimal autobidding.}
Recall in previous sections we assumed that each channel adopt ``optimal autobidding'' that solves Eq. \eqref{eq:budperchannelsol} to optimality. This raises the natural question that whether our findings will still hold when channels do not procure ads optimally, perhaps because of non-stationary environments \citep{besbes2014stochastic,luo2018efficient,cheung2019learning, chen2022dynamic}, or the presence of strategic market participants who aim to manipulate the market \citep{golrezaei2019incentive,drutsa2020optimal,golrezaei2021learning,golrezaei2021bidding}. To model non-optimal autobidding behavior,  (i.e. a channel $j$ does not optimally solve for $V_{j}(\rho_{j};\bm{z}_{j})$ in Eq. \eqref{eq:budperchannelsol}), we utilize corruption factors $(\alpha_{1},\alpha_{2}) \in [0,1]^{2}$, for non-lucrative and lucrative channels, respectively, to  capture each channel's degree of sub-optimal autobidding: after submitting some per-channel budget $\rho_{j}$ to channel $j\in [M]$, the channel will report back a conversion value $\widetilde{V}_{j}(\rho_{j};\bm{z}_{j}) = \alpha_{1} V_{j}(\rho_{j};\bm{z}_{j})$ 
if $j\in \{1\ldots 5\}$ or $\widetilde{V}_{j}(\rho_{j};\bm{z}_{j}) = \alpha_{2} V_{j}(\rho_{j},\bm{z}_{j})$ 
if $j\in \{6\ldots 10\}$, given any realized value-cost pairs $\bm{z}_{j}$. Consequently, in step 4 of the  SGD-UCB algorithm, we observe $\widetilde{V}_{j}(\rho_{j};\bm{z}_{t})$ instead of the optimal conversion $V_{j}(\rho_{j};\bm{z}_{t})$. In words, we assume channel $j$ can only achieve some $\alpha_{1}$ or $\alpha_{2}$ portion of the optimal conversion, where $\alpha_{1} = 1$ or $\alpha_{2} = 1$  corresponds to optimal autobidding. 

\textbf{Experiment procedure. }
We repeat the following procedure for trial $\ell\in [L]$, corruption factors $\alpha_{1},\alpha_{2} \in \{0.2, 0.4,0.6,0.8,1\}$, and $T \in \{100,200, 500, 1000\}$. We run the SGD-UCB algorithm over $T$ periods, where in each period $t\in [T]$, value-cost pairs for all $\sum_{j\in[M]}m_{j} = 1000$ auctions, namely $\bm{z}_{t} = (\bm{v}_{t}, \bm{d}_{t}) \in \R^{\sum_{j\in[M]}m_{j}}\times \R^{\sum_{j\in[M]}m_{j}}$, are sampled from support $F^{(\ell)}$ according to distribution $\bm{p}^{(\ell)}$. After submitting $(\rho_{j,t})_{j\in[M]}$ to  channels, we observe $\widetilde{V}_{j}(\rho_{j},\bm{z}_{j}) = \alpha_{1} V_{j}(\rho_{j},\bm{z}_{j})$ for $j \in [M]$. We denote the final output per-channel budget profile as $\overline{\bm{\rho}}_{T}^{(\ell,\alpha_{1},\alpha_{2})}$. 

\begin{figure}[ht]
\begin{center}
\includegraphics[width=0.95\textwidth]{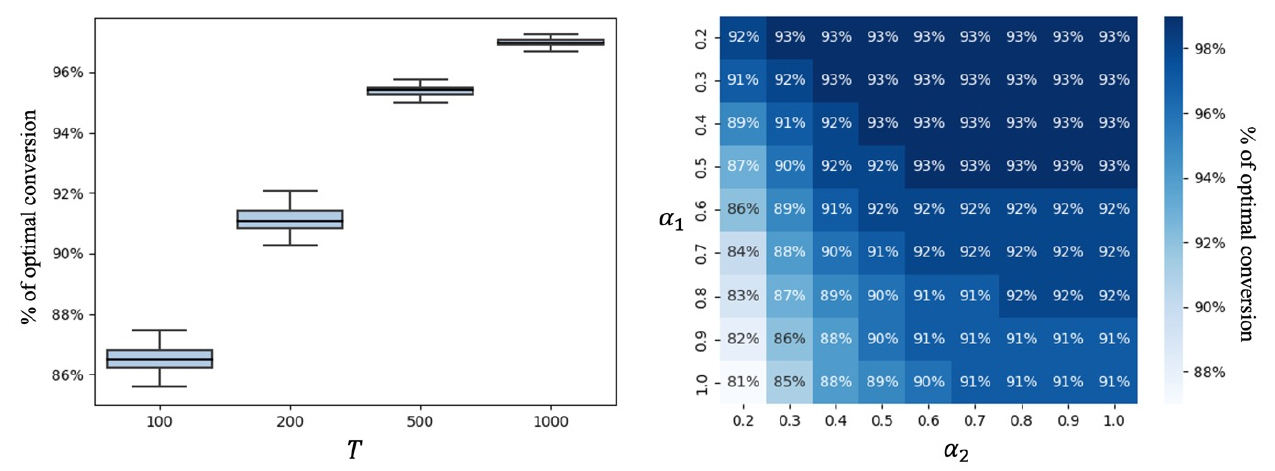}
\caption{Left: percentage of global optimal conversion under different total periods $T$ and optimal autobidding (i.e. $\alpha_{1} = \alpha_{2}=1$). Each point in box plot corresponds to the percentage conversion of a single trial $\ell\in [L]$, namely, $\frac{\sum_{j\in [M]} V_{j}\Big(\overline{\bm{\rho}}_{T}^{(\ell,\alpha_{1},\alpha_{2})}\Big)}{\textsc{GL-OPT}^{(\ell)}}$ for some $\ell\in [L]$;  Right: percentage of global optimal conversion under $T = 200$, and non-optimal autobidding as we vary $\alpha_{1},\alpha_{2}\in \{0.2,0.4,0.6,0.8,1\}$. Each point in the heatmap is the average conversion percentage over all $L$ trials, namely,  $\frac{1}{L}\sum_{\ell\in [L]} \frac{\sum_{j\in[M]}V_{j}\Big(\overline{\bm{\rho}}_{T}^{(\ell,\alpha_{1},\alpha_{2})}\Big)}{\textsc{GL-OPT}^{(\ell)}}$. }
\label{fig:sim}
\end{center}
\end{figure}

\textbf{Results. } 
To analyze the influence of number of total periods on our SGD-UCB algorithm under optimal auotbidding, or in other words assess how data hungry our algorithm is, we vary $T$ and display the corresponding percentage of conversion our algorithm's output achieves, compared to the hindsight global optimal conversion, namely \[\frac{\sum_{j\in [M]} V_{j}\Big(\overline{\bm{\rho}}_{T}^{(\ell,\alpha_{1},\alpha_{2})}\Big)}{\textsc{GL-OPT}^{(\ell)}}\] for $\alpha_{1}= \alpha_{2}=1$, in the left subgraph of Figure \ref{fig:sim}. 
Note that each box plots correspond to the variation over different trials $\ell\in [L]$. We observe that increasing the number of periods $T$ in our algorithm yields improved performance by allowing for more opportunities to collect data and learn  per-channel budgets, while simultaneously reducing the variance associated with its performance.  More interestingly, we note that for $T = 200$ (i.e. 8-9 days if one period corresponds to running an ad campaign for an hour),  our algorithm yields a per-channel budget profile that achieves more than $91\%$ of the global optimal conversion.

 In the right subgraph of  Figure \ref{fig:sim}, we further calculate the average percentage of conversion our algorithm's output achieves compared to the global optimal conversion as we vary corruption factors $\alpha_{1},\alpha_{2}$. In particular, fixing $T = 200$, for any $(\alpha_{1},\alpha_{2})$-pair,  we plot 
$$\frac{1}{L}\sum_{\ell\in [L]} \frac{\sum_{j\in[M]}V_{j}\Big(\overline{\bm{\rho}}_{T}^{(\ell,\alpha_{1},\alpha_{2})}\Big)}{\textsc{GL-OPT}^{(\ell)}}$$
in the right subgraph of  Figure \ref{fig:sim}. Note  here we are displaying the performance of the output per-channel budget profile  of our algorithm $\overline{\bm{\rho}}_{T}^{(\ell,\alpha_{1},\alpha_{2})}$ under optimal autobidding conversion $\totv_{j}(\cdot)$, instead of the observed non-optimal autobidding conversion  $\widetilde{\totv}_{j}(\cdot)$. This metric can be viewed as a normalized version of realized conversion by accounting for corruption factors $\alpha_{1}$ and $\alpha_{2}$, which allows us to assess now much our algorithm's  output per-channel budget profile deviates from the optimal per-channel budget profile.

Fixing the corruption factor $\alpha_{1}$ for the ``non-lucrative'' channels (i.e. channels with lower value-cost ratio auctions on average), the conversion performance of our algorithm improves as the other lucrative channels autobid more optimally. Nevertheless, even in the most corrupted case where both channels have corruption factors of 0.2 (i.e. both only achieve 20\% of optimal autobidding conversion),  our algorithm would still be able to output a per-channel budget profile under which we attain more than $92\%$ of the optimal (normalized) conversion. In contrast, as we fix the corruption factor  $\alpha_{2}$ for the ``lucrative'' channels, our algorithm actually achieves better performance as non-lucrative channels are more sub-optimal in its autobidding. This is primarily because with a decreasing corruption factor (more sub-optimal autobidding) for the non-lucrative channels, our algorithm tends to allocate more budget to the more lucrative channels, which inherently yield higher conversion with the same amount of spend. 

\section{Additional discussions}
\label{sec:extend}
In the following Section \ref{subsec:genutil}, we discuss extensions of our results to more general advertiser objectives; and in Section \ref{subsec:nonoptautobid}, we discuss future directions on non-optimal channel autobidding. 
\subsection{More general advertiser objectives}
\label{subsec:genutil}
In \textsc{GL-OPT}  and $\textsc{CH-OPT}(\mathcal{I})$  defined Section  \ref{sec:model}
(or similarly $\textsc{GL-OPT}^{+}$ and $\textsc{CH-OPT}^{+}(\mathcal{I})$ defined in the multi-item setting in Section \ref{sec:multi}),  we can also consider more general objectives, namely $\max_{\bm{x}_{1},\ldots, \bm{x}_{M}} ~ \sum_{j\in [M]}\E\left[\bm{v}_{j}^{\top} \bm{x}_{j} - \alpha\bm{d}_{j}^{\top} \bm{x}_{j} \right] $ and $\max_{(\gamma_{j},\rho_{j})_{j\in [M]}\in \mathcal{I}} ~ \sum_{j \in M} \E\left[\totv_{j}(\gamma_{j},\rho_{j};\bm{z}_{j})-\alpha \totv_{j}(\gamma_{j},\rho_{j};\bm{z}_{j})\right]$ for some private cost $\alpha\in [0,\gamma]$\footnote{If $\alpha > \gamma$ the ROI constraints in \textsc{GL-OPT} as well as $\textsc{CH-OPT}(\mathcal{I})$ become redundant.} in \textsc{GL-OPT} and $\textsc{CH-OPT}(\mathcal{I})$, respectively. When $\alpha = 0$, we recover our considered models in the previous section, whereas in when $\alpha = 1$, we obtain the classic quasi-linear utility. We remark that this private cost model has been introduced and studied in related literature; see \cite{balseiro2019black} and references therein. Nevertheless, when each channel's autobidding problem remains as is in  Eq.\eqref{eq:budperchannelsol}, i.e. channels still aim to maximize conversion which causes a misalignment between advertiser objectives and channel behavior, it is not difficult to see in our proofs that all our results still hold in Section \ref{sec:futil}, and our UCB-SGD algorithm still produces estimates of the same order of accuracy via introducing $\alpha$ into the Lagrangian. 
In other words, even if channels aim to maximize total conversion for advertisers, advertisers can optimize for \textsc{GL-OPT} with a private cost $\alpha$ through optimizing  $\textsc{CH-OPT}(\mathcal{I})$ that also incorporates the same private cost.

\subsection{Non-optimal autobidding in channels}
\label{subsec:nonoptautobid}
In Section \ref{sec:num} we addressed the practical setting where each channel non-optimally autobids (i.e. Eq. \eqref{eq:budperchannelsol} is not solved to optimality) from a numerical perspective. Here, we briefly discuss potential  frameworks under which we can analyze the impact of non-optimal autobidding to our findings from a theoretical standpoint.  In such a scenario, non-optimal bidding regime, an advertiser's  (bandit) conversion feedback in a channel $j$ would be $\totv(\rho_{j};\bm{z}_{j}) - \epsilon_{j}$ for some channel-specific and possibly adversarial loss $\epsilon_{j} > 0$. 
One potential resolution is to treat such $\epsilon_{j}$ as adversarial corruptions to bandit rewards, and instead of integrating vanilla UCB with SGD as in Algorithm \ref{alg:omducb}, augment SGD with bandit algorithms that are robust to robust corruptions; see e.g. \cite{lykouris2018stochastic,gupta2019better,golrezaei2021learning}. Nevertheless, it remains an open question to prove how such augmentation would perform in our specific bandit-feedback constrained optimization setup. This leads to potential research directions of both practical and theoretical significance.

\bibliographystyle{informs2014}
\bibliography{ref}

\newpage
\begin{center}
\vspace{0.8cm}
    \Large Appendices for\\
    \vspace{0.2cm}
    \Large \textbf{Multi-channel Autobidding with Budget and ROI constraints}
    \noindent\makebox[\linewidth]{\rule{1\linewidth}{0.9pt}}
\end{center}
\setcounter{page}{1}

\begin{APPENDICES}

\section{Proofs for Section \ref{sec:futil}}
\subsection{Proof of Lemma \ref{lem:glbest}}
\label{pf:lem:glbest}

Fix any option $\mathcal{I} \in \{ \mathcal{I}_{\budg}, \mathcal{I}_{\roi},\mathcal{I}_{G}\}$ defined in Eq. \eqref{eq:option}, and let $(\widetilde{\bm{\gamma}},\widetilde{\bm{\rho}} )\in \mathcal{I}$ be the optimal solution to $\textsc{CH-OPT}(\mathcal{I})$. Note that for the per-channel ROI only option $\mathcal{I}_{\roi}$, we have $\widetilde{\rho}_{j} = \infty$ and for the per-channel budget only we have $\widetilde{\gamma}_{j} = 0$ for all $j\in[M]$. Further, for any realization of value-cost pairs over all auctions 
$\bm{z} = (\bm{v}_{j},\bm{d}_{j})_{j\in [M]}$, recall the optimal solution $\bm{x}_{j}^{*}(\widetilde{\gamma}_{j}, \widetilde{\rho}_{j};\bm{z}_{j}) $
to $\totv_{j}(\widetilde{\gamma}_{j}, \widetilde{\rho}_{j};\bm{z}_{j})$ for each channel $j \in [M]$ as defined in Eq. \eqref{eq:budperchannelsol}. 

Due to feasibility of  $(\widetilde{\bm{\gamma}},\widetilde{\bm{\rho}} )\in \mathcal{I}$ for $\textsc{CH-OPT}(\mathcal{I})$, we have
\begin{align*}
    \sum_{j \in M} \E\left[\totv_{j}(\widetilde{\gamma}_{j},\widetilde{\rho}_{j}; \bm{z}_{j})\right] \geq \gamma \sum_{j \in M} \E\left[\totc_{j}(\widetilde{\gamma}_{j},\widetilde{\rho}_{j}; \bm{z}_{j})\right] \Longrightarrow \sum_{j\in [M]} \E\left[\bm{v}_{j}^{\top} \bm{x}_{j}^{*}(\widetilde{\gamma}_{j},\widetilde{\rho}_{j}; \bm{z}_{j})\right] \geq \gamma \sum_{j\in [M]} \left[\bm{d}_{j}^{\top}\bm{x}_{j}^{*}(\widetilde{\gamma}_{j},\widetilde{\rho}_{j}; \bm{z}_{j})\right]
\end{align*}
where we used the definitions $\totv_{j}(\widetilde{\gamma}_{j},\widetilde{\rho}_{j}; \bm{z}_{j}) = \bm{v}_{j}^{\top} \bm{x}_{j}^{*}(\widetilde{\gamma}_{j},\widetilde{\rho}_{j}; \bm{z}_{j})$ and $\totc_{j}(\widetilde{\gamma}_{j},\widetilde{\rho}_{j}; \bm{z}_{j}) = \bm{d}_{j}^{\top} \bm{x}_{j}^{*}(\widetilde{\gamma}_{j},\widetilde{\rho}_{j}; \bm{z}_{j})$ in Eq. \eqref{eq:budperchannel}.
This implies $\left(\bm{x}_{j}^{*}(\widetilde{\gamma}_{j},\widetilde{\rho}_{j}; \bm{z}_{j})\right)_{j\in[M]}$ satisfies the ROI constraint in \textsc{GL-OPT}. A similar analysis implies $\left(\bm{x}_{j}^{*}(\widetilde{\gamma}_{j},\widetilde{\rho}_{j}; \bm{z}_{j})\right)_{j\in[M]}$ also satisfies the budget constraint in \textsc{GL-OPT}. Therefore, 
$\left(\bm{x}_{j}^{*}(\widetilde{\gamma}_{j},\widetilde{\rho}_{j}; \bm{z}_{j})\right)_{j\in[M]}$ is feasible to \textsc{GL-OPT}. So
\begin{align}
    \textsc{GL-OPT}\geq  \sum_{j\in [M]} \E\left[\bm{v}_{j}^{\top} \bm{x}_{j}^{*}(\widetilde{\gamma}_{j},\widetilde{\rho}_{j}; \bm{z}_{j})\right] =  \sum_{j \in M}\left[ \totv_{j}(\widetilde{\gamma}_{j},\widetilde{\rho}_{j}; \bm{z}_{j})\right] = \textsc{CH-OPT}(\mathcal{I})\,.
\end{align}
where the final equality follows from the assumption that $(\widetilde{\bm{\gamma}},\widetilde{\bm{\rho}} )\in \mathcal{I}$ is the optimal solution to $\textsc{CH-OPT}(\mathcal{I})$.
\halmos

\subsection{Proof of Theorem \ref{thm:roifail}}
\label{pf:thm:roifail}

Let $\widetilde{\bm{\gamma}} = (\widetilde{\gamma}_{1},\widetilde{\gamma}_{2})$ be the optimal solution to $\textsc{CH-OPT}(\mathcal{I}_{\roi})$ and recall under the option $\mathcal{I}_{\roi}$, we let per-channel budgets to be infinity. It is easy to see that $\widetilde{\gamma}_{1}$ can be any arbitrary nonnegative number because the advertiser always wins auction 1, and $\widetilde{\gamma}_{2} > \frac{X}{1+X}$: if otherwise $\widetilde{\gamma}_{2} \leq \frac{X}{1+X}$, then the optimal outcome of channel 2 is to win both auctions 2 and 3.
However, in this case, the advertiser wins all auctions and acquires total value $1+X+2X = 1+3X$, and incurs total spend $0 + (1+X) + 2(1+X) = 3 + 3X$, which violates the ROI constraint in $\textsc{CH-OPT}(\mathcal{I}_{\roi})$ because $\frac{1+3X}{3 + 3X} < 1$. Therefore the advertiser can only win auction 1, or in other words $\widetilde{\gamma}_{2} > \frac{X}{1+X}$. This implies that the optimal objective to $\textsc{CH-OPT}(\mathcal{I}_{\roi})$ is $1$. On the other hand, it is easy to see that  the optimal solution to \textsc{GL-OPT} is to only win auctions 1 and 2, yielding an optimal value of $1+X$. Therefore $\frac{\textsc{CH-OPT}(\mathcal{I}_{\roi})}{\textsc{GL-OPT}} = \frac{1}{1+X}$.  Taking $X\to \infty$ yeilds the desired result.
\halmos

\subsection{Proof of Theorem \ref{thm:budgonlysuff}}
\label{pf:thm:budgonlysuff}
In light of Lemma \ref{lem:glbest}, we only
need to show $\textsc{CH-OPT}(\mathcal{I}_{\budg}) \geq \textsc{GL-OPT}$.  Let $\widetilde{\bm{x}}(\bm{z}) = \{\widetilde{\bm{x}}_{j}(\bm{z}_{j}))\}_{j\in [N]}$
 be the optimal solution to \textsc{GL-OPT}, and define $\widetilde{\gamma}_{j} = 0$ and $\widetilde{\rho}_{j}= \E\left[\bm{d}_{j}^{\top}\widetilde{\bm{x}}_{j}(\bm{z}_{j}))\right]$ 
to be the corresponding expected spend for each channel $j$ under the optimal solution $\widetilde{\bm{x}}(\bm{z})$ to \textsc{GL-OPT}, respectively.

We first argue that $(\widetilde{\gamma}_{j}, \widetilde{\rho}_{j})_{j\in[M]}$ is feasible to  $\textsc{CH-OPT}(\mathcal{I}_{\budg})$. Recall the optimal solution $\bm{x}_{j}^{*}(\widetilde{\gamma}_{j}, \widetilde{\rho}_{j};\bm{z}_{j}) $ to $\totv_{j}(\widetilde{\gamma}_{j}, \widetilde{\rho}_{j};\bm{z}_{j})$
for each channel $j \in [M]$ as defined in Eq. \eqref{eq:budperchannelsol}, as well as the definitions $\totv_{j}(\widetilde{\gamma}_{j},\widetilde{\rho}_{j}; \bm{z}_{j}) = \bm{v}_{j}^{\top} \bm{x}_{j}^{*}(\widetilde{\gamma}_{j},\widetilde{\rho}_{j}; \bm{z}_{j})$ and $\totc_{j}(\widetilde{\gamma}_{j},\widetilde{\rho}_{j}; \bm{z}_{j}) = \bm{d}_{j}^{\top} \bm{x}_{j}^{*}(\widetilde{\gamma}_{j},\widetilde{\rho}_{j}; \bm{z}_{j})$ in Eq. \eqref{eq:budperchannel}.
Then,
we have 
\begin{align}
\label{eq:budgsuffbound1}
    \E\left[\totc_{j}(\widetilde{\gamma}_{j},\widetilde{\rho}_{j}; \bm{z}_{j})\right] ~=~ \E\left[\bm{d}_{j}^{\top}\bm{x}_{j}^{*}(\widetilde{\gamma}_{j},\widetilde{\rho}_{j};\bm{z}_{j})\right]~\overset{(i)}{\leq}~ \widetilde{\rho}_{j} ~=~ \E\left[\bm{d}_{j}^{\top}\widetilde{\bm{x}}_{j}(\bm{z}_{j})\right]\,,
\end{align}
where (i) follows from feasibility of  $\bm{x}_{j}^{*}(\widetilde{\gamma}_{j}, \widetilde{\rho}_{j};\bm{z}_{j}) $ to $\totv_{j}(\widetilde{\gamma}_{j},\widetilde{\rho}_{j}; \bm{z}_{j})$. Summing over $j \in [M]$ we conclude that $(\bm{\gamma}_{j}, \bm{\rho}_{j})_{j\in[M]}$ satisfies the budget constraint in 
$\textsc{CH-OPT}(\mathcal{I}_{\budg})$:
\begin{align}
\label{eq:budgsuffbound0}
    \sum_{j \in [M]} \E\left[\totc_{j}(\widetilde{\gamma}_{j},\widetilde{\rho}_{j}; \bm{z}_{j})\right] \leq \sum_{j \in [M]} \E\left[\bm{d}_{j}^{\top}\widetilde{\bm{x}}_{j}(\bm{z}_{j})\right] \overset{(i)}{\leq} \rho\,.
\end{align}
Here (i) follows from feasibility of $\widetilde{\bm{x}}(\bm{z}) = \{\widetilde{\bm{x}}_{j}(\bm{z}_{j}))\}_{j\in [N]}$
 to \textsc{GL-OPT} since it is the optimal solution.
 
On the other hand, we have
\begin{align}
     \totv_{j}(\widetilde{\gamma}_{j},\widetilde{\rho}_{j};\bm{z}_{j})~=~ \bm{v}_{j}^{\top}\bm{x}_{j}^{*}(\widetilde{\gamma}_{j},\widetilde{\rho}_{j};\bm{z}_{j})~\overset{(i)}{\geq}~ \bm{v}_{j}^{\top}\widetilde{\bm{x}}_{j}(\bm{z}_{j})
\end{align}
where (i) follows from optimality of $\bm{x}_{j}^{*}(\widetilde{\gamma}_{j},\widetilde{\rho}_{j};\bm{z}_{j})$ to $ \totv_{j}(\widetilde{\gamma}_{j},\widetilde{\rho}_{j};\bm{z}_{j})$. Hence, we have
\begin{align}
\label{eq:budgsuffbound2}
     \sum_{j \in M} \E\left[\totv_{j}(\widetilde{\gamma}_{j},\widetilde{\rho}_{j};\bm{z}_{j})\right]~\geq~  \sum_{j \in M}\E\left[\bm{v}_{j}^{\top}\widetilde{\bm{x}}_{j}(\bm{z}_{j})\right]
     ~\overset{(i)}{\geq}~ \gamma \sum_{j \in M} \E\left[\bm{d}_{j}^{\top}\widetilde{\bm{x}}_{j}(\bm{z}_{j})\right] ~\overset{(ii)}{\geq}~ \gamma  \sum_{j \in [M]} \E\left[\totc_{j}(\widetilde{\gamma}_{j},\widetilde{\rho}_{j}; \bm{z}_{j})\right] 
\end{align}
where (i) follows from feasibility of $\widetilde{\bm{x}}(\bm{z}) = \{\widetilde{\bm{x}}_{j}(\bm{z}_{j}))\}_{j\in [N]}$
 to \textsc{GL-OPT} since it is the optimal solution; (ii) follows from Eq. \eqref{eq:budgsuffbound1}.  Hence combining Eq. \eqref{eq:budgsuffbound0} \eqref{eq:budgsuffbound2} we can conclude that $(\widetilde{\gamma}_{j}, \widetilde{\rho}_{j})_{j\in[M]}$  is feasible to  $\textsc{CH-OPT}(\mathcal{I}_{\budg})$. 
 
Finally, we have $ \textsc{CH-OPT}(\mathcal{I}_{\budg}) \geq  \sum_{j \in M} \E\left[\totv_{j}(\widetilde{\gamma}_{j},\widetilde{\rho}_{j}; \bm{z}_{j})\right]  \geq \sum_{j \in M} \E\left[\bm{v}_{j}^{\top}\widetilde{\bm{x}}_{j}(\bm{z}_{j})\right] = \textsc{GL-OPT}$, where the last inequality follows from \eqref{eq:budgsuffbound2}, and the final equality is because we assumed $\widetilde{\bm{x}}(\bm{z}) = \{\widetilde{\bm{x}}_{j}(\bm{z}_{j}))\}_{j\in [N]}$
 is the optimal solution to \textsc{GL-OPT}.
\halmos

\subsection{Proof of Corollary \ref{cor:roiredund}}
\label{pf:cor:roiredund}
In light of Lemma \ref{lem:glbest}, we only
need to show $\textsc{CH-OPT}(\mathcal{I}_{G}) \geq \textsc{GL-OPT}$. Let $(\widetilde{\bm{\gamma}},\widetilde{\bm{\rho}})\in \mathcal{I}_{\budg}$, and by definition of $\mathcal{I}_{\budg}$ in Eq. \eqref{eq:option} we have $\widetilde{\gamma}_{j} = 0$ for all $j\in [M]$. Since $(\widetilde{\bm{\gamma}},\widetilde{\bm{\rho}})$ is feasible to $\textsc{CH-OPT}(\mathcal{I}_{\budg})$, it is also feasible to $\textsc{CH-OPT}(\mathcal{I}_{G})$ since these two problems share the same ROI and budget constraints. Because they also share the same objectives, we have
\begin{align}
   \textsc{CH-OPT}(\mathcal{I}_{G}) \geq \textsc{CH-OPT}(\mathcal{I}_{\budg}) = \textsc{GL-OPT}
\end{align}
where the final equality follows from Theorem \ref{thm:budgonlysuff}. 
\halmos

\section{Proofs for Section \ref{sec:online}}

\subsection{Proof of Proposition \ref{prop:ubbenchlag}}
\label{pf:prop:ubbenchlag}
Let $(\rho_{j}^{*})_{j\in[M]}$ be the optimal per-channel budgets to $ \textsc{CH-OPT}(\mathcal{I}_{\budg})$, and define $\Bar{\mu}_{T} = \frac{1}{\tau_{A}}\sum_{t\in[\tau_{A}]}\mu_{t}$ as well as $\Bar{\lambda}_{T} = \frac{1}{\tau_{A}}\sum_{t\in[\tau_{A}]}\lambda_{t}$ . Then
\begin{align}
\begin{aligned}
     & T\cdot \textsc{GL-OPT} - \sum_{t\in [T]}\sum_{j\in[M]}\totv_{j}(\rho_{j,t})\\
    ~\overset{(i)}{\leq}~ & M\overline{\totv}(T-\tau_{A}) + \tau_{A} \textsc{CH-OPT}(\mathcal{I}_{\budg}) - \sum_{t\in[\tau_{A}]}\sum_{j\in[M]}\totv_{j}(\rho_{j,t})\\
     ~\leq~ &  M\overline{\totv}(T-\tau_{A})  + \tau_{A}\cdot \Big(\mathcal{L}_{j}(\rho_{j}^{*}, \Bar{\lambda}_{T} , \Bar{\mu}_{T})  + \rho \Bar{\mu}_{T} \Big)- \sum_{t\in[\tau_{A}]}\sum_{j\in[M]}\totv_{j}(\rho_{j,t})\\
     ~\overset{(ii)}{\leq}~ &  M\overline{\totv}(T-\tau_{A})  + \rho\sum_{t\in[\tau_{A}]}\mu_{t}+ \sum_{t\in[\tau_{A}]}\sum_{j\in [M]} \mathcal{L}_{j}(\rho_{j}^{*}, \lambda_{t}, \mu_{t}) -  \sum_{t\in[\tau_{A}]}\sum_{j\in [M]}\mathcal{L}_{j}(\rho_{j,t}, \cont_{t}) - \lambda_{t}\left(\totv_{j}(\rho_{j,t}) - \gamma\rho_{j,t}\right) + \mu_{t}\rho_{j,t}\\
      ~\overset{(iii)}{\leq}~ &  M\overline{\totv}(T-\tau_{A})  +  \sum_{j\in [M]} \sum_{t\in[\tau_{A}]} 
   \mathcal{L}_{j}(\rho_{j}^{*}(t), \cont_{t}) - \mathcal{L}_{j}(\rho_{j,t}, \cont_{t})+  \sum_{t\in[\tau_{A}]}\left(\lambda_{t}g_{1,t} + \mu_{t}g_{2,t}\right)\,.
    \end{aligned}
\end{align}
Here, (i) follows from Theorem \ref{thm:budgonlysuff} that states $\textsc{GL-OPT} =   \textsc{CH-OPT}(\mathcal{I}_{\budg})$ and $\textsc{CH-OPT}(\mathcal{I}_{\budg})$ is apparently upper bounded by $M\overline{\totv}$; (ii) follows from the $ \textsc{CH-OPT}(\mathcal{I}_{\budg}) = \sum_{j\in[M]}\totv_{j}(\rho_{j}^{*})$ and the definition of the Lagrangian in Eq. \eqref{eq:Lagrangian};  
in (iii) we define $\rho_{j}^{*}(t) = \arg\max_{\rho_{j}\geq 0} \mathcal{L}_{j}(\rho_{j}, \cont_{t})$ to be the optimal budget that maximizes the Lagrangian w.r.t. the dual variables $\cont_{t}=(\lambda_{t},\mu_{t})$.
\qed

\subsection{Proof for Lemma \ref{lem:boundcs}}
\label{pf:lem:boundcs}
Recall $g_{1,t} = \sum_{j\in [M]}\left(\totv_{j,t}(\rho_{j,t};\bm{z}_{j,t}) - \gamma \rho_{j,t}\right)$ and $g_{2,t} = \rho - \sum_{j\in [M]}\rho_{j,t}$ defined in Algorithm \ref{alg:omducb}. Also recall $\tau_{A} \in [T]$ defined in step 10 of Algorithm \ref{alg:omducb}. In the following, we will show 
\begin{align}
\begin{aligned}
\label{eq:cs0}
    & M\overline{\totv}(T-\tau_{A}) + \sum_{t \in [\tau_{A}]} (\lambda_{t}g_{1,t} + \mu_{t}g_{2,t}) \\
    ~\leq~&   \dualub_{F} \max\{M\overline{\totv},\rho\} + M^{2}\overline{\totv}\rho \cdot \max\Big\{\frac{1}{\safe\safebudg},\frac{1}{\rho - M\safebudg}\Big\}+  \frac{(\gamma M^{2}\Bar{\totv}^{2} + \rho^{2})}{2} \cdot \eta T +  \frac{1}{2\eta}\dualub_{F}^{2}
    ~=~   \mathcal{O}\Big(\eta T + \frac{1}{\eta}\Big)\,,
    \end{aligned}
\end{align}
where we recall $\dualub_{F} = M \overline{\totv}  \max\Big\{\frac{1}{\safe\safebudg},\frac{1}{\rho - M\safebudg}\Big\}$ defined in Eq. \eqref{def:daulvarub}.

From Lemma \ref{lem:omdtelescope}, we have for any $t\in [T]$, and $\lambda , \mu \in [0,\dualub_{F}]$,
\begin{align}
\label{eq:cs1}
\begin{aligned}
    & \sum_{\tau\in [t]} \left(\lambda_{\tau} - \lambda\right) g_{1,\tau} \leq  \frac{\eta M^{2}\Bar{\totv}^{2}}{2} \cdot t +  \frac{1}{2\eta}\lambda^{2}\\
    & \sum_{\tau \in [t]} \left(\mu_{\tau} - \mu\right) g_{2,\tau} \leq  \frac{\eta \rho^{2}}{2} \cdot t +  \frac{1}{2\eta}\mu^{2}\,,
  \end{aligned}  
\end{align}
where we used the fact that $\lambda_{1} = \mu_{1} = 0$ in Algorithm \ref{alg:omducb}.

Suppose that $\tau_{A} = T$ and thus $ M\overline{\totv}(T-\tau_{A}) = 0$. Then, considering $\lambda = \mu = 0$ in Eq. \eqref{eq:cs1}, we have
\begin{align}
\label{eq:cs2}
    \sum_{t \in [\tau_{A}]} \lambda_{t}g_{1,t}\leq \frac{\eta M^{2}\Bar{\totv}^{2}}{2} \cdot T \text{ and } \sum_{t \in [\tau_{A}]} \mu_{t}g_{2,t}\leq \frac{\eta \rho^{2}}{2} \cdot T\,. 
\end{align}
Thus, Eq. \eqref{eq:cs0} holds.

If  $\tau_{A} < T$, then according to Algorithm \ref{alg:omducb}, we either have $S_{1,\tau_{A}} - \gamma M\rho +\safe \safebudg (T-\tau_{A})  < 0$ or $S_{2,\tau_{A}} + M\rho + M\safebudg (T-\tau_{A}) > \rho T$, where we recall 
$S_{1,\tau_{A}} = \sum_{t\in [\tau_{A}-1]}g_{1,t}$ and $S_{2,\tau_{A}} =  \sum_{t\in [\tau_{A}-1]}\sum_{j\in [M]}\rho_{j,t} = \sum_{t\in [\tau_{A}-1]}(\rho - g_{2,t})$:
\begin{itemize}
    \item If $S_{1,\tau_{A}} - \gamma M\rho +\safe \safebudg (T-\tau_{A})  < 0$, then we have $ \sum_{t \in [\tau_{A}-1]} g_{1,t} < \gamma M\rho - \safe \safebudg (T-\tau_{A}) $. Hence,
    considering $\lambda = \frac{M\overline{\totv}}{\safe\safebudg} \in [0,\dualub_{F}]$ in Eq. \eqref{eq:cs1}, we have 
\begin{align}
\begin{aligned}
  & M\overline{\totv}(T-\tau_{A}) + \sum_{t \in [\tau_{A}]} \lambda_{t}g_{1,t} \\
  ~\leq~& M\overline{\totv}(T-\tau_{A})  + \lambda_{\tau_{A}} g_{1,\tau_{A}} +  \sum_{t \in [\tau_{A}-1]} \lambda g_{1,t} +  \frac{\eta M^{2}\Bar{\totv}^{2}}{2} \cdot (\tau_{A}-1) +  \frac{1}{2\eta}\lambda^{2}\\
  ~<~&   \lambda_{\tau_{A}} g_{1,\tau_{A}} + M\overline{\totv}(T-\tau_{A})  -  M\overline{\totv} (T-\tau_{A}) + \frac{\gamma M^{2}\overline{\totv}\rho}{\safe\safebudg}+  \frac{\eta M^{2}\Bar{\totv}^{2}}{2} \cdot (\tau_{A}-1) +  \frac{1}{2\eta}\lambda^{2}\\
  ~\leq~&   \dualub_{F}M\overline{\totv} + \frac{\gamma M^{2}\overline{\totv}\rho}{\safe\safebudg}+  \frac{\eta M^{2}\Bar{\totv}^{2}}{2} \cdot T +  \frac{1}{2\eta}\dualub_{F}^{2},
  \end{aligned}
\end{align}
where the final inequality uses the fact that $\tau_{A}\leq T$, $\lambda \leq \dualub_{F}$, and $g_{1,t}\leq M\overline{\totv}$ for any $t\in [T]$. Hence, similar to Eq. \eqref{eq:cs2} by further taking $\mu = 0$ in Eq.\eqref{eq:cs1} we show that Eq. \eqref{eq:cs0} holds.

\item If $S_{2,\tau_{A}} + M\rho + M\safebudg (T-\tau_{A}) > \rho T$, then we have $\sum_{t\in[\tau_{A}-1]}(\rho - g_{2,t}) > \rho T - M\rho - M\safebudg (T-\tau_{A})$, or equivalently $\sum_{t\in[\tau_{A}-1]}g_{2,t} < M\safebudg (T-\tau_{A}) + M\rho - \rho(T - \tau_{A}) \leq  -(\rho - M\safebudg)(T - \tau_{A}) + M\rho $. Hence, considering $\mu = \frac{M\overline{\totv}}{\rho - M\safebudg} \in [0,\dualub_{F}]$ in Eq.\eqref{eq:cs1} we have

\begin{align}
\begin{aligned}
  M\overline{\totv}(T-\tau_{A}) + \sum_{t \in [\tau_{A}]} \mu_{t}g_{1,t} ~\leq~& M\overline{\totv}(T-\tau_{A})  + \mu_{\tau_{A}}g_{2,\tau_{A}} + \sum_{t \in [\tau_{A}-1]} \mu g_{2,t} +  \frac{\eta \rho^{2}}{2} \cdot \tau_{A} +  \frac{1}{2\eta}\mu^{2}\\
  ~<~&  \mu_{\tau_{A}}g_{2,\tau_{A}}  + M\overline{\totv}(T-\tau_{A})  -  M\overline{\totv}(T-\tau_{A}) + \frac{M^{2}\overline{\totv}\rho}{\rho - \safebudg}+  \frac{\eta \rho^{2}}{2} \cdot \tau_{A} +  \frac{1}{2\eta}\mu^{2}\\
  ~\leq~&  \dualub_{F}\rho + \frac{M^{2}\overline{\totv}\rho}{\rho - \safebudg}+  \frac{\eta \rho^{2}}{2} \cdot T +  \frac{1}{2\eta}\dualub_{F}^{2},
  \end{aligned}
\end{align}
where the final inequality uses the fact that $\tau_{A}\leq T$, $\lambda \leq \dualub_{F}$, and $g_{2,t}\leq \rho$ for any $t\in [T]$. Hence,  similar to Eq. \eqref{eq:cs2}  by further taking $\lambda = 0$ in Eq.\eqref{eq:cs1} we show that Eq. \eqref{eq:cs0} holds.
\end{itemize}

\qed

\subsection{Proof of Lemma \ref{lem:Vstruct}}
\label{pf:lem:Vstruct}
We first show for any realization $\bm{z} = (\bm{z}_{j})_{j\in [M]} = (\bm{v}_{j},\bm{d}_{j})_{j\in [M]} $, the conversion function $\totv_{j}(\rho_{j};\bm{z}_{j})$
is piecewise linear, strictly inreasing, and concave for any $j \in [M]$. 

Fix any channel $j$ which consists of $m_{j}$ parallel auctions, and recall that we assumed the orderding
$\frac{v_{j,1}}{d_{j,1}} > \frac{v_{j,2}}{d_{j,2}} > \dots  > \frac{v_{j,m_{j}}}{d_{j,m_{j}}}$ for any realization $\bm{z}_{j}$. Then, with the option where the per-channel ROI is set to 0 (i.e. omitted) $\totv_{j}(\rho_{j};\bm{z}_{j})$ is exactly the LP relaxation of a 0-1 knapsack, whose optimal solution  $\bm{x}_{j}^{*}(\rho_{j};\bm{z}_{j})$ is well known to be unique, and takes the form for any auction index $n\in [m_{j}]$:
\begin{align}
\label{eq:knapsack}
    x_{j,n}^{*}(\rho_{j};\bm{z}_{j}) = \begin{cases}
    1 & \text{if } \sum_{n' \in [n]} d_{j,n'} \leq \rho_{j}\\
    \frac{\rho_{j} - \sum_{n' \in [n-1]} d_{j,n'}}{d_{j,n}} & \text{if } \sum_{n' \in [n]} d_{j,n'} > \rho_{j}\\
    0 & \text{otherwise}
    \end{cases}
\end{align}
where we denote $d_{j,0} =0$. With this form, it is easy to see
\begin{align}
\label{eq:pieceline}
   \totv_{j}(\rho_{j};\bm{z}_{j}) = \bm{v}_{j}^{\top} \bm{x}_{j}^{*}(\rho_{j};\bm{z}_{j})=  \sum_{n \in [m_{j}]}\left(\frac{v_{j,n}}{d_{j,n}}\rho_{j}+ b_{j,n}\right)\I\left\{d_{j,0}+ \dots +d_{j,n-1}\leq \rho_{j}\leq d_{j,0}+ \dots +d_{j,n} \right\}
   \end{align}
where we denote $d_{j,0} =0$ and also $b_{j,n} = \sum_{n' \in [n-1]} v_{j,n'} - \frac{v_{j,n}}{d_{j,n}}\cdot \left(\sum_{n' \in [n-1]} d_{j,n'}\right)$ and $ v_{j,0} = 0$. It is easy to check that any two line segments, say $[X_{n-1},X_{n}]$ and $[X_{n},X_{n+1}]$ where we write $X_{n} =  d_{j,0}+ \dots +d_{j,n} $, intersect at $\rho_{j} =X_{n}$, because $\frac{v_{j,n}}{d_{j,n}}\rho_{j}+ b_{j,n} = \frac{v_{j,n+1}}{d_{j,n+1}}\rho_{j}+ b_{j,n+1}$
at $\rho_{j} =X_{n}$. Hence, from Eq. \eqref{eq:pieceline} we can conclude $ \totv_{j}(\rho_{j};\bm{z}_{j})$ is continuous, which further implies it is piecewise linear and strictly increasing. Further, the ordering
$\frac{v_{j,1}}{d_{j,1}} > \frac{v_{j,2}}{d_{j,2}} > \dots  > \frac{v_{j,m_{j}}}{d_{j,m_{j}}}$ implies that the slopes on each segment $[X_{n},X_{n+1}]$ decreases as $n$ increases, which implies  $ \totv_{j}(\rho_{j};\bm{z}_{j})$  is concave.

Since $\totv_{j}(\rho_{j})= \E\left[\totv_{j}(\rho_{j};\bm{z}_{j})\right]$, where the expectation is taken w.r.t. randomness in $\bm{z}_{j}$, and since the $\bm{z}_{j}$ is sampled from some discrete distribution $\bm{p}_{j}$ on finite support $F_{j}$, $\totv_{j}(\rho_{j})$ is simply a weighted average over all $\left(\totv_{j}(\rho_{j};\bm{z}_{j})\right)_{\bm{z}_{j} \in F_{j}}$ with weights in $\bm{p}_{j}$, so $\totv_{j}(\rho_{j})$ is also continuous, piecewise linear, strictly increasing, and concave, and thus can be written as in Lemma \ref{lem:Vstruct} with parameters $\{(s_{j,n},b_{j,n},r_{j,n})\}_{n\in [S_{j}]}$ that only depend on the support $F_{j}$ and distribution $\bm{p}_{j}$. 

Finally, according to the definition of $\mathcal{L}_{j}(\rho_{j},\cont)= \E\left[\mathcal{L}_{j}(\rho_{j},\cont;\bm{z}_{j})\right]$ and $  \mathcal{L}_{j}(\rho_{j}, \cont; \bm{z}_{j}) =  (1+\lambda)\totv_{j}(\rho_{j};\bm{z}_{j}) - (\lambda \gamma  + \mu) \rho_{j} $
as defined in Eq. \eqref{eq:Lagrangian}, we have
\begin{align}
\label{eq:Elagr}
    \mathcal{L}_{j}(\rho_{j},\cont) = (1+\lambda)\totv_{j}(\rho_{j}) - (\lambda \gamma  + \mu) \rho_{j}
\end{align}
which implies $ \mathcal{L}_{j}(\rho_{j},\cont)$ is continuous, piecewise linear, and concave because $\totv_{j}(\rho_{j})$ is continuous, piecewise linear, and concave as shown above. Combining Eq. \eqref{eq:Elagr} and the representation of $\totv_{j}(\rho_{j})$ in Lemma \eqref{lem:Vstruct}, we have
\begin{align}
        \mathcal{L}_{j}(\rho_{j}, \cont) =  \sum_{n\in [S_{j}]} \left( \slope_{j,n}(\cont) \rho_{j} + (1+\lambda)b_{j,n}\right)\I\{r_{j,n-1}\leq \rho_{j} \leq r_{j,n}\}\,.
\end{align}
where the slope $\slope_{j,n}(\cont) = (1+\lambda) s{j,n} -(\mu + \gamma \lambda)$ decreases in $n$. Thus at the point $r_{j,n^{*}} = \max\{r_{j,n}:n=0,1\dots, S_{j}~,\slope_{j,n}(\cont) \geq 0\}$ in which the slope to the right turns negative for the first time, $ \mathcal{L}_{j}(\rho_{j}, \cont) $ takes its maximum value
$\max_{\rho_{j}\geq 0} \mathcal{L}_{j}(\rho_{j}, \cont)$, because to the left of  $r_{j,n^{*}}$, namely the region
$[0,r_{j,n^{*}}]$, $\mathcal{L}_{j}(\rho_{j}, \cont) $ strictly increases because slopes are positive; and to the right of  $r_{j,n^{*}}$, namely the region $[r_{j,n^{*}},\rho]$, $\mathcal{L}_{j}(\rho_{j}, \cont) $ strictly decreases because slopes are negative.
\qed

\subsection{Proof for Lemma \ref{lem:dualvarub}}
\label{pf:lem:dualvarub}
Recall the definition of the Lagrangian function $\mathcal{L}_{j}(\rho_{j}, \cont; \bm{z}_{j}) =  (1+\lambda)\totv_{j}(\rho_{j};\bm{z}_{j}) - (\lambda \gamma  + \mu) \rho_{j}$ in Eq.\eqref{eq:Lagrangian}. Then, since $\totv_{j}(\rho_{j};\bm{z}_{j}) \leq \overline{\totv}$ , and  $\lambda_{t},\mu_{t}\in [0,\dualub_{F}]$ for any period $t\in [T]$ and per-channel budget $\rho_{j}\in [0,\rho]$, we can conclude $-\left(1+\gamma \right)\rho\dualub_{F}
   ~\leq~ \mathcal{L}_{j}(\rho_{j},\lambda_{t},\mu_{t})~\leq~(1+\dualub_{F})\overline{\totv} $.
\qed

\subsection{Proof for Lemma \ref{lem:ucbregret}}
\label{pf:lem:ucbregret}

In the following, instead of bounding $\sum_{t\in [\tau_{A}]}\mathcal{L}_{j}(\rho_{j,t}^{*}, \cont_{t}) - \mathcal{L}_{j}(\rho_{j,t}, \cont_{t})$, we bound $\sum_{t \in [T]}\mathcal{L}_{j}(\rho_{j,t}^{*}, \cont_{t}) - \mathcal{L}_{j}(\rho_{j,t}, \cont_{t})$ where we consider the hypothetical scenario in which we ignore the termination criteria for the while loop in Algorithm \ref{alg:omducb}, and continue to set per-channel budgets based on steps 4-6 in the algorithm until the end of period $T$. This is due to the fact that $\sum_{t \in [T]}\mathcal{L}_{j}(\rho_{j,t}^{*}, \cont_{t}) - \mathcal{L}_{j}(\rho_{j,t}, \cont_{t}) \geq \sum_{t \in [\tau_{A}]}\mathcal{L}_{j}(\rho_{j,t}^{*}, \cont_{t}) - \mathcal{L}_{j}(\rho_{j,t}, \cont_{t})$.

 We fix some channel $j\in [M]$ and omit the subscript $j$ when the context is clear. Also, we first introduce some definitions that will be used throughout our proof. Fix some positive constant $\underline{\slope}>0$ whose value we choose later, and 
recall $a_{k}$ denotes the $k$th arm in the discretized budget set $\budgset(\discerr)$ as we defined in Eq. \eqref{eq:budgetdisc}. Then we define the following
\begin{align}
\label{eq:UCBdefs}
\begin{aligned}
    &  \Delta_{k}(\cont) =  \max_{\rho_{j}\in [0,\rho]}\mathcal{L}_{j}(\rho_{j}, \cont) -  \mathcal{L}_{j}(a_{k}, \cont) \\
    & \contset_{n} = \Big\{\cont \in \{\cont_{t}\}_{t\in [T]}: r_{j,n} = \arg\max_{\rho_{j}\geq 0} \mathcal{L}_{j}(\rho_{j},\cont)\Big\} \text{ for } n = 0 \dots S_{j} \\
    & \contset(\underline{\slope}) = \left\{\cont \in \{\cont_{t}\}_{t\in [T]}: \slope_{j}^{-}(\cont) > \underline{\slope},~ |\slope_{j}^{+}(\cont)| > \underline{\slope}\right\} \text{ for } n = 0, \dots,  S_{j} \\
    & m_{k}(\cont) = \frac{8\log(T)}{\Delta_{k}^{2}(\cont)} \text{ for } \forall (k,\cont) \text{ s.t. }\Delta_{k}(\cont) > 0 \,.
    \end{aligned}
\end{align}
Here, the ``adjacent slopes'' $\slope_{j}^{-}(\cont)$ and $\slope_{j}^{+}(\cont)$, which are defined in Eq.\eqref{eq:adj}, 
 represent the slopes that are adjacent to the optimal budget $\arg\max_{\rho_{j}\in[0,\rho]}\mathcal{L}_{j}(\rho_{j}, \cont) $ for any context $\cont = (\lambda,\mu) $. Further, $S_{j}$ and $\{r_{j,n}\}_{j\in [S_{j}]}$ are defined in Lemma \ref{lem:Vstruct}.  Here we state in words the meanings of $\Delta_{k}(\cont)$, $\contset(\underline{\slope}) $ and $\contset_{n} $, respectively.
\begin{itemize}
    \item $\Delta_{k}(\cont)$ denotes the loss in contextual bandit rewards when pulling arm $a_{k}$ under context $\cont$.  
    \item $\contset_{n} $ is the set including all context $\cont_{t}$ under which the optimal per-channel budget $\arg\max_{\rho_{j}\geq 0} \mathcal{L}_{j}(\rho_{j},\cont_{t})$ is taken at the $n$th ``turning point'' $r_{j,n}$ (see Lemma \ref{lem:Vstruct}).
     \item $\contset(\underline{\slope}) $ is the set of all contexts,  in which the adjacent slopes to the optimal point w.r.t. the context $\cont$, namely $\arg\max_{\rho_{j}\geq 0} \mathcal{L}_{j}(\rho_{j},\cont)$, have magnitude greater than $\underline{\slope}$, or in other words, the adjacent slopes are steep.
\end{itemize}
On a related note, for any context $\cont$, we define the following ``adjacent regions''  that sandwich  the optimal budget w.r.t.$\cont$
\begin{align}
\label{eq:adjregion}
    \adj_{j}^{-}(\cont) = [r_{j,n-1},r_{j,n}]    ~\text{ and }~ \adj_{j}^{+}(\cont) = [r_{j,n},r_{j,n+1}] \quad \text{ if }\quad \cont \in \contset_{n}\,.
\end{align}
In other words, if $\cont\in \contset_{n}$, per the definition of $\contset_{n}$ above, $\arg\max_{\rho_{j}\in[0,\rho]} \mathcal{L}_{j}(\rho_{j},\cont)$ is located at the $n$th ``turning point'' $r_{j,n}$, then $\adj_{j}^{-}(\cont)$ and $\adj_{j}^{-}(\cont)$ are respectively  the left and right regions surrounding $r_{j,n}$.

With the above definitions, we demonstrate how to bound the UCB-error. Define $N_{k,t} = \sum_{\tau\leq t-1}\I\{\rho_{j,\tau} = a_{k}\}$ to be the number of times arm $k$ is pulled up to time $t$, then we can decompose the UCB error as followed
\begin{align}
\begin{aligned}
    &  \sum_{ t > K}\mathcal{L}_{j}(\rho_{j}^{*}(t), \cont_{t}) - \mathcal{L}_{j}(\rho_{j,t}, \cont_{t}) = X_{1}+X_{2} + X_{3} \quad \text{ where }\\
    & X_{1} = \sum_{t > K:\cont_{t} \notin \contset(\underline{\slope})}  ~ \sum_{k\in [K]} \Delta_{k}(\cont_{t}) \I\{\rho_{j,t} = a_{k}, N_{k,t} \leq m_{k}(\cont_{t})\}\\
    & X_{2} = \sum_{t > K:\cont_{t} \in \contset(\underline{\slope})}   ~ \sum_{k\in [K]}\Delta_{k}(\cont_{t}) \I\{\rho_{j,t} = a_{k}, N_{k,t} \leq m_{k}(\cont_{t})\}\\
    & X_{3} = \sum_{k\in [K]}~ \sum_{t > K} \Delta_{k}(\cont_{t}) \I\{\rho_{j,t} = a_{k}, N_{k,t} > m_{k}(\cont_{t})\}\,.
    \end{aligned}
\end{align}
In Section \ref{subsubsec:X1}, we show that $X_{1} \leq \widetilde{\mathcal{O}}(\discerr T + \underline{\slope}T + \frac{1}{\discerr})$; in Section \ref{subsubsec:X2}  we show that $X_{2} \leq \widetilde{\mathcal{O}}(\discerr T + \frac{1}{\discerr \underline{\slope}})$;
 in Section \ref{subsubsec:X3}  we show that $X_{3} \leq \widetilde{\mathcal{O}}(\frac{1}{\delta T})$. 

 \begin{remark}
 \label{rmk:bestarm}
 In the following sections  \ref{subsubsec:X1}, \ref{subsubsec:X2} and \ref{subsubsec:X3} where we bound $X_{1}$, $X_{2}$, and $X_{3}$, respectively,  we assume the optimal per-channel $\rho_{j}^{*}(t) = \arg\max_{\rho_{j}\in [0,\rho]}\mathcal{L}_{j}(\rho_{j}, \cont_{t})$ lies in the arm set $\budgset(\discerr)$ for all $t$. This is because otherwise, we can consider the following decomposition of the UCB error in period $t$ as followed:
  \begin{align*}
     \mathcal{L}_{j}(\rho_{j}^{*}(t), \cont_{t}) - \mathcal{L}_{j}(\rho_{j,t}, \cont_{t}) = \mathcal{L}_{j}(\rho_{j}^{*}(t), \cont_{t}) - \mathcal{L}_{j}(a_{t}^{*}, \cont_{t}) + \mathcal{L}_{j}(a_{t}^{*}, \cont_{t}) - \mathcal{L}_{j}(\rho_{j,t}, \cont_{t}) \quad \text{where }a_{t}^{*}=\arg\max_{a_{k}\in \budgset(\discerr)}\mathcal{L}_{j}(a_{k}, \cont_{t})
  \end{align*}
 The first term will yield an error in the order of $\mathcal{O}(\discerr)$ due to the Lagrangian function being unimodal, piecewise linear liner, which implies 
$|a_{t}^{*} - \rho_{j}^{*}(t)|\leq \discerr$ so that $\mathcal{L}_{j}(\rho_{j}^{*}(t), \cont_{t}) - \mathcal{L}_{j}(a_{t}^{*}, \cont_{t}) = \mathcal{O}(\discerr)$. Hence, this ``discretization error'' will accumulate to a magnitude of $\mathcal{O}(\discerr T)$ over $T$ periods, which leads to an additional error that is already accounted for in the statement of the lemma.
 \end{remark}

\subsubsection{Bounding $X_{1}$.}
\label{subsubsec:X1} 
 Our strategy to bound $X_{1}$ consists of 4 steps, namely bounding the 
loss of arm $a_{k}$ at each context $\cont\notin \contset(\underline{\slope})$ when $a_{k} \in \adj_{j}^{-}(\cont)$ lies on the left adjacent region of the optimal budget; $a_{k} < \min~ \adj_{j}^{-}(\cont)$
lies to the left of the left adjacent region; $a_{k} \in \adj_{j}^{+}(\cont)$ lies on the right adjacent region of the optimal budget; and $a_{k} > \max~ \adj_{j}^{+}(\cont)$
lies to the right of the right adjacent region. Here we recall the adjacent regions are defined in Eq.\eqref{eq:adjregion}.


\textbf{Step 1: $a_{k} \in \adj_{j}^{-}(\cont_{t})$. } For arm $k$ such that $a_{k} \in \adj_{j}^{-}(\cont_{t})$, recall Lemma \ref{lem:Vstruct} that $\mathcal{L}_{j}(a,\cont_{t})$ is linear in $a$ for $a \in \adj_{j}^{-}(\cont_{t})$, so $\Delta_{k}(\cont_{t}) = \slope_{j}^{-}(\cont_{t})\cdot (\rho_{j}^{*}(t) - a_{k}) \leq  \underline{\slope} \rho$
 where we used the condition
that $\cont_{t}\notin \contset(\underline{\slope})$ so the adjacent slopes have magnitude at most $\underline{\slope}$, and $\rho_{j}^{*}(t)\leq \rho$.
Thus, summing over all such $k$  we get 
\begin{align}
\label{eq:X1near}
\begin{aligned}
   &  \sum_{t > K:\cont_{t} \notin \contset(\underline{\slope})}  ~\sum_{k\in[K]:a_{k} \in \adj_{j}^{-}(\cont_{t})}  \Delta_{k}(\cont_{t}) \I\{\rho_{j,t} = a_{k}, N_{k,t} \leq m_{k}(\cont_{t})\} \\
   ~\leq~& \sum_{t > K:\cont_{t} \notin \contset(\underline{\slope})}   ~\sum_{k\in[K]:a_{k} \in \adj_{j}^{-}(\cont_{t})}  \underline{\slope} \rho \cdot  \I\{\rho_{j,t} = a_{k}, N_{k,t} \leq m_{k}(\cont_{t})\}  ~\leq~ \underline{\slope} \rho T = \mathcal{O}(\underline{\slope}T )\,.
   \end{aligned}
\end{align}

\textbf{Step 2: $a_{k} < \min~ \adj_{j}^{-}(\cont_{t})$. } For arm $k$ such that $a_{k} < \min ~ \adj_{j}^{-}(\cont_{t})$, we further split contexts into groups $\contset_{n}$ for $n = 0 \dots S_{j}$ (defined in Eq. \eqref{eq:UCBdefs}) based on whether the corresponding optimal budget w.r.t. the Lagrangian at the context is taken at the $n$th ``turning point'' (see Figure \ref{fig:lagr} of illustration). Then, for each context group $n$ by defining  $k':=\max\{k: a_{k} < r_{j,n-1}\}$ to be the arm closest to and less than $r_{j,n-1}$, we have
\begin{align}
\label{eq:nbadslope}
\begin{aligned}
   & \sum_{t > K:\cont_{t} \in \contset_{n}/\contset(\underline{\slope})} ~ \sum_{k\in[K]:a_{k} < \min \adj_{j}^{-}(\cont_{t})} ~  \Delta_{k}(\cont_{t}) \I\{\rho_{j,t} = a_{k}, N_{k,t} \leq m_{k}(\cont_{t})\} \\
    ~\overset{(i)}{=}~ & \sum_{t > K:\cont_{t} \in \contset_{n}/\contset(\underline{\slope})} ~ \sum_{k\in[K]:a_{k} < r_{j,n-1}} ~  \Delta_{k}(\cont_{t}) \I\{\rho_{j,t} = a_{k}, N_{k,t} \leq m_{k}(\cont_{t})\} \\
    ~=~ &  \sum_{t > K} \sum_{\cont \in \contset_{n}/\contset(\underline{\slope})} ~ \sum_{k\in[K]:a_{k} < r_{j,n-1}} \Delta_{k}(\cont) \I\{\cont_{t} = \cont, \rho_{j,t} = a_{k}, N_{k,t} \leq m_{k}(\cont)\}\\
    ~\overset{(ii)}{\leq}~ &  \sum_{t > K} \sum_{\cont \in \contset_{n}/\contset(\underline{\slope})} ~ \left(\Delta_{k'}(\cont) \I\{\cont_{t}=\cont\} + \sum_{k\in[K]:a_{k} < r_{j,n-1}-\discerr} \Delta_{k}(\cont) \I\{\cont_{t} = \cont, \rho_{j,t} = a_{k}, N_{k,t} \leq m_{k}(\cont)\}\right)\\
   ~\overset{(iii)}{\leq}~ & \left((1+\dualub_{F})s_{j,n-1} \discerr + \rho \underline{\slope}\right) T + \sum_{k\in[K]:a_{k} < r_{j,n-1}-\discerr} ~ \sum_{\cont \in \contset_{n}/\contset(\underline{\slope})} \Delta_{k}(\cont) Y_{k}(\cont)
\end{aligned}
\end{align}
where in the final equality we defined $ Y_{k}(\cont)=  \sum_{t > K} \I\{\cont_{t} = \cont, \rho_{j,t} = a_{k}, N_{k,t} \leq m_{k}(\cont)\}$. In (i) we used the fact that the left end of the left adjacent region, i.e. $
\min \adj_{j}^{-}(\cont_{t})$ is exactly $r_{j,n-1}$ because for context $\cont_{t} \in \contset_{n}$ the optimal budget $\arg\max_{\rho_{j}\in[0,\rho]} \mathcal{L}_{j}(\rho_{j},\cont_{t})$ is at the $n$th turning point; in (ii) we used the definition $k':=\max\{k: a_{k} < r_{j,n-1}\}$  where we recall arms are indexed such that $a_{1} < a_{2}<\dots < a_{K}$. Note that in (ii) we separate out the arm $a_{k'}$ because its distance to
the optimal per-channel may be less than $\discerr$ since it is the closest arm, and thus we ensure  all other arms indexed by $k\in[K]:a_{k} < r_{j,n-1}-\discerr$, 
 are at least $\discerr$ away from the optimal per-channel budget; (iii) follows from the fact that under a context $\cont \in \contset_{n}/\contset(\underline{\slope})$,  we have $\arg\max_{\rho_{j}\in [0,\rho]} \mathcal{L}_{j}(\rho_{j}, \cont) = r_{j,n}$ so
\begin{align*}
    \Delta_{k'}(\cont) ~=~& \mathcal{L}_{j}(r_{j,n},\cont) -   \mathcal{L}_{j}(r_{j,n-1},\cont) + \mathcal{L}_{j}(r_{j,n-1},\cont) - \mathcal{L}_{j}(a_{k'},\cont)\\
     ~=~&  \slope_{j}^{-}(\cont) (r_{j,n} - r_{j,n-1}) + \slope_{j,n-1}(\cont)(r_{j,n-1} - a_{k'}) \\
     ~\overset{(iv)}{\leq}~ & \underline{\slope}\rho + \slope_{j,n-1}(\cont) \discerr\\
       ~\overset{(v)}{\leq}~ & \underline{\slope}\rho + (1+\dualub_{F})s_{j,n-1}\discerr
     \,,
\end{align*}
where in (iv) we used  $\cont \in \contset_{n}/\contset(\underline{\slope})$ implies $ \slope_{j}^{-}(\cont) \leq \underline{\slope}$,   as well as all $r_{j,n}\leq \rho$ for any $n$ and the fact that $k'$ lies on the line segment between points $r_{j,n-2}$ and $r_{j,n-1}$ since $\discerr < \min_{n' \in [S_{j}]} r_{j,n'} - r_{j,n'-1}$; in (v) we recall $\slope_{j,n-1}(\cont) = (1+\lambda)s_{j,n-1} - (\mu + \gamma\lambda) \leq (1+\dualub_{F})s_{j,n-1}$ where $\dualub_{F}$ is defined in Lemma \ref{lem:dualvarub}.

We now bound $\sum_{\cont \in \contset_{n}/\contset(\underline{\slope})} \Delta_{k}(\cont) Y_{k}(\cont)$ in Eq. \eqref{eq:nbadslope}.  It is easy to see the following inequality for any sequence of context $\cont_{(1)}, \dots , \cont_{(\ell)} \in \{\cont_{t}\}_{t\in[T]}$ (This is a slight generalization of an inequality result shown in \cite{balseiro2019contextual}):
\begin{align}
\label{eq:numbadarms}
    Y_{k}(\cont_{(1)}) + \dots +  Y_{k}(\cont_{(\ell)}) \leq \max_{\ell'=1\dots \ell} m_{k}(\cont_{(\ell')})\,. 
\end{align}
This is because 
\begin{align*}
   \sum_{\ell'\in [\ell]}  Y_{k}(\cont_{(\ell')}) ~=~ &\sum_{t > K} \sum_{\ell'\in [\ell]}  \I\{\cont_{t} =\cont_{(\ell')}, \rho_{j,t} = a_{k}, N_{k,t} \leq m_{k}(\cont_{(\ell')})\}\\
     ~\leq~& \sum_{t > K}  \sum_{\ell'\in [\ell]}  \I\{\cont_{t}= \cont_{(\ell')}, \rho_{j,t} = a_{k}, N_{k,t} \leq \max_{\ell'=1\dots \ell} m_{k}(\cont_{(\ell')})\}\\
      ~=~& \sum_{t > K} \I\{\cont_{t} \in \{\cont_{(\ell')}\}_{\ell'\in[\ell]}, \rho_{j,t} = a_{k}, N_{k,t} \leq \max_{\ell'=1\dots \ell} m_{k}(\cont_{(\ell')})\}\\
        ~\leq~& \max_{\ell'=1\dots \ell} m_{k}(\cont_{(\ell')})\,.
\end{align*}
For simplicity denote $L = |\contset_{n}/\contset(\underline{\slope})|$, and order contexts in $\cont \in \contset_{n}/\contset(\underline{\slope})$ as $\{\cont_{(\ell)}\}_{\ell \in [L]}$ s.t. 
$\Delta_{k}(\cont_{(1)}) > \Delta_{k}(\cont_{(2)}) > \dots >\Delta_{k}(\cont_{(L)})$, or equivalently 
$m_{k}(\cont_{(1)}) < m_{k}(\cont_{(2)}) < \dots  < m_{k}(\cont_{(L)})$ according to Eq.\eqref{eq:UCBdefs}. Then multiplying Eq. \eqref{eq:numbadarms} by 
by $\Delta_{k}(\cont_{(\ell)}) - \Delta_{k}(\cont_{(\ell+1)})$ (which is strictly positive due to the ordering of contexts), and summing $\ell = 1 \dots L$  we get 
\begin{align}
\begin{aligned}
\label{eq:slope}
    \sum_{\cont \in \contset_{n}/\contset(\underline{\slope})} \Delta_{k}(\cont) Y_{k}(\cont) &= \sum_{\ell \in [L]} \Delta_{k}(\cont_{(\ell)}) Y_{k}(\cont_{(\ell)})
    ~\leq~  \sum_{\ell \in [L]} m_{k}(\cont_{(\ell)}) \left( \Delta_{k}(\cont_{(\ell)}) -  \Delta_{k}(\cont_{(\ell+1)})\right)\\
    ~\overset{(i)}{=}~& 8\log(T)\sum_{\ell \in [L-1]} \frac{ \Delta_{k}(\cont_{(\ell)}) -  \Delta_{k}(\cont_{(\ell+1)})}{\Delta_{k}^{2}(\cont_{(\ell)})}
    ~\overset{(ii)}{\leq}~8\log(T) \int_{\Delta_{k}(\cont_{(L)})}^{\infty}\frac{dz}{z^{2}}\\
     ~=~& \frac{8\log(T)}{\Delta_{k}(\cont_{(L)})} ~\overset{(iii)}{=} \frac{8\log(T)}{\min_{\cont \in \contset_{n}/\contset(\underline{\slope})} \Delta_{k}(\cont)}\,.
     \end{aligned}
\end{align}
Here  (i) follows from the definition of $m_k(\bm{c})$ in Eq. \eqref{eq:UCBdefs} where $m_{k}(\cont) = \frac{8\log(T)}{\Delta_{k}^{2}(\cont)}$;  both (ii) and (iii) follow from the ordering of contexts so that $\Delta_{k}(\cont_{(1)}) > \Delta_{k}(\cont_{(2)}) > \dots >\Delta_{k}(\cont_{(L)})$. 
Note that for any $\cont \in \contset_{n}/\contset(\underline{\slope})$ and arm $k$ such that $a_{k} < r_{j,n-1}$, we have
\begin{align}
\label{eq:slope2}
\begin{aligned}
  \Delta_{k}(\cont) ~=~& \mathcal{L}_{j}(r_{j,n},\cont) -   \mathcal{L}_{j}(r_{j,n-1},\cont) + \mathcal{L}_{j}(r_{j,n-1},\cont) - \mathcal{L}_{j}(a_{k},\cont)\\
  ~>~&  \mathcal{L}_{j}(r_{j,n-1},\cont) - \mathcal{L}_{j}(a_{k},\cont)\\
  ~\overset{(i)}{\geq} ~&  \slope_{j,n-1}(\cont) (r_{j,n-1} - a_{k})\\
   ~\overset{(ii)}{\geq} ~& \left( \slope_{j,n-1}(\cont) -  \slope_{j,n}(\cont)\right) (r_{j,n-1} - a_{k})\\
    ~\overset{(iii)}{=} ~& (1+\lambda)\left(s_{j,n-1}-s_{j,n}\right) (r_{j,n-1} - a_{k})\\
     ~> ~& \left(s_{j,n-1}-s_{j,n}\right) (r_{j,n-1} - a_{k})\,,
\end{aligned}
\end{align}
where in (i) we recall the slope $\slope_{j,n-1}(\cont)$ is defined in Lemma \ref{lem:Vstruct} and further (i) follows from concavity of $ \mathcal{L}_{j}(\rho{j},\cont)$ in the first argument $\rho_{j}$; in (ii) we used the fact that $ \slope_{j,n}(\cont)\geq 0$ since the optimal budget 
$\arg\max_{\rho_{j}\in [0,\rho]} \mathcal{L}_{j}(\rho_{j}, \cont)$ is taken at the $n$th turning point, and is the largest turning point whose left slope is non-negative from  Lemma \ref{lem:Vstruct}; (iii) follows from the definition $\slope_{j,n'}(\cont) = (1+\lambda)s_{j,n'} - (\mu+\gamma\lambda)$ for any $n'$.

Finally combining Eqs. \eqref{eq:nbadslope}, \eqref{eq:slope} and \eqref{eq:slope2}, and summing over $n = 1 \dots S_{j}$ we get
\begin{align}
\begin{aligned}
\label{eq:X1secbound}
  & \sum_{t > K:\cont_{t} \notin \contset(\underline{\slope})} ~ \sum_{k\in[K]:a_{k} < \min \adj_{j}^{-}(\cont_{t})} ~  \Delta_{k}(\cont_{t}) \I\{\rho_{j,t} = a_{k}, N_{k,t} \leq m_{k}(\cont_{t})\}   \\
 ~=~& \sum_{n\in [S_{j}]} \sum_{t > K:\cont_{t} \in \contset_{n}/\contset(\underline{\slope})} ~ \sum_{k\in[K]:a_{k} < \min \adj_{j}^{-}(\cont_{t})} ~  \Delta_{k}(\cont_{t}) \I\{\rho_{j,t} = a_{k}, N_{k,t} \leq m_{k}(\cont_{t})\} \\
 ~\leq~& \sum_{n\in [S_{j}]}\left((1+\dualub_{F})s_{j,n-1} \discerr + \rho \underline{\slope}\right) T  + \sum_{n\in [S_{j}]}~\sum_{k\in[K]:a_{k} < r_{j,n-1}-\discerr} \frac{8\log(T)}{\left(s_{j,n-1}-s_{j,n}\right) (r_{j,n-1} - a_{k})}\\
 ~\overset{(i)}{\leq}~& \sum_{n\in [S_{j}]}\left((1+\dualub_{F})s_{j,n-1} \discerr + \rho \underline{\slope}\right) T  + \sum_{n\in [S_{j}]}~\sum_{\ell=1}^{K} \frac{8\log(T)}{\left(s_{j,n-1}-s_{j,n}\right) \ell \discerr}\\
 ~\leq~& \sum_{n\in [S_{j}]}\left((1+\dualub_{F})s_{j,n-1} \discerr + \rho \underline{\slope}\right) T  + \frac{8\log(T)\log(K)}{\discerr} \sum_{n\in [S_{j}]}~\frac{1}{\left(s_{j,n-1}-s_{j,n}\right)}\\
 ~=~& \widetilde{\mathcal{O}}(\discerr T + \underline{\slope}T + \frac{1}{\discerr})\,.
 \end{aligned}
\end{align}
Note that (i) follows because for all $a_{k} < r_{j,n-1} - \discerr$, the $a_{k}$'s distances from $r_{j,n-1}$ are at least $\discerr, 3\discerr, 2\discerr \dots $.  In the last equation, we hide all logarithmic factors using the notation $\widetilde{\mathcal{O}}$, and note that the constants $\dualub_{F}$, $(s_{j,n})_{n\in S_{j}}$, $S_{j}$ are all absolute constants that depend only on the support $F_{j}$ and corresponding sampling distribution $\bm{p}_{j}$ for value-cost pairs; see definitions of these absolute constants in Lemmas \ref{lem:Vstruct} and \ref{lem:dualvarub}.

\textbf{Step 3 and 4: $a_{k} \in \adj_{j}^{+}(\cont_{t})$ or $a_{k} > \max~ \adj_{j}^{+}(\cont_{t})$. } The cases where arm $a_k \in \adj_{j}^{+}(\cont_{t})$ and $a_{k} > \max \adj_{j}^{+}(\cont_{t})$
are symmetric to $a_k \in \adj_{j}^{-}(\cont_{t})$ and $a_{k} < \min \adj_{j}^{+}(\cont_{t})$, respectively, and we omit from this paper. 

Therefore, combining Eqs. \eqref{eq:X1near} and \eqref{eq:X1secbound} we  can conclude 
\begin{align}
    X_{1} \leq  \widetilde{\mathcal{O}}(\discerr T + \underline{\slope}T + \frac{1}{\discerr})\,.
\end{align}

\subsubsection{Bounding $X_{2}$.}
\label{subsubsec:X2}
We first rewrite $X_{2}$ as followed
\begin{align}
\label{eq:X2all}
\begin{aligned}
    X_{2} ~=~& \sum_{t > K:\cont_{t} \in \contset(\underline{\slope})} \sum_{k\in [K]} \Delta_{k}(\cont_{t}) \I\{\rho_{j,t} = a_{k}, N_{k,t} \leq m_{k}(\cont_{t})\} \\
     ~=~& \sum_{t > K}\sum_{n\in[S_{j}]}\sum_{k\in [K]}\sum_{\cont \in \contset_{n}\cap \contset(\underline{\slope})}  \Delta_{k}(\cont) \I\{\cont_{t}=\cont, \rho_{j,t} = a_{k}, N_{k,t} \leq m_{k}(\cont)\} \\
     ~\overset{(i)}{=}~& 
   \sum_{n\in[S_{j}]}\sum_{k\in [K]}\sum_{\cont \in \contset_{n}\cap \contset(\underline{\slope})} \Delta_{k}(\cont) Y_{k}(\cont)\\
    ~\overset{(ii)}{=}~&  \sum_{n\in[S_{j}]}~\sum_{\cont \in \contset_{n}\cap \contset(\underline{\slope})}~\sum_{k\in \{k_{n}^{-},k_{n}^{+}\}} \Delta_{k}(\cont) Y_{k}(\cont)+ \sum_{n\in[S_{j}]}~\sum_{\cont \in \contset_{n}\cap \contset(\underline{\slope})}~\sum_{k\in [K]/\{k_{n}^{-},k_{n}^{+}\}}\Delta_{k}(\cont) Y_{k}(\cont)\\
    ~\overset{(iii)}{\leq}~&  T \discerr (1+\dualub_{F}) \sum_{n\in[S_{j}]} \left(s_{j,n} + s_{j,n+1}\right) + 
   \sum_{n\in[S_{j}]}~\sum_{\cont \in \contset_{n}\cap \contset(\underline{\slope})}~\sum_{k\in [K]/\{k_{n}^{-},k_{n}^{+}\}} \Delta_{k}(\cont) Y_{k}(\cont)\,.
    \end{aligned}
\end{align}
where in (i) we define 
$Y_{k}(\cont) =  \sum_{t > K}\I\{\cont_{t} = \cont, \rho_{j,t} = a_{k}, N_{k,t} \leq m_{k}(\cont)\}$;  in (ii)
we separate out two arms  $k_{n}^{-}$ and $k_{n}^{+}$ defined as followed: recall for context $\cont \in \contset_{n}\cap \contset(\underline{\slope})$, the optimal budget $\arg\max_{\rho_{j}\in [0,\rho]} \mathcal{L}_{j}(\rho_{j}, \cont) = r_{j,n}$ is taken at the $n$th turning point per the definition of $\contset_{n}$ in Eq. \eqref{eq:UCBdefs}, and thereby we defined $k_{n}^{-}:= \max\{k\in [K]: a_{k} < r_{j,n}\}$ to be the arm closest to and no greater than  $r_{j,n}$, whereas  $k_{n}^{+}:= \min\{k\in [K]: a_{k} > r_{j,n}\}$ to be the arm closest to and no less than  $r_{j,n}$; in (iii), for small enough $\discerr < \min_{n'\in[S_{j}]}r_{j,n'} - r_{j,n'-1} $, we know that $k_{n}^{-}$ lies on the line segment between $r_{j,n-1}$ and $r_{j,n}$, so $\Delta_{k_{n}^{-}}(\cont) = \slope_{j}^{-}(\cont)(r_{j,n} - a_{k_{n}^{-}}) \leq \slope_{j}^{-}(\cont) \discerr \leq (1+\dualub_{F})s_{j,n-1} \discerr$, where in the final inequality follows from the definition of  $\slope_{j}^{-}(\cont) = \slope_{j,n}(\cont) = (1+\lambda)s_{j,n} - (\mu + \gamma\lambda) \leq (1+\lambda)s_{j,n} \leq (1+\dualub_{F})s_{j,n}$ where $\dualub_{F}$ is defined in Eq. \eqref{lem:dualvarub}. A similar bound holds for 
$\Delta_{k_{n}^{+}}(\cont)$.

Then, following the same logic as Eqs. \eqref{eq:numbadarms}, \eqref{eq:slope}, \eqref{eq:slope2} in Section \ref{subsubsec:X1} where we bound $X_{1}$, we can bound $\sum_{\cont \in \contset_{n}\cap \contset(\underline{\slope})} \Delta_{k}(\cont) Y_{k}(\cont)$ as followed for any arm $k\in [K]/\{k_{n}^{-},k_{n}^{+}\}$, i.e. arms who are at least $\discerr$ away from the optimal per-channel budget w.r.t. $\cont$:
\begin{align}
\label{eq:X2slope}
    \sum_{\cont \in \contset_{n}\cap \contset(\underline{\slope})} \Delta_{k}(\cont) Y_{k}(\cont) \leq \frac{8\log(T)}{\min_{\cont \in \contset_{n}\cap \contset(\underline{\slope})}\Delta_{k}(\cont)}\,.
\end{align}

Now, the set $k\in [K]/\{k_{n}^{-},k_{n}^{+}\}$ in Eq. \eqref{eq:X2all} can be further split into two subsets, namely $\{k\in [K]: a_{k} < r_{j,n} - \discerr\}$ and $\{k\in [K]: a_{k} > r_{j,n} + \discerr\}$ due to the definitions $k_{n}^{-}:= \max\{k\in [K]: a_{k} < r_{j,n}\}$ and $k_{n}^{+}:= \min\{k\in [K]: a_{k} > r_{j,n}\}$. Therefore, for any $k$ s.t. $ a_{k} < r_{j,n} - \discerr$ and any $\cont \in  \contset_{n}\cap \contset(\underline{\slope})$, 
\begin{align*}
    \Delta_{k}(\cont) = \mathcal{L}_{j}(r_{j,n},\cont) - \mathcal{L}_{j}(a_{k},\cont) \geq \slope_{j}^{-}(\cont) (r_{j,n} - a_{k}) \geq \underline{\slope}  (r_{j,n} - a_{k})\,,
\end{align*}
where the final inequality follows from the definition of $\contset(\underline{\slope})$ in Eq. \eqref{eq:UCBdefs} such that $ \slope_{j}^{-}(\cont) \geq \underline(\slope) $ for $\cont \in \contset(\underline{\slope})$. Hence combining this with Eq. \eqref{eq:X2slope} we have
\begin{align}
\label{eq:X2last}
    \sum_{k\in [K]:a_{k} < r_{j,n}- \discerr}~ \sum_{\cont \in \contset_{n}\cap \contset(\underline{\slope})} \Delta_{k}(\cont) Y_{k}(\cont) \leq  \sum_{k\in [K]:a_{k} < r_{j,n}- \discerr}\frac{8\log(T)}{ \underline{\slope}  (r_{j,n} - a_{k})} \overset{(i)}{\leq} \sum_{\ell=1}^{K}\frac{8\log(T)}{ \underline{\slope} \ell \discerr} \leq\frac{8\log(T)\log(K)}{\underline{\slope} \discerr}\,, 
\end{align}
where (i) follows because for all $a_{k} < r_{j,n} - \discerr$, the $a_{k}$'s distances from $r_{j,n-1}$ are at least $\discerr, 3\discerr, 2\discerr \dots $. Symmetrically, we can show an identical bound for the set  $\{k\in [K]: a_{k} > r_{j,n} + \discerr\}$. Hence, combining Eqs. \eqref{eq:X2all} and \eqref{eq:X2last} we can conclude 
\begin{align}
    X_{2}\leq \widetilde{\mathcal{O}}\left(\discerr T + \frac{1}{\discerr \underline{\slope}}\right).
\end{align}
Here, similar to our bound in Eq. \eqref{eq:X1secbound} for bounding $X_{1}$, we hide all logarithmic factors using the notation $\widetilde{\mathcal{O}}$, and note that the constants $\dualub_{F}$, $(s_{j,n})_{n\in S_{j}}$, $S_{j}$ are all absolute constants that depend only on the support $F_{j}$ and corresponding sampling distribution $\bm{p}_{j}$ for value-cost pairs; see definitions of these absolute constants in Lemma \ref{lem:Vstruct} and \ref{lem:dualvarub}.

\subsubsection{Bounding $X_{3}$.}
\label{subsubsec:X3}

We first define 
\begin{align}
    \Bar{\mathcal{L}} = \left(1+\gamma \right)\rho\dualub_{F} + (1+\dualub_{F})\Bar{\totv}
\end{align}
 where $\dualub_{F}$ is specified in Lemma \ref{lem:dualvarub}. Recalling the definition  $\Delta_{k}(\cont) =  \max_{\rho_{j}\in [0,\rho]}\mathcal{L}_{j}(\rho_{j}, \cont) -  \mathcal{L}_{j}(a_{k}, \cont) $ in Eq. \eqref{eq:UCBdefs}, and $-(1+\gamma)\rho \dualub_{F}\leq \mathcal{L}_{j}(\rho_{j},\cont)\leq (1+\dualub_{F})\Bar{\totv}$ for any $\rho_{j}\in[0,\rho]$ and context $\cont$
 (see Lemma \ref{lem:dualvarub}), it is easy to see 
 \begin{align}
 \label{eq:boundsuboptarmDelta}
     \Delta_{k}(\cont)\leq \Bar{\mathcal{L}}  \quad \forall k \in [K] , \forall \cont\,.
 \end{align}
 
 Then we bound $X_{3}$ as followed 
\begin{align}
\label{eq:X3first}
\begin{aligned}
   X_{3} ~=~&  \sum_{k\in [K]}\sum_{t > K} \E\left[\Delta_{k}(\cont) \I\{ \rho_{j,t} = a_{k}, N_{k,t} > m_{k}(\cont)\}\right] \\
    ~\overset{(i)}{\leq}~& \Bar{\mathcal{L}}\cdot \sum_{k\in[K]}\sum_{t > K} \prob\left(\rho_{j,t} = a_{k}, N_{k,t} > m_{k}(\cont_{t})\right)\\
    ~\overset{(ii)}{\leq}~& \Bar{\mathcal{L}} \cdot \sum_{k\in[K]} \sum_{t > K} \prob\Big(\Hat{\totv}_{j,t}(a_{k}) -\frac{\lambda_{t}\gamma + \mu_{t}}{1+\lambda_{t}} a_{k} + \ucb_{j,t}(a_{k}) \geq \Hat{\totv}_{j,t}(\rho_{j}^{*}(t))-\frac{\lambda_{t}\gamma + \mu_{t}}{1+\lambda_{t}} \rho_{j}^{*}(t)+ \ucb_{j,t}(\rho_{j}^{*}(t)),\\
    & \quad \quad N_{k,t} > m_{k}(\cont_{t})\Big)\,,
\end{aligned}
\end{align}
 where (i) follows from Eq. \eqref{eq:boundsuboptarmDelta}; in (ii), recall that we choose arm $\rho_{j,t} = a_{k}$ because the estimated UCB rewards of arm $a_{k}$ is greater than that of any other arm including $\rho_{j}^{*}(t)$ according to the UCB-SGD (Algorithm \ref{alg:omducb}), or mathematically,
 $\Hat{\totv}_{j,t}(a_{k}) -\frac{\lambda_{t}\gamma + \mu_{t}}{1+\lambda_{t}} a_{k} + \ucb_{j,t}(a_{k}) \geq \Hat{\totv}_{j,t}(\rho_{j}^{*}(t))-\frac{\lambda_{t}\gamma + \mu_{t}}{1+\lambda_{t}} \rho_{j}^{*}(t)+ \ucb_{j,t}(\rho_{j}^{*}(t))$.  Here we also used the fact that $\rho_{j}^{*}(t)$ lies in the arm set $\budgset(\discerr)$ for all $t$ (see Remark \ref{rmk:bestarm}).
 
Now let $\Hat{\ucbtot}_{n}(a_{k})$ denote the average conversion of arm $k$ over its first $n$ pulls, i.e. 
\begin{align}
    \Hat{\ucbtot}_{n}(a_{k}) =\Hat{\totv}_{j,\tau}(a_{k}) \text{ for }\tau = \min\{t\in [T]:N_{k,t} = n\}
\end{align}
 where we recall $\Hat{\totv}_{j,\tau}(a_{k})$ is the estimated conversion for arm $a_{k}$ in channel $j$ during period $\tau$ as defined in Algorithm \ref{alg:omducb}.
 In other words, $\tau$ is the period during which arm $a_{k}$ is pulled for the $n$th time so $ \Hat{\ucbtot}_{n}(a_{k})  = \Hat{\totv}_{j,\tau}(a_{k}) $.
 
 Hence, we continue with Eq. \eqref{eq:X3first} as followed:
\begin{align}
\label{eq:X3sec}
\begin{aligned}
&\prob\Big(\Hat{\totv}_{j,t}(a_{k}) -\frac{\lambda_{t}\gamma + \mu_{t}}{1+\lambda_{t}} a_{k} + \ucb_{j,t}(a_{k}) \geq \Hat{\totv}_{j,t}(\rho_{j}^{*}(t))- \frac{\lambda_{t}\gamma + \mu_{t}}{1+\lambda_{t}} \rho_{j}^{*}(t)+ \ucb_{j,t}(\rho_{j}^{*}(t)),~ N_{k,t} > m_{k}(\cont_{t})\Big)\\
    ~\leq~&  \prob \Big(\max_{n: m_{k}(\cont_{t}) < n \leq t} \Big\{\Hat{\ucbtot}_{n}(a_{k}) + \ucb_{n}(a_{k}) - \frac{\lambda_{t} \gamma + \mu_{t}}{1+\lambda_{t}}a_{k}\Big\} \\
    & \quad \quad \geq \min_{n': 1\leq n'\leq t} \Big\{\Hat{\ucbtot}_{n'}(\rho_{j}^{*}(t)) + \ucb_{n'}(\rho_{j}^{*}(t)) - \frac{\lambda_{t} \gamma + \mu_{t} }{1+\lambda_{t}}\rho_{j}^{*}(t)\Big\}
    \Big)\\
    ~\leq~& \sum_{n = \left\lceil m_{k}(\cont_{t}) \right\rceil+1}^{t}\sum_{n'=1}^{t}\prob\left(\Hat{\ucbtot}_{n}(a_{k}) + \ucb_{n}(a_{k}) - \frac{\lambda_{t} \gamma + \mu_{t} }{1+\lambda_{t}}a_{k} > \Hat{\ucbtot}_{n'}(\rho_{j}^{*}(t)) + \ucb_{n'}(\rho_{j}^{*}(t)) - \frac{\lambda_{t} \gamma + \mu_{t} }{1+\lambda_{t}}\rho_{j}^{*}(t)
    \right)
    \end{aligned}
\end{align}
Now, when the event $\left\{\Hat{\ucbtot}_{n}(a_{k}) + \ucb_{n}(a_{k}) - \frac{\lambda_{t} \gamma + \mu_{t} }{1+\lambda_{t}}a_{k} > \Hat{\ucbtot}_{n'}(\rho_{j}^{*}(t)) + \ucb_{n'}(\rho_{j}^{*}(t)) - \frac{\lambda_{t} \gamma + \mu_{t} }{1+\lambda_{t}}\rho_{j}^{*}(t)
    \right\}$ occurs, it is easy to see that one of the following events must also occur: 
\begin{align}
\label{eq:event}
\begin{aligned}
    & \mathcal{G}_{1,n} =\left\{  \Bar{\ucbtot}_{n}(a_{k})  \geq \totv(a_{k}) + \ucb_{n}(a_{k})\right\} \quad \text{ for } n \text{ s.t. } m_{k}(\cont_{t}) < n \leq t\\
    & \mathcal{G}_{2,n'} =\left\{  \Bar{\ucbtot}_{n'}(\rho_{j}^{*}(t))  \leq \totv(\rho_{j}^{*}(t)) -\ucb_{n}(\rho_{j}^{*}(t)) \right\}\quad  \text{ for } n' \text{ s.t. } 1\leq n' \leq t\\
    & \mathcal{G}_{3} =\left\{ \totv_{j}(\rho_{j}^{*}(t)) -  \frac{\lambda_{t} \gamma + \mu_{t} }{1+\lambda_{t}}\rho_{j}^{*}(t)~<~ \totv_{j}(a_{k}) - \frac{\lambda_{t} \gamma + \mu_{t} }{1+\lambda_{t}}a_{k}  + 2 \cdot \ucb_{n}(a_{k}) \right\}\\
    \end{aligned}
\end{align}
Note that for $n > m_{k}(\cont_{t})$, we have 
$\ucb_{n}(a_{k}) = \sqrt{\frac{2\log(T)}{n}} < \sqrt{\frac{2\log(T)}{m_{k}(\cont_{t})}} = \frac{\Delta_{k}(\cont_{t})}{2}$ since we defined $m_{k}(\cont) = \frac{8\log(T)}{\Delta_{k}^{2}(\cont)}$ in Eq. \eqref{eq:UCBdefs}. Therefore
\begin{align*}
    \totv_{j}(a_{k}) - \frac{\lambda_{t} \gamma + \mu_{t}}{1+\lambda_{t}}a_{k} + 2 \cdot \ucb_{n}(a_{k}) <  \underbrace{\totv_{j}(a_{k}) - \frac{\lambda_{t} \gamma + \mu_{t}}{1+\lambda_{t}}a_{k}}_{=\mathcal{L}(a_{k},\cont_{t})} + \Delta_{k}(\cont_{t}) ~\overset{(i)}{=}~ \underbrace{\totv_{j}(\rho_{j}^{*}(t)) -  \frac{\lambda_{t} \gamma + \mu_{t} }{1+\lambda_{t}}\rho_{j}^{*}(t)}_{=\mathcal{L}(\rho_{j}^{*}(t),\cont_{t}) = \max_{a\in \budgset(\discerr)}\mathcal{L}(a,\cont_{t})}
\end{align*}
where (i) follows from the definition of $\Delta_{k}(\cont) = \max_{a\in \budgset(\discerr)}\mathcal{L}(a,\cont) - \mathcal{L}(a_{k},\cont)$ in Eq. \eqref{eq:UCBdefs} for any context $\cont$. This implies that event $\mathcal{G}_{3}$ in Eq. \eqref{eq:event} cannot hold for $n >  m_{k}(\cont_{t})$. Therefore
\begin{align}
\label{eq:X3third}
    \prob\left(\Hat{\ucbtot}_{n}(a_{k}) + \ucb_{n}(a_{k}) - \frac{\lambda_{t} \gamma + \mu_{t} }{1+\lambda_{t}}a_{k} > \Hat{\ucbtot}_{n'}(\rho_{j}^{*}(t)) + \ucb_{n'}(\rho_{j}^{*}(t)) - \frac{\lambda_{t} \gamma + \mu_{t} }{1+\lambda_{t}}\rho_{j}^{*}(t)\right)~\leq~\prob\left(\mathcal{G}_{1,n} \cup \mathcal{G}_{2,n'}
    \right)\,.
\end{align}

From the standard UCB analysis and the Azuma Hoeffding's inequality, we have $\prob(\mathcal{G}_{1,n} )\leq \frac{\Bar{\totv}}{T^{4}}$ and  $\prob(\mathcal{G}_{2,n'} )\leq \frac{\Bar{\totv}}{T^{4}}$. Hence combining Eqs. \eqref{eq:X3first} \eqref{eq:X3sec}, \eqref{eq:X3third} we can conclude 

\begin{align}
\begin{aligned}
 X_{3} ~\leq~& \sum_{k\in[K]}\sum_{t > K}\sum_{n = \left\lceil m_{k}(\cont_{t}) \right\rceil+1}^{t}\sum_{n'=1}^{t}\left(\prob\left(\mathcal{G}_{1,n}\right) + \prob\left(\mathcal{G}_{2,n'}
    \right)\right)\\
    ~\leq~& \sum_{k\in[K]}\sum_{t > K}\ \sum_{n = \left\lceil m_{k}(\cont_{t}) \right\rceil+1}^{t}\sum_{n'=1}^{t} \frac{2\Bar{\totv}}{T^{4}}\\
     ~\leq~&  \frac{2K\Bar{\totv}}{T} = \mathcal{O}\left(\frac{1}{\discerr T}\right).
    \end{aligned}
\end{align}
\qed

\subsection{Proof for Theorem \ref{thm:regret}}
\label{pf:thm:regret}
Starting from Proposition \ref{prop:ubbenchlag}, we get
\begin{align}
\begin{aligned}
     & T\cdot \textsc{GL-OPT} - \E\left[\sum_{t\in [T]}\sum_{j\in[M]}\totv_{j}(\rho_{j,t})\right]\\
    ~\leq~& M\overline{\totv}(T-\tau_{A}) +  \sum_{j\in [M]}  
     \E\left[\sum_{t\in[\tau_{A}]}\mathcal{L}_{j}(\rho_{j}^{*}(t), \cont_{t}) - \mathcal{L}_{j}(\rho_{j,t}, \cont_{t})\right] +  \E\Big[\sum_{t\in[\tau_{A}]}\left(\lambda_{t}g_{1,t} + \mu_{t}g_{2,t}\right)\Big]\\
     ~\overset{(i)}{\leq}~&  M\overline{\totv}(T-\tau_{A}) + \mathcal{O}\left(\underline{\slope}T + \discerr T + \frac{1}{\underline{\slope} \discerr}\right) + \mathcal{O}\left(\eta T + \frac{1}{\eta}\right)
     \end{aligned}
\end{align}
where in (i) we applied Lemma \ref{lem:ucbregret} and \ref{lem:boundcs}. Taking $\eta = 1/\sqrt{T}$, $\discerr = \underline{\slope} = T^{-1/3}$ (i.e. $K = \mathcal{O}(T^{1/3})$ yields $ T\cdot \textsc{GL-OPT} - \E\left[\sum_{t\in [T]}\sum_{j\in[M]}\totv_{j}(\rho_{j,t})\right] \leq \mathcal{O}(T^{2/3})$. According to Lemma \ref{lem:Vstruct}, $\totv_{j}(\rho_{j})$ is concave for all $j \in [M]$, so
\begin{align}
\begin{aligned}
    \mathcal{O}(T^{-1/3})~\geq~& \textsc{GL-OPT} -  \frac{1}{T}\sum_{t\in [T]} \E\left[\sum_{j\in[M]}\totv_{j}(\rho_{j,t})\right] \\
    ~\geq~&  \textsc{GL-OPT}  - \E\left[ \sum_{j\in[M]}  \totv_{j}\left(\frac{1}{T}\sum_{t\in [T]} \rho_{j,t}\right)\right]\\
     ~\geq~&  \textsc{GL-OPT}  - \E\left[ \sum_{j\in[M]}\totv_{j}(\overline{\rho}_{j,T})\right]
     \end{aligned}
\end{align}
where in the final equality we used the definition $\Bar{\bm{\rho}}_{T}$ as defined in Algorithm \ref{alg:omducb}.

Regarding ROI constraint satisfaction, consider 
\begin{align}
\begin{aligned}
  0 ~\overset{(i)}{\leq}~& \frac{1}{T}\sum_{t\in[T]} \E\left[g_{1,t}\right]\\
    ~=~& \frac{1}{T}\sum_{t\in[T]}\sum_{j\in [M]}\E\left[\totv_{j}(\rho_{j,t};\bm{z}_{j,t}) - \gamma \rho_{j,t}\right]\\
     ~=~& \frac{1}{T}\sum_{t\in[T]}\sum_{j\in [M]}\E\left[\totv_{j}(\rho_{j,t}) - \gamma \rho_{j,t}\right]\\
    ~\overset{(ii)}{\leq}~&\sum_{j\in [M]}\E\left[\totv_{j}\left( \frac{1}{T}\sum_{t\in[T]}\rho_{j,t}\right) - \gamma \cdot \frac{1}{T}\sum_{t\in[T]}\rho_{j,t}\right]\\
     ~=~&\sum_{j\in [M]}\E[\totv_{j}\left(\overline{\rho}_{j,T}\right) - \gamma \overline{\rho}_{j,T}]\,.
     \end{aligned}
\end{align}
where (i) follows from Lemma \ref{lem:approxconst}; 
in (ii) we again applied concavity of $\totv_{j}(\rho_{j})$. We omit the analysis for the budget constraint as it is similar to the above.
\qed

\subsection{Additional Results for Section \ref{sec:online}}
\label{app:onlineadd}

\begin{proposition}
\label{prop:slater}
Assume Assumption \ref{ass:feas} holds, and recall $\bm{z}_{j}= (\bm{v}_{j},\bm{d}_{j}) \in F_{j}$ is any realization of values and costs for channel $j\in [M]$. 
Then,  for any channel $j \in [M]$, we have $\min_{\bm{z}_{j} \in F_{j}}\frac{v_{j,1}}{d_{j,1}} > \gamma  $, where we recall the ordering $\frac{v_{j,1}}{d_{j,1}} > \frac{v_{j,2}}{d_{j,2}} > \dots  > \frac{v_{j,m_{j}}}{d_{j,m_{j}}}$ for any element $\bm{z}_{j} = (\bm{v}_{j},\bm{d}_{j})\in F_{j}$ (see Section \ref{sec:online}). Further,  there exists some $\widetilde{\rho} \in (0,\rho)$ s.t. for any per-channel budget $\rho_{j} \leq \widetilde{\rho}$, we have  $ \totv_{j}(\rho_{j};\bm{z}_{j}) = \frac{v_{j,1}}{d_{j,1}} \rho_{j}  > \gamma  \rho_{j} $ for any $j\in [M]$.
\end{proposition}
\textit{Proof.} Under Assumption \ref{ass:feas}, it is easy to see for any realization of value-cost pairs  $\bm{z}_{j} = (\bm{v}_{j},\bm{d}_{j})$ there always exists an auction $n\in [m_{j}]$ whose value-to-cost ratio is at least $\gamma$, i.e. $v_{j,n} >  \gamma d_{j,n}$. Hence we know that 
$\frac{v_{j,1}}{d_{j,1}}\geq \frac{v_{j,n}}{d_{j,n}} > \gamma$. Now, in Eq. \eqref{eq:pieceline} within the proof of Lemma \ref{lem:Vstruct}, we showed 
\begin{align*}
   \totv_{j}(\rho_{j};\bm{z}_{j}) = \bm{v}_{j}^{\top} \bm{x}_{j}^{*}(\rho_{j};\bm{z}_{j})=  \sum_{n \in [m_{j}]}\left(\frac{v_{j,n}}{d_{j,n}}\rho_{j}+ b_{j,n}\right)\I\left\{d_{j,0}+ \dots +d_{j,n-1}\leq \rho_{j}\leq d_{j,0}+ \dots +d_{j,n} \right\}\,,
\end{align*}
where  $  d_{j,0} = v_{j,0} = b_{j,1} = 0$. This implies that for any $\rho_{j} < d_{j,1}$,  we have $\totv_{j}(\rho_{j} ; \bm{z}_{j}) = \frac{v_{j,1}}{d_{j,1}}\rho_{j} > \gamma \rho_{j} $. Therefore, we can take $\widetilde{\rho} = \min_{j\in [M]} \min_{\bm{z}_{j}\in F_{j}}d_{j,1}$, which ensures that for any $\rho_{j} \leq \widetilde{\rho}$ and realization $\bm{z}_{j}\in F_{j}$ we have  $\totv_{j}(\rho_{j} ; \bm{z}_{j}) = \frac{v_{j,1}}{d_{j,1}}\rho_{j} > \gamma \rho_{j}$ for any channel $j\in [M]$. 
\qed

\begin{lemma}[Constraint satisfaction]
\label{lem:approxconst}
Assume Assumption \ref{ass:feas} holds, and consider $\safe = \safebudg =  \frac{1}{\log(T)}$ in Algorithm \ref{alg:omducb}. Then, for large enough $T$ we have
\begin{align*}
     \frac{1}{T}\sum_{t\in[T]}g_{1,t} \geq 0 \quad \text{ and } \quad 
  \frac{1}{T}\sum_{t\in[T]}\sum_{j\in [M]}\rho_{j,t} \leq \rho\,, 
\end{align*}
where we recall $g_{1,t} = \sum_{j\in [M]}\left(\totv_{j}(\rho_{j,t};\bm{z}_{j,t}) - \gamma \rho_{j,t}\right)$. 
\end{lemma}

\textit{Proof.} Recall $\tau_{A} \in [T]$ defined in step 10 of Algorithm \ref{alg:omducb}.

If $\tau_{A} = T$, then we know that Algorithm \ref{alg:omducb} does not exit the while loop, and therefore $S_{1,t} - \gamma M\rho +\safe \safebudg (T-t)  \geq 0$ for $t = T$, or equivalently $S_{1,T}  \geq  \gamma M\rho  > 0$. Since we recall $S_{1,T} = \sum_{t\in[T-1]}g_{1,t}$, we can conclude that $\sum_{t\in[T]}g_{1,t} = S_{1,T} + g_{1,T} \geq  M\rho  + g_{1,T} \geq 0$ since $g_{1,T} \geq - \gamma M\rho$. Similarly, 
we also have $S_{2,t} + M\rho + \safebudg (T-t) \leq \rho T$ for $t = T$, or equivalently $S_{2,T} \leq \rho T -  M\rho $ where we used the fact that $\safebudg = 1/\log(T)< \rho$  for large enough $T$ and $M \geq 2$. Hence, recalling $S_{2,T} = \sum_{t\in[T-1]} \sum_{j\in [M]}\rho_{j,t}$, we can conclude that $ \sum_{t\in[T]} \sum_{j\in [M]}\rho_{j,t}  = S_{2,T} + \sum_{j\in [M]}\rho_{j,T}\leq \rho T -  M\rho + \sum_{j\in [M]}\rho_{j,T} \leq \rho T$ since $\sum_{j\in [M]}\rho_{j,T}\leq M\rho$.

If $\tau_{A} < T$, then we know that at the ``stopping time'' $\tau_{A}$, the while loop in  Algorithm \ref{alg:omducb} has  not yet exited, so we have 
\begin{align}
\label{eq:stopping}
        S_{1,\tau_{A}} - \gamma M\rho +\safe \safebudg (T-\tau_{A})  \geq 0 \text{ and } S_{2,\tau_{A}} + M\rho + M\safebudg (T-\tau_{A}) \leq \rho T
\end{align}
Hence, 
\begin{align}
\begin{aligned}
    \sum_{t\in[T]}g_{1,t} ~=~&   \sum_{t\in[\tau_{A}-1]}g_{1,t} + g_{1,\tau_{A}} + \sum_{t= \tau_{A}+1}^{T}g_{1,t} \\
    ~\overset{(i)}{\geq}~&  \gamma M\rho -\safe \safebudg (T-\tau_{A}) + g_{1,\tau_{A}} + \sum_{t= \tau_{A}+1}^{T}g_{1,t} \\
    ~\geq~&  \gamma M\rho -\safe \safebudg (T-\tau_{A}) - \gamma M\rho + \sum_{t= \tau_{A}+1}^{T}g_{1,t} \\
      ~\overset{(ii)}{=}~& -\safe \safebudg (T-\tau_{A}) + \sum_{t= \tau_{A}+1}^{T}\sum_{j \in [M]} \left(\totv_{j}(\safebudg;\bm{z}_{j,t}) -  \gamma \safebudg\right)\\
       ~\overset{(iii)}{\geq}~&   -\safe \safebudg (T-\tau_{A}) + \sum_{t= \tau_{A}+1}^{T}\sum_{j \in [M]} \left(\safebudg\cdot \min_{\bm{z}_{j}\in F_{j}} \frac{v_{j,1}}{d_{j,1}}-  \gamma \safebudg\right)\\
       ~=~&  -\safe \safebudg (T-\tau_{A}) +  (T-\tau_{A}) M   \left(\safebudg\cdot \min_{\bm{z}_{j}\in F_{j}} \frac{v_{j,1}}{d_{j,1}}-  \gamma \safebudg\right)\\
        ~\overset{(iv)}{\geq}~&  0
    \end{aligned}
\end{align}
where (i) follows from $ S_{1,\tau_{A}-1} = \sum_{t\in[\tau_{A}-2]}g_{1,t}$ and Eq. \eqref{eq:stopping}; (ii) follows from Algorithm \ref{alg:omducb} where we set $\rho_{j,t} = \safebudg$ for all $j \in [M]$ and $t = \tau_{A}+1 \ldots T$; for (iii), assuming the  $j$th channel's realized value cost pairs $\bm{z}_{j,t}$ is the element $\bm{z}_{j}\in F_{j}$, then Proposition \ref{prop:slater} says $\totv_{j}(\safebudg;\bm{z}_{j,t}) \ge  \frac{v_{j,1}}{d_{j,1}} \safebudg$ since $\safebudg = \frac{1}{\log(T)} < \widetilde{\rho}$ for large enough $T$. Hence $\totv_{j}(\safebudg;\bm{z}_{j,t}) \geq\min_{\bm{z}_{j}\in F_{j}} \frac{v_{j,1}}{d_{j,1}} \safebudg$; 
(iv) follows from the fact that $ \min_{\bm{z}_{j}\in F_{j}} \frac{v_{j,1}}{d_{j,1}} >  \gamma$  according to Proposition \ref{prop:slater}, so $ M\min_{\bm{z}_{j}\in F_{j}} \frac{v_{j,1}}{d_{j,1}} \geq M\gamma + \safe$ since $\safe = \frac{1}{\log(T)} \leq  M\min_{\bm{z}_{j}\in F_{j}} \frac{v_{j,1}}{d_{j,1}}- M\gamma $ for large enough $T$.

Similarly, we have 
\begin{align}
    \begin{aligned}
         \sum_{t\in[T]} \sum_{j\in [M]}\rho_{j,t} ~=~&  \sum_{t\in[\tau_{A}-1]} \sum_{j\in [M]}\rho_{j,t}+\sum_{j\in [M]}\rho_{j,\tau_{A}} +   \sum_{t= \tau_{A}+1}^{T} \sum_{j\in [M]}\rho_{j,t}\\
         ~\overset{(i)}{\leq}~& \rho T - M\rho - M \safebudg (T-\tau_{A})  +\sum_{j\in [M]}\rho_{j,\tau_{A}} +   M (T-\tau_{A}) \safebudg \\
           ~\leq~& \rho T - M\rho - M \safebudg (T-\tau_{A})  +M\rho +   M (T-\tau_{A}) \safebudg \\
         ~=~& \rho T 
    \end{aligned}
\end{align}
where (i) follows from $ S_{2,\tau_{A}} = \sum_{t\in[\tau_{A}-1]} \sum_{j\in [M]}\rho_{j,t}  $ and Eq. \eqref{eq:stopping}, as well as in Algorithm \ref{alg:omducb} we set $\rho_{j,t} = \safebudg$ for all $j \in [M]$ and $t = \tau_{A}, \tau_{A}+1 \ldots T$.
\qed

\begin{lemma}
\label{lem:omdtelescope}
    Let $(\lambda_{t},\mu_{t})_{t\in[T]}$ be the dual variables generated by Algorithm \ref{alg:omducb}. Then for any $\lambda,\mu \in [0,\dualub_{F}]$ and $t\in [T]$ we have
\begin{align}
\label{eq:CSbound}
\begin{aligned}
    & \sum_{\tau\in [t]} \left(\lambda_{\tau} - \lambda\right) g_{1,\tau} \leq  \frac{\eta M^{2}\Bar{\totv}^{2}}{2} \cdot t +  \frac{1}{2\eta}(\lambda- \lambda_{1})^{2}\\
    & \sum_{\tau \in [t]} \left(\mu_{\tau} - \mu\right) g_{2,\tau} \leq  \frac{\eta \rho^{2}}{2} \cdot t +  \frac{1}{2\eta}(\mu- \mu_{1})^{2}\,.
  \end{aligned}  
\end{align}
where we  recall $g_{1,\tau} = \sum_{j\in [M]}\left(\totv_{j,\tau}(\rho_{j,\tau}) - \gamma \rho_{j,\tau}\right)$ and $g_{2,\tau} = \rho - \sum_{j\in [M]}\rho_{j,\tau}$. 
\end{lemma}
\textit{Proof.}
We will show Eq. \eqref{eq:CSbound}. Starting with the first inequality w.r.t. $\lambda_{\tau}$'s, we have
\begin{align}
\label{eq:decompCS}
    \left(\lambda_{\tau} - \lambda\right) g_{1,\tau} =  \left(\lambda_{\tau+1} - \lambda\right) g_{1,\tau}  +  \left(\lambda_{\tau} - \lambda_{\tau+1}\right) g_{1,\tau} 
\end{align}
Since $\lambda_{\tau+1} = \Pi_{[0,\dualub_{F}]}\left(\lambda_{\tau} - \eta g_{1,\tau}\right)_{+} = \arg\min_{\lambda\in[0,\dualub_{F}]}\left(\lambda - \left(\lambda_{\tau} - \eta g_{1,\tau}\right)\right)^{2}$, we have
\begin{align}
    \left(\lambda_{\tau+1} - \left(\lambda_{\tau} - \eta g_{1,\tau}\right)\right)\cdot (\lambda - \lambda_{\tau+1}) \geq 0 \quad  \forall \lambda \in [0,\dualub_{F}]\,.
\end{align}
So we have
\begin{align}
\begin{aligned}
     \left(\lambda_{\tau+1} - \lambda\right) g_{1,\tau}~\leq~ & \frac{1}{\eta}(\lambda_{\tau+1} - \lambda_{\tau})\cdot(\lambda - \lambda_{\tau+1})\\
     ~=~& \frac{1}{2\eta}\left((\lambda- \lambda_{\tau})^{2} - (\lambda- \lambda_{\tau+1})^{2}-(\lambda_{\tau+1}- \lambda_{\tau})^{2}\right)\,.
     \end{aligned}
\end{align}
Plugging the above back into Eq. \eqref{eq:decompCS} we get
\begin{align}
\label{eq:omdtelescope}
\begin{aligned}
     \left(\lambda_{\tau} - \lambda\right) g_{1,\tau} ~\leq~&  \left(\lambda_{\tau} - \lambda_{\tau+1}\right) g_{1,\tau}  +  \frac{1}{2\eta}\left((\lambda- \lambda_{\tau})^{2} - (\lambda- \lambda_{\tau+1})^{2}-(\lambda_{\tau+1}- \lambda_{\tau})^{2}\right)\\
     ~\leq~&  
   \frac{\eta}{2}g_{1,\tau}^{2}  +  \frac{1}{2\eta}\left((\lambda- \lambda_{\tau})^{2} - (\lambda- \lambda_{\tau+1})^{2}\right)\\
    ~\leq~&  
   \frac{\eta M^{2}\Bar{\totv}^{2}}{2}  +  \frac{1}{2\eta}\left((\lambda- \lambda_{\tau})^{2} - (\lambda- \lambda_{\tau+1})^{2}\right)\,,
   \end{aligned}
\end{align}
where the final inequality follows from the fact that $\totv_{j,\tau}(\rho_{j,\tau})\leq \Bar{\totv}$ for any $j\in [M]$ and $\tau\in [t]$ so $g_{1,\tau}\leq M\Bar{\totv}$. Summing the above over $\tau =1\dots t$ and telescoping we get
\begin{align*}
    \sum_{\tau \in [t]} \left(\lambda_{\tau} - \lambda\right) g_{1,\tau} \leq  \frac{\eta M^{2}\Bar{\totv}^{2}}{2} \cdot t +  \frac{1}{2\eta}(\lambda- \lambda_{1})^{2}
\quad \text{ for } \forall \lambda \in [0,\dualub_{F}]\,.
\end{align*}
Following the same arguments above we can show 
\begin{align*}
   \sum_{\tau \in [t]} \left(\mu_{\tau} - \mu\right) g_{2,\tau} \leq  \frac{\eta \rho^{2} }{2} \cdot T +  \frac{1}{2\eta}(\mu- \mu_{1})^{2}
   \quad \text{ for } \forall \mu \in [0,\dualub_{F}]\,.
\end{align*} 
\qed

\begin{proposition}
\label{prop:convexprob}
    Under Assumption \ref{ass:feas}, the advertiser's per-channel only budget optimization problem, namely $\textsc{CH-OPT}(\mathcal{I}_{\budg})$ is a convex problem.
\end{proposition}
\text{Proof.}
Recalling the $\textsc{CH-OPT}(\mathcal{I}_{\budg})$ in Eq. \eqref{eq:chopt} and the definition of $\mathcal{I}_{\budg}$ in Eq. \eqref{eq:option}, we can write $\textsc{CH-OPT}(\mathcal{I}_{\budg})$ as
\begin{align}
\begin{aligned}
   \textsc{CH-OPT}(\mathcal{I}_{\budg})=  \max_{(\gamma_{j})_{j\in [M]}\in \mathcal{I}} ~&~ \sum_{j \in M} \totv_{j}(\rho_{j})\\
    \text{s.t. } &  \sum_{j \in M} \totv_{j}(\rho_{j}) \geq \gamma \sum_{j \in M} \rho_{j}\\
   &   \sum_{j\in [M]}\rho_{j}\leq \rho\,,
    \end{aligned}
\end{align}
Here we used the definition $\totv_{j}(\rho_{j}) = \E\left[\totv_{j}(\rho_{j};\bm{z}_{j})\right]$ in Eq. \eqref{eq:budperchannel}, and $\totc_{j}(\rho_{j};\bm{z}_{j}) = \rho_{j}$ for any $\bm{z}_{j}$ under Assumption \ref{ass:feas}. According to Lemma \ref{lem:Vstruct}, $\totv_{j}(\rho_{j})$ is concave in $\rho_{j}$ for any $j$, so the objective of $\textsc{CH-OPT}(\mathcal{I}_{\budg})$ maximizes a concave function. For the feasibility region, assume $\rho_{j}$ and $\rho_{j}'$ are feasible, then defining $\rho_{j}'' = \theta \rho_{j} +  (1-\theta)\rho_{j}'$ for any $\theta \in [0,1]$, we know that 
\begin{align*}
\begin{aligned}
     \sum_{j \in M} \left(\totv_{j}(\rho_{j}'') - \gamma  \rho_{j}''\right) 
     ~\overset{(i)}{\geq}~ &  \sum_{j \in M} \left(\theta\totv_{j}(\rho_{j})  +(1-\theta)\totv_{j}(\rho_{j}')  - \gamma  \rho_{j}''\right)\\
      ~=~ &  \theta\sum_{j \in M} \left(\totv_{j}(\rho_{j}) - \gamma  \rho_{j}\right) + (1-\theta)
     \sum_{j \in M} \left(\totv_{j}(\rho_{j}') - \gamma  \rho_{j}'\right)\\
       ~\overset{(ii)}{\geq}~ & 0
       \end{aligned}
\end{align*}
where (i) follows from concavity of $\totv_{j}(\rho_{j})$ and (ii) follows from feasiblity of $\rho_{j}$ and $\rho_{j}'$. On the other hand it is apparent that $\sum_{j\in [M]}\rho_{j}'' \leq\rho$. Hence we conclude that for any $\rho_{j}$ and $\rho_{j}'$  feasible,  $\rho_{j}'' = \theta \rho_{j} +  (1-\theta)\rho_{j}'$  is also feasible, so the feasible region of 
$\textsc{CH-OPT}(\mathcal{I}_{\budg})$ is convex. This concludes the statement of the proposition. \qed

\section{Proofs for Section \ref{sec:multi}}
\subsection{Proof of Lemma \ref{lem:Vstructmulti}}
\label{pf:lem:Vstructmulti}
Before we show the lemma, we first show the following claim is true:

\begin{claim}
\label{cl:ordering}
 Recall $v_{j,n}(1) > \ldots > v_{j,n}(L_{j,n}) > 0$  and $d_{j,n}(1) > \ldots > d_{j,n}(L_{j,n}) > 0$ for any channel $j\in [M]$ and auction $n\in [m_{j}]$. If auction $n$ in channel $j$ has increasing marginal values, i.e. for any realization $\bm{z}_{j} = (\bm{v}_{j},\bm{d}_{j})$,  for any $n\in [m_{j}]$ we have $\frac{v_{j,n}(\ell-1) - v_{j,n}(\ell)}{d_{j,n}(\ell-1) - d_{j,n}(\ell)}$ decreases in $\ell$ then $\frac{v_{j,n}(\ell)}{d_{j,n}(\ell)}$ also decreases in $\ell$.
\end{claim}
\textit{Proof.}
We prove this claim by induction. The base case is $\ell = L_{j,n}$:
it is easy to see 
\begin{align*}
    \frac{v_{j,n}(L_{j,n}-1) - v_{j,n}(L_{j,n})}{d_{j,n}(L_{j,n}-1) - d_{j,n}(L_{j,n})} > \frac{v_{j,n}(L_{j,n})}{d_{j,n}(L_{j,n})} \Longrightarrow  \frac{v_{j,n}(L_{j,n}-1)}{d_{j,n}(L_{j,n}-1)} > \frac{v_{j,n}(L_{j,n})}{d_{j,n}(L_{j,n})}\,. 
\end{align*}
Now assume the induction hypothesis $\frac{v_{j,n}(\ell)}{d_{j,n}(\ell)} > \frac{v_{j,n}(\ell+1)}{d_{j,n}(\ell+1)} > \dots > \frac{v_{j,n}(L_{j,n})}{d_{j,n}(L_{j,n})} $. Then, we have 
\begin{align}
\label{eq:ordering}
\begin{aligned}
   \frac{v_{j,n}(\ell)}{d_{j,n}(\ell)} > \frac{v_{j,n}(\ell+1)}{d_{j,n}(\ell+1)} \Longrightarrow  & \frac{d_{j,n}(\ell+1) - d_{j,n}(\ell)}{d_{j,n}(\ell)} > \frac{v_{j,n}(\ell+1)-v_{j,n}(\ell)}{v_{j,n}(\ell)}\\
   \Longrightarrow  & \frac{d_{j,n}(\ell) - d_{j,n}(\ell+1)}{d_{j,n}(\ell)} < \frac{v_{j,n}(\ell)-v_{j,n}(\ell+1)}{v_{j,n}(\ell)}\\
    \Longrightarrow  & \frac{v_{j,n}(\ell)}{d_{j,n}(\ell)} < \frac{v_{j,n}(\ell)-v_{j,n}(\ell+1)}{d_{j,n}(\ell) - d_{j,n}(\ell+1) }\,.
    \end{aligned}
\end{align}
Since $\frac{v_{j,n}(\ell-1) - v_{j,n}(\ell)}{d_{j,n}(\ell-1) - d_{j,n}(\ell)}$ decreases in $\ell$  we have 
\begin{align*}
    & \frac{v_{j,n}(\ell-1) - v_{j,n}(\ell)}{d_{j,n}(\ell-1) - d_{j,n}(\ell)} > \frac{v_{j,n}(\ell) - v_{j,n}(\ell+1)}{d_{j,n}(\ell) - d_{j,n}(\ell+1)} \overset{(i)}{>} \frac{v_{j,n}(\ell)}{d_{j,n}(\ell)}\\
      \Longrightarrow  & \frac{v_{j,n}(\ell-1)}{d_{j,n}(\ell-1)} \overset{(ii)}{>} \frac{v_{j,n}(\ell)}{d_{j,n}(\ell)}\,,
\end{align*}
where (i) follows from Eq. \eqref{eq:ordering}, and (ii) follows from the fact that $\frac{A}{B} > \frac{C}{D}$ for $A,B,C,D > 0$ implies $\frac{A+C}{B+D} > \frac{C}{D}$ where we let $A = v_{j,n}(\ell-1) - v_{j,n}(\ell)$, $B = d_{j,n}(\ell-1) - d_{j,n}(\ell)$, $C = v_{j,n}(\ell)$ and $D = d_{j,n}(\ell)$. This concludes the proof.
\halmos

Now we prove Lemma \ref{lem:Vstructmulti}. Similar to the proof of Lemma \ref{lem:Vstruct}, we only need to show for any realization $\bm{z}_{j} = (\bm{v}_{j},\bm{d}_{j})_{j\in [M]}$, the conversion function $\totv_{j}^{+}(\rho_{j};\bm{z}_{j}) = \bm{v}_{j}^{\top} \bm{x}_{j}^{*,+}(\rho_{j};\bm{z}_{j})$ where $\bm{x}_{j}^{*,+}(\rho_{j};\bm{z}_{j})$ is defined as Eq. \eqref{eq:budperchannelsolmulti} is piecewise linear, continuous, strictly increasing and concave.

For simplicity we use the shorthand notation $\bm{x}_{j}^{*} = \bm{x}_{j}^{*,+}(\rho_{j};\bm{z}_{j})\in [0,1]^{\sum_{n\in [m_{j}]}L_{j,n}}$ as the optimal solution to $\totv_{j}^{+}(\rho_{j};\bm{z}_{j})$, defined in Eq. \eqref{eq:budperchannelsolmulti}. By re-labeling the auction indices in channel $j\in [M]$ such that
$\frac{v_{j,1}(1)}{d_{j,1}(1)} > \frac{v_{j,2}(1)}{d_{j,2}(1)} > \dots  > \frac{v_{j,m_{j}}(1)}{d_{j,m_{j}}(1)}$, we claim that $\bm{x}_{j}^{*} $ takes the following form:
\begin{align}
\label{eq:multilprelax}
    x_{j,n}^{*}(\ell) = 
    \begin{cases}
    1 & \text{if } \ell = 1 \text{ and } \sum_{n' \in [n]} d_{j,n'}(1) \leq \rho_{j}\\
    \frac{\rho_{j} - \sum_{n' \in [n-1]} d_{j,n'}(1)}{d_{j,n}(1)} & \text{if } \ell = 1 \text{ and } \sum_{n' \in [n]} d_{j,n'}(1) > \rho_{j}\\
    0 & \text{otherwise}
    \end{cases}
\end{align}
which is analogous to that of Eq. \eqref{eq:knapsack} in the proof of  Lemma \ref{lem:Vstruct}. In other words, in the optimal solution, an advertiser would only procure impressions who are in the first position in each auction, and also those with high value-to-cost ratios.
With the above representation of $\bm{x}_{j}^{*} $, the rest of the proof follows exactly from that for Lemma \ref{lem:Vstruct}.

It now remains to show that Eq. \eqref{eq:multilprelax} holds. 
We first argue by contradiction that in any auction, no impression other than the first would get procured, i.e. $x_{j,n}^{*}(\ell) = 0$  for any $\ell \in 2 \ldots L_{j,n}$.  Assume there exists some auction $n \in [m_{j}]$ and impression slot $\ell' \in 2 \ldots L_{j,n}$ such that
$x_{j,n}^{*}(\ell') > 0$, then by the constraint that at most 1 impression can be procured, i.e. $\sum_{\ell \in [L_{j,n}]}x_{j,n}^{*}(\ell) \leq 1$ in Eq. \eqref{eq:budperchannelsolmulti}, we know that $x_{j,n}^{*}(1) < 1$. Also, note that $x_{j,n}^{*}(\ell')$ incurs a cost of $d_{j,n}(\ell') \cdot x_{j,n}^{*}(\ell')$ amongst the total per-channel budget $\rho_{j}$. If we instead use this 
cost on the first impression,  then we will obtain a value increase of 
\begin{align*}
    v_{j,n}(1)\cdot \frac{d_{j,n}(\ell') \cdot x_{j,n}^{*}(\ell')}{d_{j,n}(1)} - v_{j,n}(\ell') \cdot x_{j,n}^{*}(\ell') = d_{j,n}(\ell')\cdot x_{j,n}^{*}(\ell') \cdot \left( \frac{ v_{j,n}(1)}{d_{j,n}(1)} - \frac{ v_{j,n}(\ell')}{d_{j,n}(\ell')}\right)
> 0,
\end{align*}
where the final inequality follows from the assumption that $x_{j,n}^{*}(\ell') > 0$, and the multi-item auction has increasing marginal values (see Definition \ref{def:incrmarginal}) so Claim \ref{cl:ordering} holds. This contradicts the optimality of $\bm{x}_{j}^{*}$, and hence $x_{j,n}^{*}(\ell) = 0$  for any $\ell \in 2 \ldots L_{j,n}$, or in other words, a channel will only procure impressions ranked first. Hence, a channel's procurement problem in Eq. \eqref{eq:budperchannelsolmulti} can be restricted to the first impression in each auction, and thus similar to the proof of Lemma \ref{lem:Vstruct}, is an LP-relaxation to the 0-1 knapsack with budget $\rho_{j}$, and $m_{j}$ items whose values are $v_{j,1}(1)\ldots v_{j,m_{j}}(1) $ with costs $d_{j,1}(1)\ldots d_{j,m_{j}}(1)$.  
\halmos

\end{APPENDICES}

\end{document}